\documentclass[A4paper, floatfix,5p,10pt,twocolumn,preprint]{elsarticle}
\usepackage{amsmath}
\usepackage{graphicx}%
\usepackage{dcolumn}
\usepackage{hyperref}
\usepackage{subcaption}
\usepackage[normalem,normalbf]{ulem}
\usepackage{enumerate}
\usepackage{tikz}
\usepackage{algpseudocode}
\usepackage{algorithm}
\usepackage{multirow}

\journal{{}}
\usetikzlibrary{arrows,chains,matrix,positioning,scopes}
\makeatletter
\tikzset{join/.code=\tikzset{after node path={%
\ifx\tikzchainprevious\pgfutil@empty\else(\tikzchainprevious)%
edge[every join]#1(\tikzchaincurrent)\fi}}}
\makeatother
\tikzset{>=stealth',every on chain/.append style={join},
         every join/.style={->}}
\tikzstyle{labeled}=[execute at begin node=$\scriptstyle,
   execute at end node=$]
\def\checkmark{\tikz\fill[scale=0.4](0,.35) -- (.25,0) -- (1,.7) -- (.25,.15) -- cycle;} 

\newcommand{\be}{\begin{equation}}
\newcommand{\ee}{\end{equation}}
\newcommand{\ve}[1]{\boldsymbol{#1}}
\newcommand{\Eqn}[1]{Eqn. (\ref{#1})}
\newcommand{\E}{\mathrm{E}}
\newcommand*{\figuretitle}[1]{%
    {\centering
    \textbf{#1}
    \par\medskip}
}

\begin{document}
\title{Learning zero-cost portfolio selection with pattern matching}
\author[csam-wits,querilab]{Tim Gebbie}
\ead{tim.gebbie@wits.ac.za}
\author[csam-wits,querilab]{Fayyaaz Loonat}
\address[csam-wits]{School of Computer Science and Applied Mathematics, University of the Witwatersrand, Johannesburg, South Africa}
\address[querilab]{QuERILab - Quantifying Emergence, Risk and Information}

\begin{abstract}
We consider and extend the adversarial agent-based learning approach of  Gy{\"o}rfi {\it et al} to the situation of zero-cost portfolio selection implemented with a quadratic approximation derived from the mutual fund separation theorems. The algorithm is applied to daily sampled sequential Open-High-Low-Close data and sequential intraday 5-minute bar-data from the Johannesburg Stock Exchange (JSE). Statistical tests of the algorithms are considered. The algorithms are directly compared to standard NYSE test cases from prior literature. The learning algorithm is used to select parameters for agents (or experts) generated by pattern matching past dynamics using a simple nearest-neighbour search algorithm. It is shown that there is a speed advantage associated with using an analytic solution of the mutual fund separation theorems. It is argued that the expected loss in performance does not undermine the potential application to intraday quantitative trading and that when transactions costs and slippage are considered the strategies can still remain profitable when unleveraged. The paper demonstrates that patterns in financial time-series on the JSE can be systematically exploited in collective but that this does not imply predictability of the individual asset time-series themselves.
\end{abstract}

\begin{keyword}
online learning \sep pattern matching \sep portfolio control \sep algorithmic portfolio selection
\PACS 89.65.Gh \sep 02.50.Ey
\MSC 91-04 \sep 91G10 \sep 91G80
\JEL G11 G14 G17 055
\end{keyword}

\maketitle

\section{Introduction}	
Sequential investment strategies aim to facilitate portfolio control decisions by collecting information from past behaviour and states of the market and using this information to deploy capital across a selection of assets in a manner the can generate consistent wealth maximization over the long-term \cite{AC1988,C1991,GUW2008}. 

The intention of the paper is not to find a profitable trading strategy for quantitative trading but to show that such strategies exists by providing a simple, transparent and easily recoverable example in the domain of unleveraged zero-cost portfolio selection for statistical arbitrage.

Here we make no specific assumptions relating to the nature of price processes for the sake of the algorithms, however, the approach is broadly based on prior mathematical analysis that use assumptions of stationarity and ergodicity of the price increments in order to allow the study of asymptotic growth rates. In particular to ensure that such growth rates have well-defined maxima when full knowledge of the distribution and its process have been achieved \cite{AC1988,C1991,CO1996,GUW2008}.  

We investigate the idea that by using pattern matching algorithms (where the patterns are unspecified) combined with learning algorithms, based on some purpose, such as wealth maximisation irrespective or risk \cite{K1956,AC1988}, we can:
\begin{enumerate}
  \item Beat a cash portfolio in the context of a self-funding strategy, a zero-cost portfolio strategy, and that 
  \item We can beat the best stock in the market \cite{BEG2004}.
\end{enumerate} 
The latter has been shown to be the case in prior literature, by investigating daily sampled stock data from the NYSE for long-only (fully invested) portfolio strategies \cite{AC1988,C1991,BK1999,BEG2004,GUV2007}.  Here we consider both of these cases: zero-cost, and fully invested strategies, in the context of the South African stock market, the Johannesburg Stock Exchange (JSE), and do so for both daily sampled data and intraday data.

The approach here should not be confused with questioning the value of technical analysis where pre-specified patterns, in the form of some sort of library or set of rules, are used to try to generate systematic wealth \cite{LMW2000}. We are considering the problem of probing phenomenology aimed at understanding financial markets as a complex adaptive system \cite{A1995, WGHC2014}. More specifically, we are considering the modelling of time-series arising from complex adaptive systems, something more closely aligned with the context of nonlinear dynamical systems thinking \cite{C2011}. The question of finding evidence of structure, as opposed to randomness, in financial time-series data, but beyond evidence of long-term memory or typical stylised facts \cite{WG2008}. We argue that we are not trying to show that specific patterns exist, that such pattens are predictable, but rather that the interaction of a purposeful agent with a stock market using pattern-matching can generate wealth that would not be expected from a typical null-hypothesis of geometric Brownian motion. 

We are specifically not looking for statistically preserved properties of time-series, in the sense of time-series models, but are rather looking for evidence of statistical repeating structures in time-series, but without {\it a-priori} ability to know the form that the structure will take, perhaps because of the nonlinearity of the system in question \cite{B1998,SSC2002}. 

We are seeking indirect evidence of structure by showing that a purposeful agent can learn to make investment decisions \cite{WGHC2014}, in a positivist manner, by looking for {\it a-priori} unspecified and unknown patterns in the data, that can be purposefully exploited, sequentially and systematically, to generate wealth in excess of that expected by randomness and the related normative perspectives of the functioning of financial markets. This is not in itself new, there is a rich literature on attempts at probing the predictability of this or that financial time-series. What can be considered controversial is the view that fairly naive computational learning agents can generate wealth within the system without special insights or understanding of the system itself\footnote{This view benefited from conversations with D Hendricks and D Wilcox}. 

By extracting positive growth rates in the excess of the performance of the best stock by using unleveraged combinations of underlying stocks over long periods of time this can be taken as building the case that there are indeed patterns, or some sort of structure, that almost repeat though time in a manner that their occurrence can be treated as exploitable information in collective. This has been shown to be the case for long-only portfolio's \cite{AC1988,C1991,BK1999,BEG2004,GUV2007}. We show this for self-funding strategies; zero-cost portfolio's.

To achieve this we construct sequential investment strategies based on pattern matching and demonstrate that these strategies can generate positive growth rates in excess of the best stocks in the investment universe, and substantial positive growth rates for zero-cost strategies in excess of that expected from investment in cash or risk-free assets.

We do not address the question of whether it is risk that the investor is being compensated for, or even whether the strategies we are isolating are in fact statistical arbitrages, in the sense that the strategies long-term volatility tending to zero in conjunction with an always positive probability of positive performance at zero initial cost \cite{JTTW2012}.

The appearance of patterns and organisation is a fundamental property of complex adaptive systems \cite{C2011}. Looking directly for pockets of predictability in complex dynamical systems \cite{PCFS1980} as an approximation to modelling complexity adaptive systems \cite{C2011} is notoriously difficult given the intricacies of noise and nonlinearity \cite{CEFG1991, SYC1991}. Coupling purpose, via a learning criterion, here wealth maximisation irrespective of risk, to the selection for patterns, in order to achieve the stated purpose, is the approach promoted here. 

It is in this sense that we built a framework that extracts pockets of predictability, if they exist, via pattern searching, ideally in an online manner, in order to increase our agents wealth irrespective of risk, but specifically in the situation where the form of the patterns are always unknown, changing and dynamic, but are represented in collective past histories of the system components. 

In Section \ref{sec:online-learning} we present the agent-based learning algorithm as an extension of prior work  \cite{K1956,AC1988,C1991,CO1996} and \cite{GUV2007,GUW2008}. The contributions here are: (i) the algorithm is explicitly re-written in online form in order to make near-real-time applications tractable, (ii) the algorithms are modified for application to the zero-cost portfolio selection problem using the mutual fund separation theorems \cite{M1995,L2000}, (iii) the algorithms are explicitly tested, using synthetic data, real daily data both from the NYSE and JSE, and for JSE intraday 5-minute bar-data.

Section \ref{sec:agent-algorithms} describes the approach we have adopted for the generation of experts or agents modified for use in zero-cost portfolio strategies. The algorithm parameters are not tuned prior to use but are left to the online-learning algorithm to select. 

In Section \ref{ssec:pattern-matching}, we consider strategies that target predictable patterns using a simple modified version of the nearest-neighbour pattern-matching strategy developed by \cite{GUW2008}. 

As in the case of the learning algorithm, the agent-generation algorithms have been modified in principle: (i) to support offline and online algorithm use, (ii) they are explicitly framed for use with zero-cost portfolio selection problems, and (iii) portfolio optimizations have been replaced with analytic quadratic approximations in order to improve execution times. 

In order to have true online pattern matching the algorithms would have to replaced with either look-up-tables built off-line or a hybrid method that combines offline building of the history of the agents performance and then an almost online method that updates that cached history of agents performance across parameters as the data arrives sequentially in real-time. 

Section \ref{sec:data} provides an overview of the data used in the various numerical experiments. 

The data is sequential and uniformly sampled and takes on the form of open-high-low-close (OHLC) data, this is described in Section \ref{ssec:ohlc-data}. The use of open, high, low and close data combinations for the daily data testing can be carried over for intraday studies, and the use of close prices is a special case.

The synthetic data is described in Section \ref{ssec:synthetic-data} along with the algorithm testing strategy. Briefly, a simple Kolmogorov-Smirnov test is adopted to assess algorithm behaviour across 4 test cases: 
\begin{enumerate}
\item SDC1: log-normal random data with zero-means, where no learning should be possible, 
\item SDC2: log-normal random data where all assets have the same positive mean and as such basic learning is not possible for zero-cost portfolios (portfolios that have long and short positions that sum to zero),  
\item SDC3: log-normal random data with varying positives means, and 
\item SDC4: where we have log-normal data with both positive and negative means with the same fixed variance.
\end{enumerate}

The synthetic data is used to understand and prove the behaviour of the zero-cost portfolio strategy (which we will call active portfolio's) and the fully-invested portfolio strategy (which we will call absolute portfolio's).  

The four real-world data sets are described in Section \ref{ssec:real-data}:
\begin{enumerate}
  \item The standard daily sampled test-data set for the NYSE \cite{AC1988,C1991,BK1999,BEG2004,GUV2007},
  \item A more extensive, merged, daily sampled test-data set for the NYSE \cite{NYSE2006}, 
  \item A daily sampled test-data set for the JSE, and
  \item An intraday test-data set for the JSE.
 \end{enumerate}

A general overview of the implementation of the numerical experiments is addressed in Section \ref{sec:imp-issues}.

Section \ref{sec:results-analysis} describes the results and analysis of the results, first the synthetic data in section \ref{ssec:SDC1234} and then for the real-world data, in Sections \ref{ssec:NYSE}, \ref{ssec:mergedNYSE}, \ref{ssec:dailyJSE}, and \ref{ssec:intradayJSE}, respectively for the four real world case studies: NYSE, extended merge NYSE, daily sampled JSE and intraday JSE.

\section{An online-learning algorithm for portfolio selection} 
\label{sec:online-learning}
The application is for a set of stocks ordered in time where each agent will consider different combinations of stocks for each time-period based on features and strategy parameters. These different agents compete in an adversarial manner in competition for capital allocations \cite{K1956,C1991,CO1996,GUW2008}. Here agents with poor performance will have incremental capital allocations reduced and agents with robust performance will have incremental increases in capital allocation. Better performing agents will over time have their relative contribution to the aggregate portfolio increased so that their decisions are preferentially selected for trade at the onset of each trading or investment period based on information available at the end of the prior trading period.

The online learning algorithm takes as inputs: a set of agents controls, and performances. These are enumerated over features (here price-relatives) and free-parameters of the temporally ordered objects (here stocks). 

The key feature used will be price relatives which are defined for the $m$-th object as:
\be
x_{m,t}= \frac{p_{m,t}}{p_{m,t-1}}
\ee
In vector notation we will write this equivalently as $\ve x_t$ where the $m$-th component is $x_{m,t}$.

The controls that represent the agents' are the portfolio weights by which each agent's decision will contribute to the final aggregate decision at a particular time. 

Agent performance is represented by factor (agent) mimicking portfolios that are formed from the portfolio controls at each time period. The controls are estimated and implemented at the beginning of each period. The relative changes in asset performance will then modify the relative weights of the asset over the investment period and the performance of a given agent is then determined at the end of the investment period. 

This is determined both by the controls, and selecting for the collection of objects the agent is holding, their weights, and the performance of those objects as determined by price relatives.  

Agents do not have to hold the same number of objects. Agents can hold all or small groups of objects, they can short-sell objects and hold long positions in objects\footnote{Short-selling is when an asset is borrowed for a small fee, and the capital raised from the sale can then be used for other trading or investment activities, for example, the raised capital can be used to buy another asset by taking a long-position. The combination of long and short positions can be cash-neutral where the total value of the initial portfolio is zero. Such a portfolio is called a zero-cost or cash-neutral portfolio.}.  The collection of objects a particular agent holds will be called the agent's object cluster.

The parameters that denote agents are typically a parameter that is an index of the cluster of objects an agent has decided to use, and the algorithm specific parameters; typically a data window parameter $k$ determining how much past data to include, and a parameter more specific to a given algorithm if it is required, such as a partition parameter $\ell$, and a forecast horizon dependent parameter, $\tau$.
 
Any four useful parameters can be used in the learning algorithm that was implemented in this paper. The number of agents is then a function of these four free-parameters. The learning algorithm will then carry out the weighted averaging process based on agent past performance over the agents enumerated by these four parameters. 

The parameters are denoted $\tau$, $w$, $k$ and $\ell$ respectively. We reserved parameters $k$ and $\ell$ for algorithm specific parameters - this is done in order to try to align with their usage in the prior literature \cite{GUV2007}. There are at most $W$ values of $w$, $K$ values of $k$, $L$ values of $\ell$ and $\tau_n$ values for the horizon parameter $\tau$. 

The default value of the horizon parameter is 1: $\tau=1$. For simplicity and computational speed the results presented in this paper have used the default value \footnote{It is anecdotally noted that there is an advantage in learning for the horizon parameter but this does not change the basic point made in this paper}. The choice of these parameters will determine the number of agents in the system. The number of agents is denoted by $n$ where the total number of agents will then be $N=\tau_n WKL$. 

The $n$-th agent is represented by a tuple containing the controls at a given time and its performance $(H_{nm,t},S_{n,t})$. This tuple will usually be represented in vector notation as $(\ve H_{n,t}, S_{n,t})$ where the object index $m$ is suppressed. 

For discrete values of sequential time running from $t=1$ until some maximal time $T$ the agent controls $\ve H$ are then collection of $T$ time-ordered $(N,M)$-dimensional matrices that are represented as multi-dimensional double precision matrices in the software.

The value of the $n$-th agents controls for the $m$-th object at time $t$ is $H_{nm,t}$ for discrete values of time. The performance of the agents is represented as a $(N,T)$-dimensional matrix where the $n$-th agent has its performance over the $t$-th time interval as $S_{n,t}$. 

There are at most $M$ objects. So $m$ can take on values on the integer interval $[1,M]$ that would enumerate the objects. The number of objects remain static for a given agent even though they may be able to achieve zero positions in a particular agent.

From the perspective of the learning algorithm the mechanism of agent generation is not important, it is required that all $N$ agents are correctly enumerated at each time increment. At the beginning of each time increment the controls determined at the end of the previous time increment are implemented and then held to the end of the time period at which time the agent performance is determined and the agent controls are then adjusted using the learning algorithm. 

The learning algorithm updates the agent mixture control $q_{n,t}$ which is a measure of how much a given agent will contribute to the aggregate portfolio. The $q$ variables control the relative mixture of agents through time as they compete based on their past performance. The mixture controls cannot in general be thought of as probabilities, which makes their use and notation different to some of the prior literature \cite{GUV2007}.

\subsection{Online-learning algorithm} \label{ssec:online-learn}

The learning algorithm is inspired by the universal portfolio approach developed by \cite{C1991,CO1996} and refined by \cite{GUW2008}. The learning agent can be thought of as a multi-manager, using asset management language, where the multi-manager is selecting and aggregating underlying strategies from a collection of portfolios $\ve H_{n,t}$ and then aggregating using some selection method to a single portfolio $\ve b_t$ that is implemented at each investment or trading period $t$.

The basic learning algorithm was incrementally implemented online, but offline it can be easily parallelized across agents. The learning algorithm has five key steps:
\begin{enumerate}
\item {\bf Update the portfolio wealth}:
The portfolio controls $b_{m,t}$ for the $m$-th asset are used to update the portfolio returns for the $t$-th period
\begin{eqnarray}
\Delta S_t &=& \left[ {\sum_{m} b_{m,t} (x_{m,t}-1)} \right] +1\\
S_t &=& S_{t-1} \Delta S_t.
\end{eqnarray}
Here the price relatives for the $t$-th period and $m$-th asset, $x_{m,t}$, are combined with the portfolio controls for the period just ending to compute the realised portfolio returns for this period, period $t$. The portfolio controls were computed at the end of the prior period and implemented at the beginning of the current period. The relative amounts of each object in the portfolio will have changed by the relative price changes assuming no cash-flows into or out of the portfolio during this investment period.

\item {\bf Update agent wealth}:
The agent controls $H_{nm,t}$ were determined at the end of time-period $t-1$ for time period $t$ by some agent generating algorithm for $N$ agents and $M$ objects about which the agents make expert capital allocation decisions. At the end of the $t$-th time period the performance of each agent, $S_{n,t}$, can be computed from the change in the price relatives $x_{m,t}$ for the each of the $M$ objects in the investment universe considered using the prices at the start, $p_{m,t-1}$, and the end of the $t$-th time increment, $p_{m,t}$, using the agent controls. 
\begin{eqnarray}
\Delta S_{n,t} &=& \left[ {\sum_{m} H_{nm,t}(x_{m,t}-1)} \right] +1. \\
S_{n,t} &=& S_{n,t-1} \Delta S_{n,t}.
\end{eqnarray}
\item {\bf Update agent mixtures}:
We considered three different agent mixture update rules: 1.) the universally consistent choice, and 2.) an exponential gradient choice \cite{HSSW1998} and 3.) an exponentially weighted moving average. We generically refer to these online updates as rule $g$. In practice one would select one of the three update rules once for the duration of the offline training, if one seeks to initialise the algorithm prior to deployment, or for use online during the system implementation in real-time. For the numerical experiments presented here we adopted the universal consistent approach inspired by \cite{CO1996,GUW2008} as this demonstrates the principle. We can define the mixture of controls as the accumulated agent wealth is used as the update feature for the next unrealised increment with some normalisation, as such, the agent mixture control for the $n$-th agent for the next time increment, $t+1$, is proportional to the measure of wealth:
\be
q_{n,t+1} = S_{n,t}.
\ee
the alternative choices can include the Exponential Gradient (EG)\footnote{{\it Exponential Gradient} (EG)  based learning: $$q_{n,t+1}= q_{n,t} e^{\left( { \frac{\eta S_{n,t}}{\sum_n q_{n,t} S_{n,t}} } \right) }$$}  approach of \cite{HSSW1998} or an Exponential Weighted Moving Average (EWMA)\footnote{{\it Exponential Weighted Moving Average} (EWMA) based learning: $$q_{n,t+1} = \lambda q_{n,t} + (1-\lambda) \left( {\frac{q_{n,t}S_{n,t}}{\sum_n q_{n,t} S_{n,t}}} \right)$$} based learning strategy. We adopt the simplest update rule for the mixture of controls, it should be noted that there can be practical advantages to using more adaptive methods such as EG and EWMA learning where the learning rates can be used as additional parameters to be learnt using a thick modelling framework \cite{BK1999}.
\item {\bf Re-normalise agent mixtures}:
If the agent mixture is to be considered a positive probability then we require that $\sum_n q_n = 1$ and that all $q_n\ge0$. This is the case of fully-invested agents where no shorting is allowed. We will call these types of agents {\it absolute} agents: 
\begin{eqnarray}
q_{n,t+1} =\frac{q_{n,t+1}}{\sum_{n} q_{n,t+1}}.
\end{eqnarray}
For agents that we will consider {\it active} the leverage is set to unity for zero-cost portfolios: (1.) $\sum_n q_n = 0$ and (2.) $\nu = \sum_n |q_n|=1$. Here the mixture controls allow for shorting of one agent against another and the portfolio becomes self-funding. The mixture controls can no-longer be thought of as positive probabilities.
\begin{eqnarray}
q_{n,t+1} = \frac{q_{n,t+1}-\frac{1}{N}\sum_{n} q_{n,t+1}}{\sum_n |q_{n,t+1}-\frac{1}{N}\sum_{n} q_{n,t+1}|}
\end{eqnarray}
The leverage is normalised in order to ensure consistency between the learning algorithms and agent generating algorithms.  
\item {\bf Update portfolio controls}:
The portfolio controls $b_{m,t}$ are updated at the end of time period $t$ for time period $t+1$ using the agent mixture controls $q_{n,t+1}$ from the updated learning algorithm and the agent controls $H_{nm,t+1}$ from the agent generating algorithms using information from time period $t$ and averaged over all $n$ agents.
\begin{eqnarray}
b_{m,t+1} = \sum_n q_{n,t+1} H_{nm,t+1}.
\end{eqnarray}  
\end{enumerate}
The strategy is to implement the portfolio controls, wait until the end of the increment, measure the features, update the agents and then re-apply the learning algorithm to compute the agent mixtures and portfolio controls for the next time increment.

\begin{algorithm}
\begin{algorithmic}
\Require {\\ \begin{enumerate} 
\item updated agent-controls $\ve H_{n,t+1}$
\item current feature realisation $\ve x_t$ 
\item current portfolio controls $\ve b_{t}$
\item current agent-controls $\ve H_{n,t}$
\item past agent-wealth $S_{n,t-1}$
\item past portfolio wealth $S_{t-1}$
\end{enumerate}}
\For {$t$-state} 
\State{Step 1: The portfolio wealth is updated}
\State $S_t = S_{t-1} ( \ve b_t (\ve x_t^{_T}-1) +1)$
\State{Step 2: The agent wealth is updated} 
\State $S_{n,t} = S_{n,t-1} (\ve H_{n,t} (\ve x_t^{_T} -1) +1)$
\State{Step 3: The agent mixture is updated for rule $g$}
 \State $q_{n,t+1} = g(q_{n,t},S_{n,t})$
\State{Step 4: The agent mixtures are re-normalised}
\State $ q_{n,t+1} = \begin{cases} \sum_n q_{n,t+1}=1,~q_{n,t+1} \ge 0 \\
\sum_{n} |q_{n,t+1}| = 1, \sum_n q_{n,t+1}=0.\end{cases}$
\State{Step 5: The portfolio controls are updated}
\State {$\ve b_{t+1} = \sum_{n} q_{n,t+1} \ve H_{n,t+1}$}
\State{Leverage corrections}
\If{ ($\nu = \sum_{m} |b_{m,t}|) \neq 1$}
\State{renormalise controls}
\State{$\ve b_{n,t+1} = \frac{1}{\nu} \ve b _{n,t+1}$}
\State{renormalise mixtures}
\State{$q_{n,t+1} = \frac{1}{\nu} q_{n,t+1}$}
\EndIf
\EndFor
\State \Return{($\ve b_{t+1}$,$S_{n,t}$,$S_t$,$q_{n,t+1}$)}
\end{algorithmic} \caption{Online-Learning Algorithm (OLA)}
\label{alg: OLA}
\end{algorithm}

\section{Agent generating algorithms} \label{sec:agent-algorithms}

The purpose of the agent generating algorithms are to sequentially generate the agent controls $H_{nm,t}$ for the $n$-th agent for the $m$-th object for implementation at the start of the $t$-th time period. These will be denoted in vector notation as $\ve H_{n,t}$.  

We initially considered three different agent-generating algorithms over which the thick modelling was carried out in order to learn the various algorithms' free-parameters: 1.) a pattern-matching algorithm \cite{GUV2007}, 2.) a contrarian mean-variance portfolio algorithm we called anti-BCRP (as it trades against the Best Constant Rebalanced Portfolio for a given k-tuple of data) \footnote{The anti-BCRP algorithm can be used to learn for mean-reversion by directly using k past realisations of performance of each object, for a given partition, by finding the mean-variance wealth minimizing portfolio (in order to be contrarian), either fully-invested or zero-cost, and using the resulting portfolio weights for the agents with the specific window and partition parameters: $\ve H_{n,t+1}=\ve H_{n,t+1} (\gamma, - \ve \mu(\ve x_{n,t}), \Sigma(\ve x_{n,t}))$ comparing with Eqn. (\ref{eqn:agentcontrols}) and (\ref{eqn:agentweights}).}, and 3.) the ANTICOR algorithm \cite{BEG2004}. The various free-parameters of these algorithms, such as the window sizes k and partitions $\ell$ were then used to enumerate the agents that would compete for capital allocations in the learning algorithm.  

We adopted the pattern-matching approach \cite{GUV2007} for the numerical experiments in this paper as we found a performance advantage in looking for more general patterns rather than merely targeting mean-reversion effects, and more importantly, the pattern-matching algorithms are more generic as they do not require any {\it a-priori} choices for the structures that are learnt for. This was considered to be more faithful to the intent of the paper - where we seek to show that unspecified patterns can be learnt for in a manner that can both beat the best single stock in a universe of stocks and can beat a cash portfolio in a self-funding strategy.

\subsection{Comments on Notation}

The feature realisations at time $t$ for the $m$-th object, $x_{m,t}$, are also denoted in vector notation as $\ve x_t$. The agent controls and the feature time-series are the key inputs in the online-learning algorithm to determine the agent mixtures $q_{n,t}$ through time. The online learning algorithm is path-dependent and as such both a function of the history of agent controls as well as the feature time-series history.

Following prior work we denote random feature variables as $\ve X$ and their realisations as $\ve x$ \cite{GUV2007,GUW2008} where for some vector valued stationary and ergodic process $\{ \ve X_t\}_{-\infty}^{+\infty}$ with realisations denoted as $\ve x_1,\ve x_2, \ldots, \ve x_t$ and their corresponding random variables as $\ve X_1, \ve X_2,\ldots,\ve X_t$. However, we will refine the notation further in order to more effectively enumerate the agents for our specific implementation.

The strategies are based on constructing a $k$-tuple of the selected feature for $m$-objects. We will denote the agent-tuple by $\ve x_{k \ell w,t}$ and the $k$-tuple as $\ve x_{t}^{t-k}$. The $k$-tuple is a slice of data of length $k$ from the current time $t$, of width $m$ enumerating all the objects. We will modify the $k$-tuple notation to $\{ {\ve x}_t^{t-k} \}_{s(n),\ell}$  to denote a $k$-tuple taken from an $\ell$-partition of the data for a given cluster of objects $w=s(n)$. Here $s$ is the cluster index of the $n$-th agent. We are suppressing the $m$ index and using vector notation to write the $k$-tuple as $\ve x$. The agent-tuple will be unique to the $n$-th agent where $n$ is the unique agent index enumerating a particular combination of $k$,$\ell$ and $w$.

A $k$-tuple is used to determine agent controls $\ve H_{n,t}$. The initial features used are historical prices sequences which are assumed to be realisation $\ve x$ from some random process $\ve X$. The pattern-matching algorithm will then refine the $k$-tuple to groups of nearest-neighbours that are expected to reflect historical selected outcomes that better reflect future outcomes than merely the last price change or price change sequence. This is done by comparing the current realisation $\ve x_{t}^{t-k}$ with the past.

In this way, given a set of parameters enumerating the $n$-th agent we will select the required tuple from the existing data realisations depending on the algorithm parameters using some selection function $f$
\begin{equation}
\ve x _{n,t} = \ve x_{n(k,\ell,w),t} = \ve x_{k \ell w,t} = f_{\ell,w}(\ve x_{1}^t,\ve x_{t}^{t-k})
\end{equation}
where the $m$-th component of the $k$-tuple is $x^m_{n,t}$. 

\subsection{The Log-optimal strategy} \label{ssec:log-opt}

The log-optimal strategy under the assumptions of stationarity and ergodicity has been shown to be the best possible choice of strategy over the long term \cite{AC1988}. This type of analysis has been extended to the semi-log-optimal case \cite{GUV2007} where weakened conditions have been derived. 

The surprising result is that even with this weaker formulation the loss of optimality is such that log-optimality has, for all practical purposes, equivalent performance to portfolios selected using semi-log-optimality \cite{GUV2007}. This provides an argument for the use of competing sequences of mean-variance portfolios in the framework of agent-based competition for capital.

With an initial investment wealth of $S_0$ using a sequence of portfolio controls $\ve B = \{ \ve b_i \}_{i=1}^{t-1}$ from time $i=1$ until the current time $t$ the portfolio wealth for a fully-invested portfolio is \cite{GUV2007}
\begin{equation}
S_t = S_0 \Pi_{i=1}^T {\ve b}(\ve x_{1}^{i-1}) {\ve x}_{i}^{_T} = S_0 e^{\sum_{i=1}^{T}\log ({\ve b}(\ve x_{1}^{i-1}) {\ve x}_{i}^{_T})}. 
\end{equation}
This gives an average portfolio growth rate $W_t(\ve B) = \frac{1}{T} {\sum_{i=1}^{T}\log ({\ve b}(\ve x_{1}^{i-1}) {\ve x}_{i}^{_T})}$. The log-optimal portfolio selection problem is thus
\begin{equation}
\ve b^*(X_{1}^{t-1}) = \arg \max \limits_{\ve b} \E \left[ {\log (\ve b(X_{1}^{t-1}) \ve X_t)} | \ve X_{1}^{t-1} \right]. \label{eqn:logoptimal}
\end{equation}
Here one is aiming to maximize the overall wealth through the incremental selection of the sequence of fully-invested portfolio controls $\ve B$. 

\subsection{Universally consistent strategies} \label{ssec:univ-cons}

The fundamental result of universal log-optimality is that no investment strategy can have a faster average rate of growth than that arising from the log-optimal portfolio \cite{K1956,AC1988,C1991,CO1996}. However, full knowledge of the distribution of the process is required. Strategies achieving an equivalent growth rate without knowing the distribution are called {\it universally consistent} \cite{AC1988,GUV2007} strategies. 

In principle one could via simulation enumerate all the possible controls and find via brute-force the set of controls that solve the log-optimal portfolio selection problem. 

This is ambitious given current technology constraints and that the opportunity set of stocks is typically large and the data representing the features even larger - particularly for intraday quantitative trading problems.

In the idealized situation we would define some simplex $\Lambda$ where there is a prior distribution $\mu$ on the simplex, such that some expert or agent $\ve b$ is a given realisation from this distribution of portfolios. We would then directly evaluate the $\mu$-weighted fully-invested universal portfolio at time $t$ \cite{CO1996,CB2003}
\begin{equation} 
{\ve b}^{*}_t = \frac{\int_{\Lambda} \ve b S_{t-1}(\ve b,\ve x_{t-1}) d \mu (\ve b) }{\int_{\Lambda} S_{t-1}(\ve b,\ve x_{t-1}) d \mu (\ve b) } \label{eqn:universal}
\end{equation}
where $\int_{\Lambda} d \mu (\ve b)=1$ and the portfolio value $S_t$ at time t is as 
\begin{equation}
S_t(\ve b, \ve x_t) = \prod_{j=1}^t \ve b \ve x_j^{_T} = \prod_{i=1}^t \sum_{j=1}^m b_j x_{j,t}.
\end{equation}
Here the portfolio is fully-invested such that $\ve b \ve 1^{_T} = 1$ for unit vector $\ve 1$. 

Although we seek strategies that are universally consistent with respect to the class of stationary and ergodic processes. A pragmatic approach is required given both the unrealistic distributional assumptions, and the curse of dimensionality we face in enumerating control space \footnote{For each random process in the long-term limit the growth-rate of these strategies is equivalent to that of the log-optimal portfolio when full-knowledge of the distribution is available. In order to construct such universally equivalent strategies one needs to know the conditional distribution $\ve X_t$ given some past $\ve X_{1}^{t-1}$. }.

The strategy is to reduce the problem by finding a more informed subset of controls that can be used to approximate the required sequence of portfolio controls that are used to represent a universally consistent strategy. In addition to reducing the set of applicable controls one also aims to streamline the evaluation of these controls and their adaption through time, this can be achieved by reducing the log-optimality criterion to semi-log-optimality.

\subsection{Semi-log optimality}

We choose to focus on the first two moments of the price relative distributions: the mean and covariance. This will allow enhanced performance speed of the algorithms (see Figure \ref{fig: JSE Intraday Timings}) but with some loss in long-term optimality \cite{V2006,GUV2007} and as such a deviation from the universally consistent strategies.  

First, we have reduced the opportunity space in the simplex of all possible portfolios in order to make the problem of finding a portfolio that is optimal over the entire feature space computationally tractable, this is achieve by using agent-generating algorithms and learning over the free-parameters for those agents generating algorithms.

Second, we replaced the optimization with a quadratic approximation that will give us analytic solutions to replace optimizations that we would otherwise have to solve numerically. In addition to a performance advantage, using the quadratic approximation this will also provide a straight-forward method for considering both fully-invested and zero-cost portfolio's in a single framework.

Streamlining the algorithms for performance was approached in two steps, first, to separate the problem into that of an online-learning algorithm and the agent generating algorithms, then, second, to reduce the log-optimality criterion to semi-log-optimality.

The semi-log-optimal portfolio selection takes on the form
\begin{equation}
\ve b^*(X_{1}^{t-1}) = \arg \max \limits_{\ve b} \E \left[ {h (\ve b(X_{1}^{t-1}) \ve X_t)} | \ve X_{1}^{t-1} \right]. \label{eqn:semilog}
\end{equation}
where $h(z) = (z-1) - \frac{1}{2}(z-1)^2$ from the second order Taylor expansion of $\log(z)$ at $z=1$.

A related approach was taken in \cite{CB2003} where they derived an analytic approximation for an efficient universal portfolio. Our simplified mean-variance approach was motivated by their development of an analytic algorithm, the difference here is that we want an algorithm that is online, analytic, explicitly includes zero-cost portfolios, and allows for the restriction of the solution space using some agent generating algorithm directly at each step rather than via side-information. 

\subsection{Active fund separation problem}

The determination of the optimal portfolio is sequentially implemented using the exact solution to the quadratic approximation to log-optimality by solving the active fund selection problem. The active fund selection problem is a special case of the mutual fund selection problem \cite{M1995,L2000}. This will give an analytic approximation that can both cater for long-only fully-invested agents (absolute agents) as well as leverage one\footnote{$\sum_i |\omega_i| = 1$ for portfolio controls $\omega$.} zero-cost portfolio's (active agents). 
   
We therefore consider the semi-log-optimal portfolio optimization problem \cite{M1995,PS1988,L2000} for return expectation vector $\ve \mu$ and asset return covariance matrix $\Sigma$ with a portfolio control vector $\ve \omega$ in terms of the risk aversion parameter $\gamma$. The conjugate transpose of a vector is denote as $(\cdot)^{_T}$ over a single investment period to define the control problem as:
\begin{equation}
\max \limits_{\ve \omega}  \left \{ {\ve \omega^{_T} \ve \mu - \frac{\gamma}{2} \ve \omega^{_T} \Sigma \ve \omega} \right \} \text{ s.t. } \ve \omega^{_T} \ve 1 = 1. \label{eqn:mutualfund1}
\end{equation}
Here we have changed notation to denote the portfolio controls as $\ve \omega$ in order to avoid confusion with the portfolio strategy controls $\ve b$ that are the result of the online-learning algorithm which aims to approximate the semi-log-optimal portfolio selection strategy for aggregate portfolio controls $\ve b_t$ for time increment $t$. 

Here the portfolio controls $\ve \omega$ are used to generate the agents that populate the agent control set $\ve H_{n,t}$. It is the agent control set that is then used to generate the semi-log-optimal portfolio choice at each time $t$: $\ve b_t$.

\Eqn{eqn:mutualfund1} can be rewritten as the mutual-fund Lagrangian  
\begin{eqnarray}
L = \ve \omega^{_T} \ve \mu - \frac{\gamma}{2} \ve \omega^{_T} \Sigma \ve \omega - \lambda_{\omega} (\ve \omega^{_T} \ve 1 - 1 ) . 
\end{eqnarray}
and solved using elementary Kuhn-Tucker methods. Two equations are found in terms of the optimal solution for the portfolio control, $\ve \omega^*$, the first gives the quadratic optimal risk-return pay-off, and the second, the fully-invested portfolio investment constraint
\begin{eqnarray}
\ve \omega^* &=& \frac{1}{\gamma} \Sigma^{-1} \left ( {\ve \mu - \lambda_{\omega} \ve 1} \right), \label{eqn:riskreturn} \\
\ve \omega^{*_T} \ve 1 &=& 1. \label{eqn:fullyinvested} 
\end{eqnarray}
The Lagrange multiplier is determined by substituting \Eqn{eqn:riskreturn} into \Eqn{eqn:fullyinvested} to find:
\begin{eqnarray}
\lambda_{\omega} = \frac{\ve 1^{_T} \Sigma^{-1} \ve \mu}{\ve 1^{_T} \Sigma^{-1} \ve 1} - \frac{\gamma}{\ve 1 \Sigma^{-1} \ve 1}.
\end{eqnarray}
This is then used to eliminate the Lagrange multiplier from \Eqn{eqn:riskreturn} to find a formulation of the mutual fund separation theorem:
\begin{eqnarray}
\ve \omega^* = \frac{\Sigma^{-1} \ve 1}{\ve 1^{_T} \Sigma^{-1} \ve 1} + \frac{1}{\gamma} \Sigma^{-1} \left( {\ve \mu - \ve 1 \frac{\ve 1^{_T} \Sigma^{-1} \ve \mu}{\ve 1^{_T} \Sigma^{-1} \ve 1}} \right). \label{eqn:mutualfund2}
\end{eqnarray}
The first term on the right is the lowest risk portfolio and the second term is the zero-cost portfolio that encapsulates the relative views of the assets. We will typically work with the separation theorem in the form given in \Eqn{eqn:mutualfund2}. The second term will give us an efficient method of generating zero-cost portfolio's.

It is then convenient to re-write the Mutual Fund Separation theorem to an Active Fund Separation theorem explicitly from \Eqn{eqn:mutualfund2} by defining the lowest risk portfolio as the benchmark portfolio:
\begin{eqnarray}
\ve \omega^* = \ve \omega_{_B} + \ve \omega_{_A}, \label{eqn:activefund}
\end{eqnarray}
where
\begin{eqnarray}
\ve \omega_{_B} &=& \frac{\Sigma^{-1} \ve 1}{\ve 1^{_T} \Sigma^{-1} \ve 1}, \\
\ve \omega_{_A} &=& \frac{\Sigma^{-1}}{\gamma}\left( {\frac{\ve \mu \ve 1^{_T} - \ve 1 \ve \mu^{_T}}{ \ve 1{_T} \Sigma^{-1} \ve 1} } \right) \Sigma^{-1} \ve 1, \label{eqn:active}
\end{eqnarray}
The formulae for $\ve \omega_{_B}$ and $\ve \omega_{_A}$ will be directly used in the agent generating algorithms based on views encoded in the mean, $\ve \mu$, and the covariances, $\Sigma$, as a function of the various agent generating parameters. The resulting controls $\ve H_{n,t}$ will then be determined from the $m$-th component of either $\ve \omega_{_A}$ for the active agents or $\ve \omega_{_B} + \ve \omega_{_A}$ for the absolute agents for the $n$-th agent for time-increment $t$.

For situations where we want agents constructed from zero-cost portfolios we will use the tactical solution from \Eqn{eqn:active} to generate the agents for a given $k$-tuple. In situations where we need fully invested agents we will use the combination of the benchmark fund and the active (or tactical) fund. 

Suppressing indexes over the $m$ objects the agent controls for the $n$-that agent for the two possible cases: (1.) the absolute agents, and (2.) the active agents is then 
\begin{equation}
\ve H_{n,t} = \begin{cases} \ve h^{_T} \ve 1  = 1,  \ve h = \ve \omega_{_B}(\Sigma) + \ve \omega_{_A}(\gamma,\ve \mu,\Sigma) \text{ s.t. } \ve h \ge 0 \\
\ve h^{_T} \ve 1 =0,  \ve h = \ve \omega_{_A}(\gamma,\ve \mu,\Sigma) \text{ s.t. } \ve h^{_T} \ve h = 1.  \end{cases}\label{eqn:agentcontrols}
\end{equation}
Here the $m$-th component of $\ve H_{n,t}$ is $H_{nm,t}$ and the portfolio weights are dependent on the agent-tuples $\ve x_{n,t}$ for a given agent
\begin{eqnarray}
\ve \omega_{_A} &=& \ve \omega_{_A}(\gamma,\ve \mu(\ve x_{n,t}),\Sigma(\ve x_{n,t})) \label{eqn:agentweights} \\
\ve \omega_{_B} &=& \ve \omega_{_B}(\Sigma(\ve x_{n,t})).
\end{eqnarray} 
For the active agent we enforce the leverage unity constraint at the beginning of each time increment, this can be considered equivalent to setting the risk-aversion $\gamma$, at the beginning of each time increment, such that the leverage is always unity. 

This is an important feature of the algorithm as we do not enforce uniform risk-aversion through time. We rather choose to ensure that capital be fully utilized given the available information. The following sections describe how the agent-tuples are constructed for the various agent generating algorithms.

\subsection{Agent generating algorithms from patterns} \label{ssec:pattern-matching}

In order to efficiently reduce the space of portfolio controls to efficiently generate a reasonable approximation to universally consistent strategies using Eqn. (\ref{eqn:universal}) we reduce the set of applicable controls using agent-generating algorithms. The agent-generating algorithm we use in our numerical experiments will be a pattern-matching algorithm \cite{GUV2007}. One can make various decisions about how to break data up into manageable pieces for the various algorithms, the most basic decisions relate to how to break up the data in time, we call this partitioning, the other choice relates to how we break the data up in terms of the objects themselves (often called the features), this we call clustering. Partitioning is typically a more intricate task because this has implications for the algorithm and system structure.  

The pattern-matching algorithm is based on two steps subsequent to the choice of clusters $s(n)$: (1.) partitioning and (2.) pattern-matching. Clusters can be chosen by a variety of methods, we would like to promote two methods: (i) correlation matrix based methods \cite{HWG2016a}, and (ii) clusters based on economic classifications of stocks \footnote{For example, using ICB (Industry Classification Benchmark) sectors classifications \cite{ICB}}. The prior method, correlation based methods, have outputs that can be directly used as inputs into the algorithms discussed here, specifically via $s(n)$, the cluster membership parameters. It is however, the method based on fixed economic sector classifications \cite{ICB}, that will be explicitly used in this paper for the intraday experiments in Section \ref{ssec:intradayJSE}, this is both for speed and simplicity \footnote{It should be noted that using ICB sectors to generate additional agents for the daily simulations does boost algorithm wealth performance but we chose to explicitly demonstrate the value of including sector information in the context of the intraday strategies}.

In the daily numerical experiments we have ignored the impact of clustering and used the clusters $s(n)$ of the $n$-th stock as being trivial, {\it i.e.} we consider a single stock cluster that includes all $m$ objects. The inclusion of clustering indexing can be important to the practical implementation of these techniques as it is often useful to restrict trading signal decisions to similar stocks. There is a wealth advantage to this, as we have shown when we considered the impact of clustering for the numerical experiments using intraday data (see Table \ref{tab: JSE IntradayClusters}).

The pattern matching algorithm is split into two key components: First, the {\it partitioning algorithm}, which selects collection of time-order features from the full set of feature data. Second, the {\it pattern-matching algorithm}, where given a measured pattern derived from the feature data, is used by the algorithm to find similar patterns in a given partition of the feature data.

\subsubsection{Partitioning}

Subsets of time-ordered data are selected from the original time-order data for a given collection of objects. The collection of objects can in turn be a sub-collection of the original set of objects. Partitioning takes place in the time domain while clustering is in the object dimension. The purpose of partitioning is to prepare data subsets for pattern-matching \cite{GUW2008}. Four distinct approaches to data partitioning are enumerated here, however only the trivial partition is used in the experiments.

A partition is a collection $\{p_t\}_{\ell}$ represented by a logical vector of the length of a given time-series where true is represented as one and false as zero to index membership in a given partition. When a partition is determined from features that determine the state of the system at a given time we will use that partition to represent the system in that state for the sake of pattern-matching. 

For the numerical experiments presented here we will use variations of the  {\it trivial partition}: Here all the temporally ordered data is kept in a single partition as represented by a vector of ones of length of the time-series.
\begin{equation}
\{ p_t \}_1 =\{(1,\ldots,1,1,1) \}.
\end{equation}
There are wealth advantages associated with more sophisticated partitions. We considered four different partitioning approaches: the {\it trivial partition}, the {\it over-lapping partition} {\footnote{Example of length T overlapping partition of features: $$
\{ p_t \}_T = \left \{ (0, \ldots,0,0,1),  (0,\ldots,0,1,1), \ldots, (1,\ldots,1,1,1) \right \}. $$}}: were data membership in partitions is repeated in order to bias the data towards a given time, for example, the last time-increments is repeated across all $\ell$ partitions for time-series of length T, the {\it exclusive partition} where the partitions are mutually exclusive subsets of the full partition, and the {\it side-information partition} \cite{CO1996}.

The most heuristically useful partition is that of the {\it side-information partition} where partitions can be pre-selected in the partitioning algorithm based on rules conditioned on side-information \cite{CO1996}, partitioning can be both useful as a nuanced exploitation of information, for example by splitting feature data over different regimes, and thus to generate distinct agents for different regimes, and as an effective approach to parallelization of algorithms. 

Here we would partition the time-series based on side-information arising from additional features drawn from the system being observed as in \cite{CO1996}. For example, we could use a Markov-switching algorithm with $\ell$ states, assign each time in the time-series a state index and the define the partition membership based on states, or we could choose a feature as side-information and $\ell$-tile the data into $\ell$ groups and then based on whether a given time has a side-information feature in a particular group it would be assigned to a given partition.

Partitioning serves as a convenient mechanism for breaking up the feature data into distinct states. This can be useful when choosing to search for patterns when the system is in a distinct state as it will enable the algorithm to search for patterns only in historic data residing from times in the past when the system was in a similar state. By combining a partitioning algorithm with a state-detection algorithm one can both improve computational times as well as algorithm performance in terms of wealth generation \cite{HGW2016b}, this is not explored further here.

\subsubsection{Pattern-matching} \label{sssec:pattern}

The pattern-matching algorithm will take a $k$-tuple and search a given partition of the feature data for similar patterns by finding the smallest distance measure between the $k$-tuple and data in a given partition. This best matching set of data in the partition will then be used to  determine a pattern-matching time $j_{\ell}$. The matching time will then be used to select a future outcome some time period $\tau$ ahead of the matched pattern. This future outcome is used to construct a tuple of data, the agent-tuple, iteratively using the look-ahead rule: $j_{n} = j_{\ell} + \tau$. A number of such pattern-matches will be accumulated to construct the agent-tuple $\ve x_{n,t}$ and from this a mean and covariance are computed.

This mean and covariance will then serve as the input into \Eqn{eqn:activefund} to determine that agent controls $\ve H_{n,t+1}$,  the $n$-th agents controls to be held for time-period $t+1$.

The pattern-matching algorithm is split into two separate algorithms. The first algorithm, which we will call the pattern algorithm, generates patterns to be matched and partitions of data into which the pattern will be matched. The second algorithm will then take the pattern and the data partitions and generate matching times. The matching times will then be used to generate an agent-tuple $\ve x_{n,t}$. 

The pattern algorithm generates a $k$-tuples $\{ \ve x_{t-k}^{t} \}_{s(n)}$ \cite{GUW2008} for matching, and a data partition $\{ \ve x_t\}_{(p_{\ell},s(n))}$ using a predefined temporal partition $\{p_{\ell}\}$ of the data and the cross-sectional cluster for the $n$-th agent $s(n)$. This is iteratively done for each agent as enumerated by the parameters that define a given agent: the cluster membership $w=s(n)$ of the $n$-th agent, the partition variable $\ell$, the $k$-tuple variable $k$ and the look-ahead horizon variable $\tau$. 

For each set of variables that define the $n$-th agent the pattern algorithm will then call the matching algorithm.

\begin{algorithm}
\begin{algorithmic}
\Require {\\ \begin{enumerate} \item features $\ve x_t$ \item $n$-agent parameters $k$, $\ell$, $s(n)$, $\tau$ \item partitions $\{ p_{\ell} \}$ \end{enumerate}}
\For {$n$-agents} 
\State{$\ve H_{n,t+1}$}{=\Call{Matching}{$\tau$,$\{ p_{\ell} \}$,$\{ \ve x_{t-k}^{t} \}_{s(n)}$,$\{ \ve x_t \}_{(p_{\ell},s(n))}$}}
\EndFor{ agents}\\
\Return{$\ve H_{t+1}$}
\end{algorithmic} \caption{PATTERN-MATCHING Algorithm (PMA)}
\label{alg: PMA}
\end{algorithm}

The matching algorithm will find matches for the $k$-tuples, $\ve x_{t-k}^{t}$ in the partitions. If there is a single partition of data, the matching algorithm will find the $\hat{\ell}$ closest matches. We consider two rules for calculating $\hat{\ell}$ and will refer to these as rule $P$. This rule is introduced in order to easily compare our algorithms with prior literature, more specifically \cite{GUV2007,GUW2008}. The difference is related to how the partitions are defined and implemented.  

We consider the {\it trivial rule}: $\hat{\ell} = \ell$ and the rule required to recover the Nearest-Neighbour (NN) algorithm performance described in \cite{GUW2008}. The {\it  {\it Gy{\"o}rfi et al Nearest Neighbour rule}}  is where $\hat{\ell}$  is determined by a variable $p_{\ell} \in (0,1)$. The choice of $p_{\ell}$ used in the experiments is the same as in \cite{GUW2008}.
\begin{eqnarray}
\displaystyle p_{\ell} &=& 0.02 + 0.5 \frac{\ell -1}{L -1} \\
\displaystyle \hat{\ell} &=& \left \lfloor{p_{\ell} t}\right \rfloor
\end{eqnarray}
Where $t$ represents the number of time periods in the history, and the floor is taken to find the smallest partition at the given time. This modification serves primarily to allow us to recover prior results in the literature using the framework we implemented in the software for the numerical experiments. 

If there are $\ell$ partitions of data the algorithm will find the best match in each partition. The matching algorithm will find $\ell$ best matches and from those best matches extract $\ell$ matching times $j_{\ell}$ associated with the time of each $k$-tuple match. From the look-ahead rule the matching algorithm will then construct the agent-tuple $\ve x_{n,t}$. The matching algorithm will then compute the agent-control for this given agent-tuple $\ve h_{n,t}$.

The distance between tuples is the 2-norm. Although we could use the distance between two matrices as the general distance in the algorithm, we have chosen to differentiate selecting the most recent vectors of object features and the test-tuple as the vector distance between these two vectors only for the case of $k=1$, while for $k>1$ we measure the distance of each object from the same object at a difference time independently from other objects. 

This will rather allow us to search for the best fits of objects independently rather than in collective. This is an important refinement, in the original version of the algorithm we followed \cite{GUW2008} and used the 2-norm in full generality independent of the window size $k$ we found better performance by independently selecting for patterns using column-wise computed distances. 

\begin{algorithm}
\begin{algorithmic}
\Require {\\ \begin{enumerate} \item look-ahead-rule $\tau$ \item partition $\{ p_{\ell} \}$ \item $k$-tuple $\{ \ve x_{t-k}^{t} \}_{s(n)}$ \item data partition $\{ \ve x_t \}_{(p_{\ell},s(n))}$  \end{enumerate}}
\For {$t$-state}
\For{ $p_{\ell} \in \{ p_{\ell} \}$}
	\For{$j$-states $\in p_{\ell} $}
		\State{find a test-tuple}
		\State{$\ve s_{t(j)-k}^{t(j)} =  \{ \ve x_{t(j)-k}^{t(j)} \}_{(p_{\ell},s(n))}$}
		\State{find distance to $k$-tuple}
		\State{$\ve \epsilon_{k,j} =\{ \ve x_{t-k}^{k} \}_{s(n)} - \ve s_{t(j)-k}^{t(j)}$}
		\If {k=1} 		
			\State{compute the 2-norm for vector $\ve \epsilon_{1,j}$}
			\State{$\epsilon_j = \sum_{m \in \text{objects}} \sqrt{ \epsilon_{m 1,j}2} = \sqrt {{\ve \epsilon}_{1,j} {\ve \epsilon}_{1,j}^{_T}}$}
			\State{distance measure of dim(objects)}
			\State{$\{ \ve \epsilon_j \} _{p_{\ell}} \leftarrow \epsilon_{m,j} = \epsilon_j \forall m$}
		\Else
			\State{column-wise 2-norms for matrix $\ve \epsilon_{j,k}$}
			\State{$\{ \ve \epsilon_{j} \}_{p_{\ell}} \leftarrow \epsilon_{m,j} = \sum_{k'=1}^k \sqrt{\epsilon_{m k',j}^2}$}
		\EndIf 
	\EndFor{ states}
	\If {dim$( \{ p_{\ell} \})=1$}
	\State{Switch NN algorithm partition choice \cite{GUW2008}}
	\State{$\hat{\ell} = P(\ell,t)$}
	\State{Find $\hat{\ell}$ matching-times in a single partition}
	\State{$\ve j_{\ell} = \min \limits_{j \forall \dim(j)=\ell} \{ \ve \epsilon_{j}\} $}
	\Else
	\State{Find the best match in each of the $\ell$ partitions}
	\State{$\ve j_{\ell} = \min \limits_{j \forall p_{\ell} \in \{ p_{\ell} \}} \{ \ve \epsilon_{j} \}_{p_{\ell}}$}
	\EndIf
	\EndFor{ partitions}
	\State{update the look-ahead-rule}
	\State{$j_n = j_{\ell} + \tau$}
	\State{Update the agent-tuple}
\State{$\ve x_{n,t} = \{ \ve x_1^t\}_{t \in j_n}$}
\State{Update the mean and covariance}
\State{$\ve \mu=\ve \mu(\ve x_{n,t}-1)$}
\State{$\Sigma=\Sigma(\ve x_{n,t}-1)$}
\State{Update the agent-controls}
\State{$\ve H_{n,t+1}=\ve H_{n,t+1} (\gamma,\ve \mu, \Sigma)$}
\EndFor{ t-state} \\
\Return{$\ve H_{n,t+1}$}
\end{algorithmic} \caption{MATCHING Algorithm (MTA)}
\label{alg: MTA}
\end{algorithm}


\section{Data Description} \label{sec:data}

\subsection{OHLC data} \label{ssec:ohlc-data}

The data we will consider will be sequential data, but not necessarily continuously sequential. For this reason we will study OHLC (Open-High-Low-Close) bar-data where the closing price of a given bar is not necessarily the opening price of the subsequent bar of the data. We will first study daily sampled data and then intraday data. The algorithms will be initially tested using synthetic data (see Section \ref{ssec:synthetic-data}), and then the real world test data used in prior research \cite{C1991,GUW2008} (See Section \ref{ssec:real-data}) which are sequences of daily sampled closing prices. 

The data and algorithms can be easily extended to accommodate additional features as side-information \cite{CO1996}; such as volumes, spreads, and various financial indicators and asset specific and state attributes. The side-information can be trivially used to re-partition data into additional sets of agents and then used as inputs into the learning algorithm. The wealth performance enhancement relating to the side-information extension is not demonstrated in the numerical experiments presented here.

\begin{figure}[h!b]
  \centering
    \includegraphics[width=0.45\textwidth]{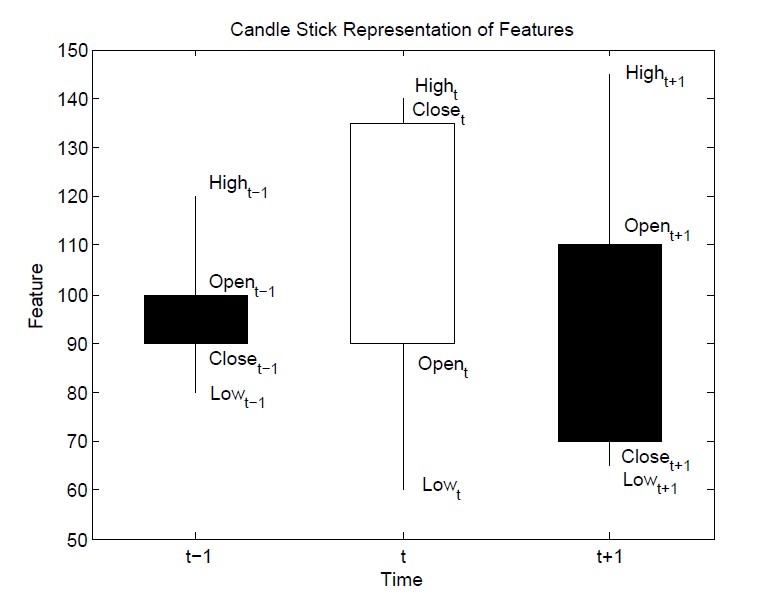}
    \caption{The feature time-series data is best thought of as OHLC (Open-High-Low-Close) bar data. The filled box in the candle chart denotes the situation where the close price is lower than the open price, conversely the unfilled box has the close price higher than the open price.}
\label{fig:candle}
\end{figure}

OHLC bar data is typically represented by a candle-stick graph as in Figure (\ref{fig:candle}).

The time-series data is such that the closing price of time-increment $t$ is not necessarily at time $t+1$ the start of time increment $t+1$. The closing price can in fact be at some time $t+\delta$ for some arbitrary data-specific time-increment $\delta$. 

A low-frequency example is that of a typical trading day on the JSE, the market opens in the morning with some opening price, $o_t$, at 9h00, the market may then close at some closing time 17h00, after a closing auction period, the official closing price $c_t$, is then printed soon after the market close (perhaps after some randomisation period). The market is then closed for some time-period over-night until the market opens again on the subsequent day. There is a period, $\delta$, when the market is close and as such information is not continuously being priced into the traded assets. 
Information that accumulates over-night will then be priced into the market prices through the process of the opening auction and subsequent trading in the various assets. 

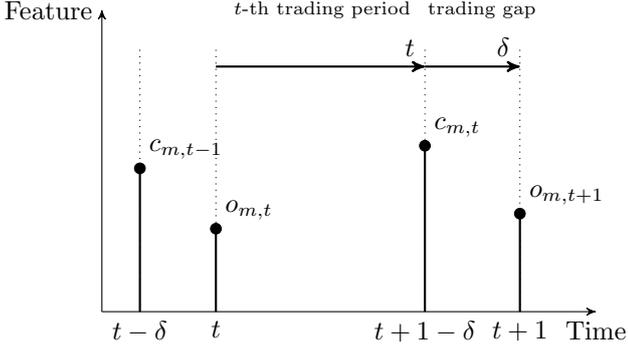
\begin{figure}
\centering
\begin{tikzpicture}]
\draw[->] (0,0) -- (6.5,0) node[anchor=north] {Time};
\draw	          (0.5,0) node[anchor=north] {$t-\delta$}
		(1.5,0) node[anchor=north] {$t$}
		(4.25,0) node[anchor=north] {$t+1-\delta$}
		(5.5,0) node[anchor=north] {$t+1$};
\draw (2.9,4) node{{\scriptsize $t$-th trading period}}
		(5,4) node{{\scriptsize trading gap}};
 \filldraw [black]  (0.5,1.9) circle (2pt) node[above right, black] {$c_{m,t-1}$};
 \draw[thick] (0.5,0) -- (0.5,1.9);
  \filldraw [black]  (1.5,1.1) circle (2pt) node[above right, black] {$o_{m,t}$};
 \draw[thick] (1.5,0) -- (1.5,1.1);
  \filldraw [black]  (4.25,2.2) circle (2pt) node[above right, black] {$c_{m,t}$};
 \draw[thick] (4.25,0) -- (4.25,2.2);
  \filldraw [black]  (5.5,1.3) circle (2pt) node[above right, black] {$o_{m,t+1}$};
 \draw[thick] (5.5,0) -- (5.5,1.3);
\draw[->] (0,0) -- (0,4) node[anchor=east] {Feature};
\draw[dotted] (0.5,0) -- (0.5,3.5);
\draw[dotted] (1.5,0) -- (1.5,3.5);
\draw[dotted] (4.25,0) -- (4.25,3.5);
\draw[dotted] (5.5,0) -- (5.5,3.5);
\draw[->,thick] (4.25,3.25) -- (5.5,3.25) node[above left]{$\delta$};
\draw[->,thick] (1.5,3.25) -- (4.25,3.25) node[above left]{$t$};
\end{tikzpicture}
\caption{Feature time-series investment period for the $t$-th time increment showing that the end of the $t$-th increment does not always have to coincide with the start of the next, here the $t$+1-th, investment period. The opening price is denote as $o_{m,t}$ and the close price for the period as $c_{m,t}$ for the $m$-th asset.}
\label{fig:timeinc}
\end{figure}

Our approach to OHLC data is applicable to a variety of synchronously sampled or re-sampled data sets, including intraday data:
\begin{enumerate}
\item {\it close-to-close}: Here the prices $p_{m,t}$ for the $m$-th assets are the time-series of close prices. The price relatives $x_{m,t}$ are then the computed from the close price time-series $c_{m,t}$
\begin{equation}
x_{m,t} = \frac{c_{m,t}}{c_{m,t-1}}.
\end{equation}
The algorithm is trying to exploit information relating to price changes from the close of trading of one time increment to the close of trading of a subsequent time increment.
\item {\it open-to-close}: Here the prices $p_{m,t}$ for the $m$-th assets are the ordered time-series pairs of open and close prices on the same data the price relatives are then computed as
\begin{equation}
x_{m,t} = \frac{c_{m,t}}{o_{m,t}}.
\end{equation}
Here one is trying to exploit price relative changes within a trade increment, for example, across a single day from the market opening to the market close ignoring the over-night price changes.
\item {\it close-to-open}: Here the prices $p_{m,t}$ for the $m$-th asset are the price changes from the close of the trade period at $t-1$ to the next trade period at time $t$ 
\begin{equation}
x_{m,t} = \frac{o_{m,t}}{c_{m,t-1}}.
\end{equation}
Here one is looking to exploit the change in prices between trade periods where the information cannot yet be fully reflected in trading until the trading commences in the next trade period.
\item {\it open-to-open}: Here the prices $p_{m,t}$ for the $m$-th asset are the time-series of opening prices. The price relatives $x_{m,t}$ are then computed from the opening prices $o_{m,t}$
\begin{equation}
x_{m,t} = \frac{o_{m,t}}{o_{m,t-1}}.
\end{equation}
This is looking for inefficiencies in the prices changes from market opening to market opening.
\end{enumerate}
The important missing component of information is that related to volume (and additional features such as spread, order-imbalance and order-book resilience for intraday data). For example, the opening price is a less reliable price when it has been determined off significantly lower volumes of trading, as compared with a typical closing price. In the case where the closing auction of a given market has more volume than the typical opening auction the relative uncertainties in the prices can be substantial. The typical time increment for a given feature is given in figure (\ref{fig:timeinc}). We promote the use of a state-detection algorithm and side-information partitioning in order to address these types of concerns. In the context of this work such issues do not change our conclusions.

It is expected that the learning algorithm will still attempt to maximise the long-term wealth given a specific agent generating algorithm for a given feature set. For both daily data and intraday data the feature set that is of most interest to us in this study will be those associated with the ``close-to-close" and ``close-to-open" price relative features.

\subsection{Synthetic Data} \label{ssec:synthetic-data}

The algorithm was tested on four synthetic data cases (SDC) for both active and absolute portfolios. The synthetic data was generated for 10 stocks over 1000 time periods. The price relatives $x_{m,t}$ for each stock at each time period was randomly generated from a lognormal distribution (\texttt{lognrnd} function in MATLAB generated using the Mersenne Twister psuedorandom number generator \cite{MT1998} and initialised using a specific seed value), each synthetic data case defines a mean, $\mu$, and variance, $v$, used to generate the dataset. The mean, $\bar{\mu}$, and standard deviation, $\bar{\sigma}$, of the associated normal distribution is given by :
\begin{equation}
\bar{\mu} = \log{\left(\frac{\mu^2}{\sqrt{v+\mu^2}}\right)} 
\end{equation}
\begin{equation}
\bar{\sigma} = \sqrt{\log{\left(\frac{v}{\mu^2} + 1\right)}}
\end{equation}

Table \ref{tab:rand data} summarises the four synthetic data cases, each case was generated 30 times and initialised with seed values $1,2,\ldots,30$ respectively.\\

\begin{enumerate}
\item \textbf{Synthetic Data Case 1 (SDC 1):}
was generated from a lognormal distribution with a mean price relative, $\mu = 1$, and a variance, $v = 0.0002$, to simulate a stock market where there is no significant increase or decrease in the value of a stock over time.

The expected outcome is that neither the active portfolio nor the absolute portfolio will be able to learn which stocks it should hold a long position or short position.

\item \textbf{Synthetic Data Case 2 (SDC 2):}
was generated from a lognormal distribution with a mean price relative, $\mu = 1.001$, and a variance, $v = 0.0002$, to simulate a stock market where the value of the stocks are increasing over time.

The expected outcome is that the absolute portfolio will learn which stocks to hold a long position on, however the active portfolio will not be able to learn which stocks to hold a short position on as no stocks decrease in value over time.

\item \textbf{Synthetic Data Case 3 (SDC 3):}
was generated from a lognormal distribution with a random mean price relative, $\mu \geq 1$, assigned to each stock and a variance, $v = 0.0002$. The random means is calculated as follows:
\begin{equation}
\mu_{m} = 1 + \max[0,\min (0.0005+0.0005 \delta,0.001)]
\label{eqn:SDC 3}
\end{equation}

where $\delta$ is a random number generated from a standard normal distribution (using the \texttt{randn} function in MATLAB with the Mersenne Twister psuedorandom number generator \cite{MT1998} and initialised using a specific seed value). This simulates a stock market where some stocks are increasing in value and some stocks are decreasing in value over time.

The expected outcome is that both the active portfolio and the absolute portfolio will learn to hold a long position on the stocks increasing in value over time and hold a short position on the stocks decreasing in value over time, however it is expected that the absolute portfolio will beat the active portfolio due to the growth rate of the stocks increasing in value over time.
\item \textbf{Synthetic Data Case 4 (SDC 4):}\label{data:SDC 4}
was generated from a lognormal distribution with mixed means assigned to the price relatives, $\mu = 0.999$ was assigned to 3 stocks and $\mu = 1.001$ was assigned to the remaining stocks, and a variance, $v = 0.0002$. This dataset simulates a stock market where the value of some stocks are increasing and the value of some stocks are decreasing.

The expected outcome is that both the active portfolio and the absolute portfolio will learn to hold a long position on the stocks increasing in value over time and hold a short position on the stocks decreasing in value over time. 
\end{enumerate}
                 
\begin{table}
\centering
\figuretitle{Summary of Random Datasets}
\begin{tabular}{ | p{1.5cm} | p{1.9cm} | p{1.9cm} | p{1.9cm} |}
\hline  \textbf{Dataset} & \textbf{$\mu$} & $v$ \\ \hline
 
 \hline SDC 1  & 1.000 & 0.0002  \\ 
 
 \hline SDC 2 & 1.001 & 0.0002  \\ 

 \hline SDC 3 & random$\geq 1$ & 0.0002 \\ 
 
 \hline SDC 4 & mixed & 0.0002 \\ \hline

\end{tabular} 
\caption{The means and variances that were chosen when generating the synthetic data sets. The random means for SDC 3 was calculated using \Eqn{eqn:SDC 3} and the means for SDC 4 was generated as described in section \ref{data:SDC 4}.}
\label{tab:rand data}
\end{table}

\subsection{Real Data} \label{ssec:real-data}

The algorithm is tested on four sets of real data, summarised in Table \ref{tab:real data}, two data sets from the New York Stock Exchange (NYSE) obtained at \cite{NYSE2006} and two data sets from the Johannesburg Stock Exchange (JSE) obtained at \cite{TRTH2015}. 

\begin{enumerate}
\item \textbf{NYSE Data:}\label{data:NYSEdata}
This is described in \cite{NYSE2006} \footnote{This data set comes from the website of Yoram Singer\cite{NYSE2006}} and contains \textit{close-to-close} price relatives for 36 stocks listed on the New York Stock Exchange from 1962-1984. This is the same data set used by Gy{\"o}rfi \textit{et al} in \cite{GUV2007,GUW2008} and Cover in \cite{C1991}.

\item \textbf{NYSE Merged Data:}
This is described in \cite{NYSE2006}\footnote{This data was original sourced from Yahoo! Finance and was cleaned and prepared by G{\'a}bor Gelencs{\'e}r and made available on his website \cite{NYSE2006}.} and the dataset contains \textit{close-to-close} price relatives data for 23 stocks listed on the New York Stock Exchange from 1962-2006. The data of the 23 stocks during 1962-1984 is identical to the data described above in point \ref{data:NYSEdata}.

\item \textbf{JSE OHLC Data:}
This was obtained from Thomson Reuters Tick History (TRTH) \cite{TRTH2015} and contains daily data for 42 stocks listed on the Johannesburg Stock Exchange (JSE) from 1995-2015 (using RIC chain \verb|0#.JTOPI|), however not all of the 42 stocks were listed in 1995 and the data for these stocks begins at a later time \footnote{The data has an implicit survivorship bias however this does not impact the results of this paper}. The data lists the open, high, low and close prices for all of the 42 stocks. This raw data was processed into four datasets containing \textit{close-to-close}, \textit{open-to-close}, \textit{close-to-open} and \textit{open-to-open} price relatives respectively. Splits, mergers\footnote{Splits and mergers were identified as having a $x_{m,t}<0.7$ and $x_{m,t}>1.3$ respectively.} and missing data were handled by assigning a price relative of 1 for that day.

\item \textbf{JSE Intraday Data:}
The transaction data was obtained from Thomson Reuters Tick History (TRTH) \cite{TRTH2015} and consisted of top-of-book and transaction updates for 40 stocks listed on the Johannesburg Stock Exchange (JSE) during 2013 in RIC chain  \verb|0#.JTOPI|. The transaction data was converted into 5-minute bar data using the trade price and volume weighted averaging. The 5-minute bar-data starts at 9h30 and ends at 16h30 for normal trading days and starts at 9h30 and ends at 11h30 for early close days. A normal trading data on the JSE starts with an opening auction between 8h30 and 9h00, continuous trading takes place between 9h00 and 16h50, and the day ends with a closing auction between 16h50 and 17h00.
\end{enumerate}
\begin{table} 
\centering
\figuretitle{Summary of Real Datasets}
\begin{tabular}{ | p{3.4cm} | p{2.4cm} | p{1.6cm} |}
\hline \textbf{Data Set} & \textbf{Time Period} & \textbf{\# Stocks}\\ \hline
 
 \hline NYSE \cite{NYSE2006} & 1962-1984 & 36  \\
 
 \hline NYSE Merged \cite{NYSE2006} & 1962-2005 & 23  \\ 

 \hline JSE daily OHLC \cite{TRTH2015} & 1995-2015 & 42  \\ 

 \hline JSE Intraday \cite{TRTH2015} & 2013 & 40  \\ \hline

\end{tabular} 
\caption{Description of the real data sets that the algorithm was tested on.}
\label{tab:real data}
\end{table}

\section{Implementation} \label{sec:imp-issues}

The wealth achieved by the portfolio and the wealth achieved by the agents is determined using Algorithm \ref{alg: OLA} (OLA). The agent controls $\ve H_{n,t}$, introduced in section \ref{sec:online-learning}, used in Algorithm \ref{alg: OLA} is determined by using Algorithm \ref{alg: PMA} (PMA), which calls up Algorithm \ref{alg: MTA} (MTA)\footnote{Calendar effects are not fully accounted for.} to determine the agent controls for each agent. Algorithm \ref{alg: MTA} updates an agents wealth as described in \Eqn{eqn:agentcontrols}. In the experiments 50 agents were used with $K=(1,2,\ldots,5)$ and $L=(1,2,\ldots,10)$, similar to choice of agents (`experts') used by Gy{\"o}rfi \textit{et al} in \cite{GUV2007,GUW2008}.

All Results and data processing was done in MATLAB. The algorithm was implemented for both the absolute and active case using a MATLAB class that we named \textit{pattern}, a MATLAB class was used instead of function because this allows the algorithm to easily be extended to a more online approach. The \textit{pattern} class was extended to include our recovered version of the Gy{\"o}rfi \textit{et al} Nearest Neighbour \cite{GUW2008} algorithm so that the running time comparisons in section \ref{sec:results-analysis} will be accurate. The Cover \textit{et al} \cite{C1991} Universal Portfolios algorithm was recovered by creating a MATLAB function that implement the algorithm. 

\section{Results and Analysis} \label{sec:results-analysis}

\subsection{Synthetic Data} \label{ssec:SDC1234}

The algorithm was tested on four synthetic data cases (SDC) to illustrate how the algorithm performs in different types of markets. Table \ref{tab: syn data wealth} displays the best and average wealth achieved by the active and absolute portfolios for  30 runs of each synthetic data case initialised with seed values $1,2,\ldots,30$ respectively.

\begin{figure}
\centering
\figuretitle{Synthetic Data Case 1}
\includegraphics[width=0.45\textwidth]{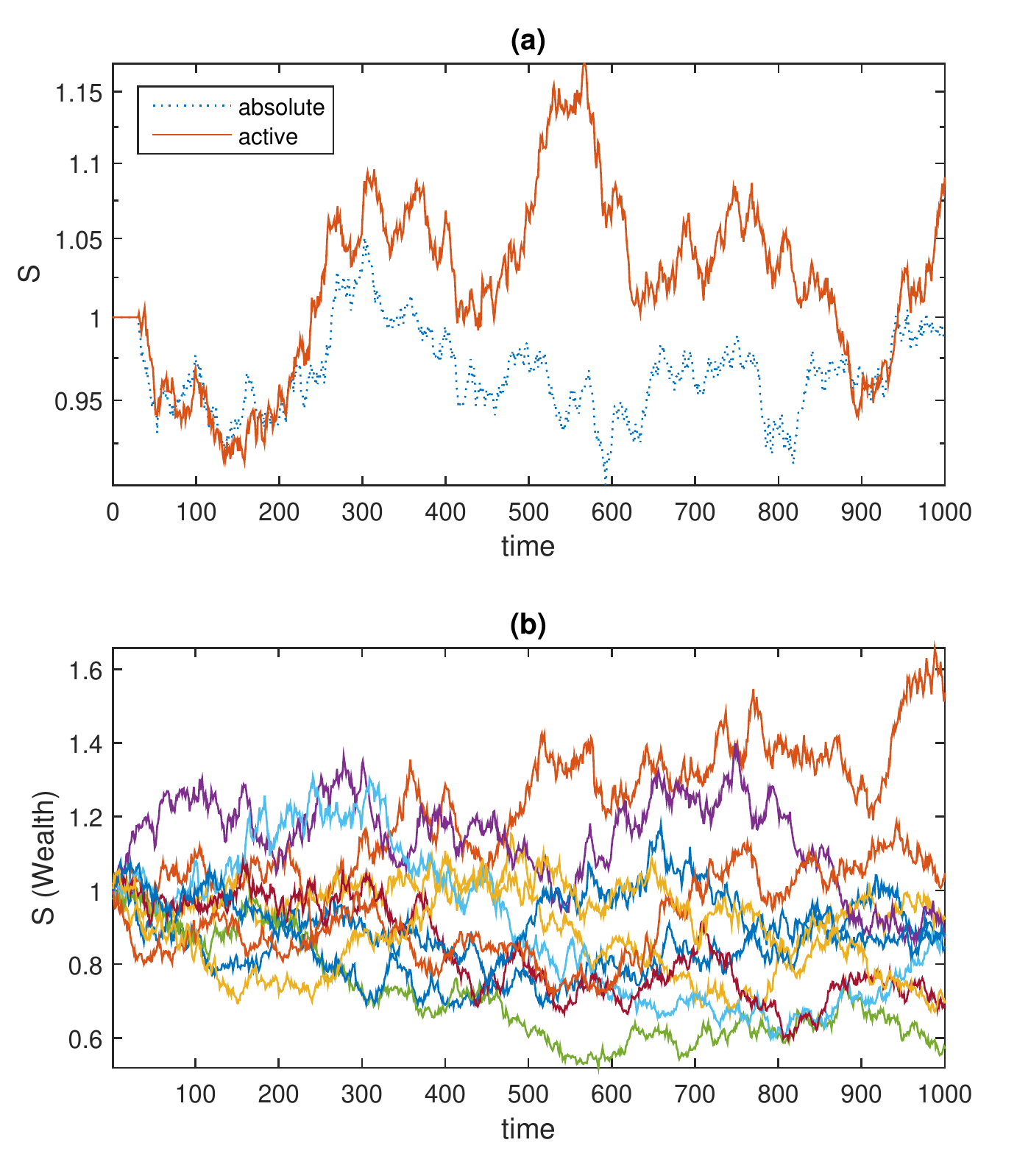}
\caption{(a) The wealth achieved by the active and absolute portfolios on SDC 1 that consists of a time period of 1000 and 10 stocks. (b) The wealth of each randomly generated stock.} 
\label{fig:SDC1_plot} 
\end{figure}

\begin{figure}
\centering
\figuretitle{Synthetic Data Case 2} 
\includegraphics[width=0.45\textwidth]{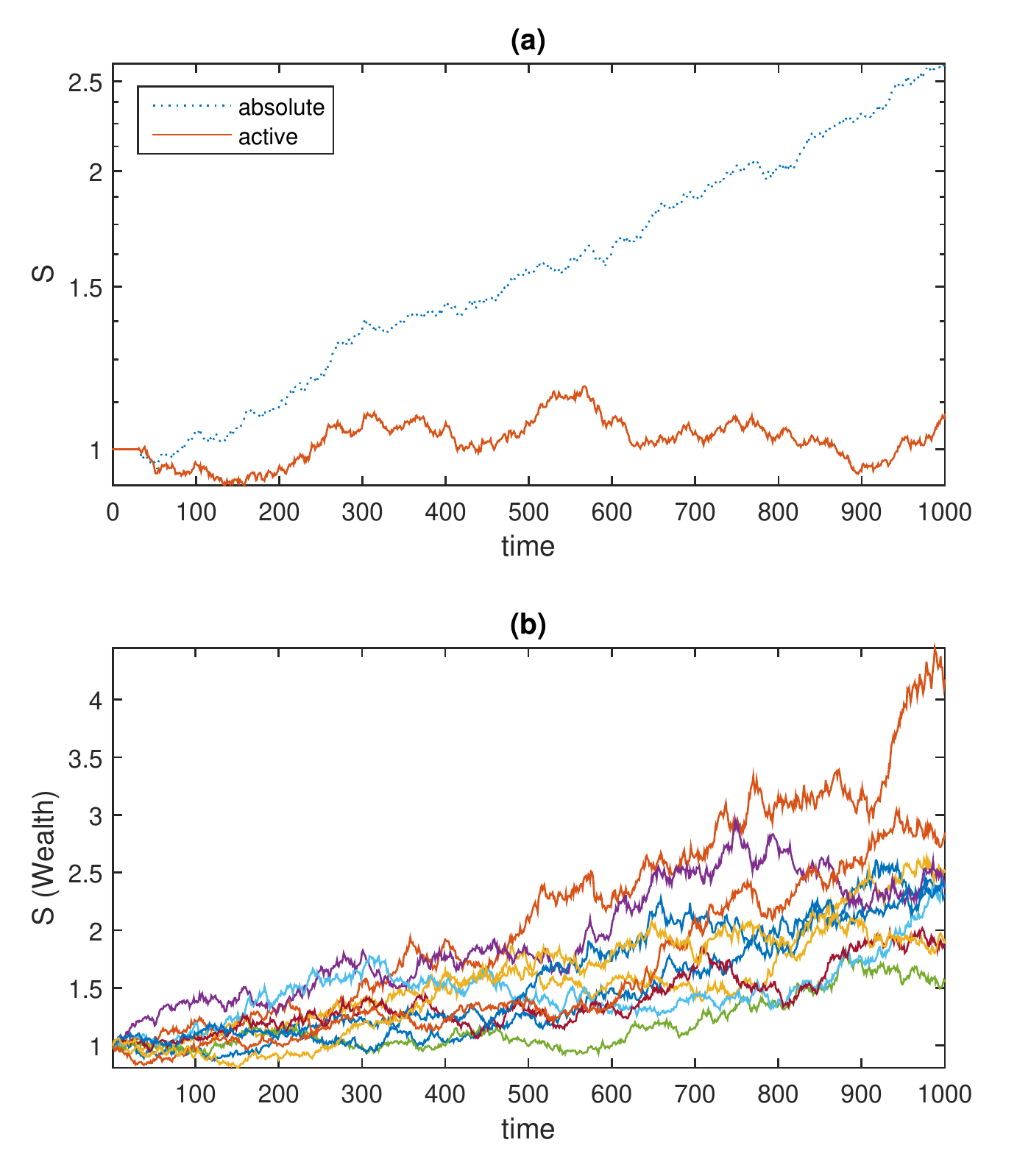}
\caption{(a) The wealth achieved by the active and absolute portfolios on SDC 2 that consists of a time period of 1000 and 10 stocks. (b) The wealth of each randomly generated stock.} 
\label{fig:SDC2_plot} 
\end{figure}

\begin{figure}%
\centering
\figuretitle{Synthetic Data Case 3}
\includegraphics[width=0.45\textwidth]{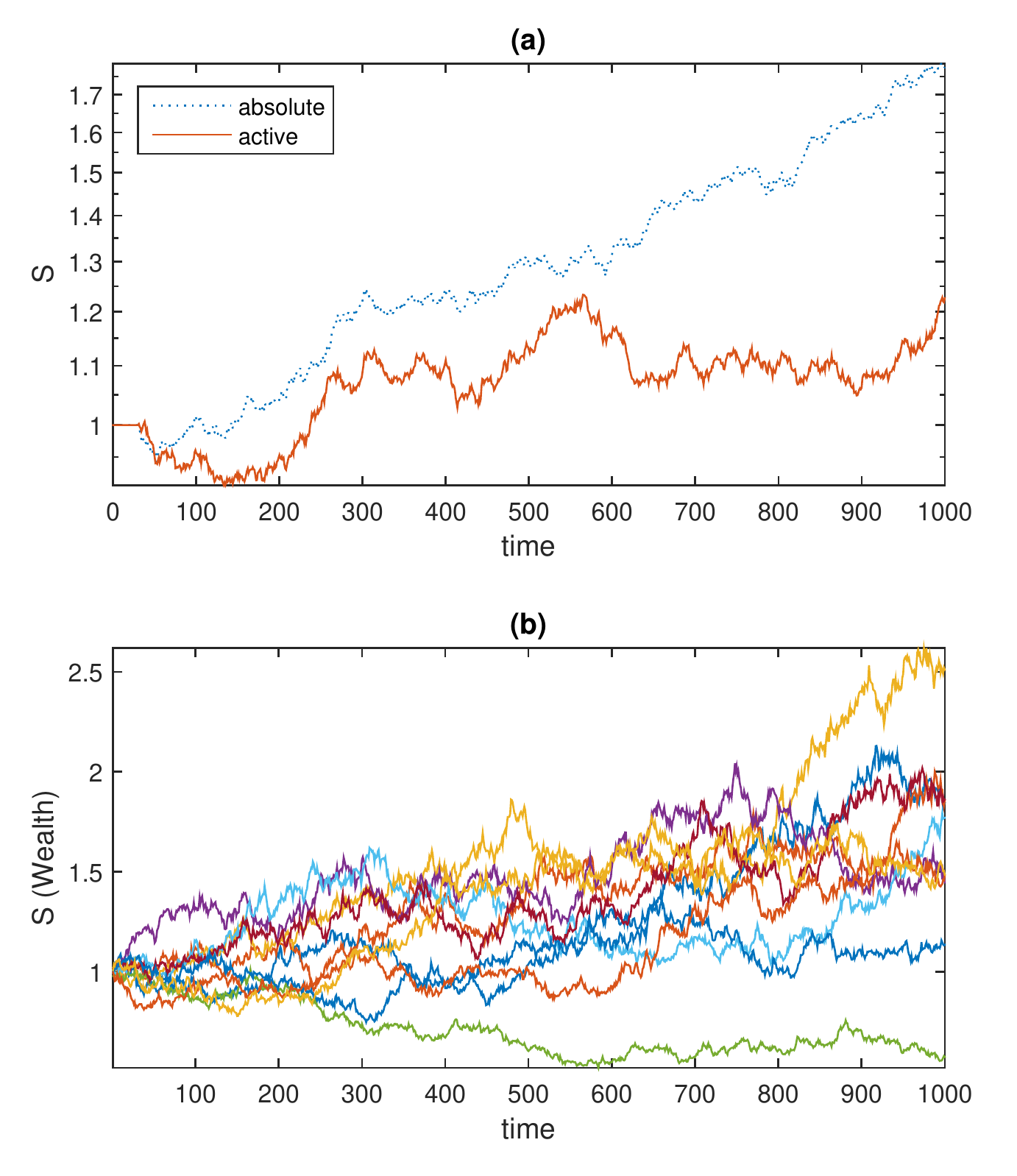}
\caption{(a) The wealth achieved by the active and absolute portfolios on SDC 3 that consists of a time period of 1000 and 10 stocks.(b) The wealth of each randomly generated stock.} 
\label{fig:SDC3_plot} 
\end{figure}

\begin{figure}%
\centering
\figuretitle{Synthetic Data Case 4}
\includegraphics[width=0.45\textwidth]{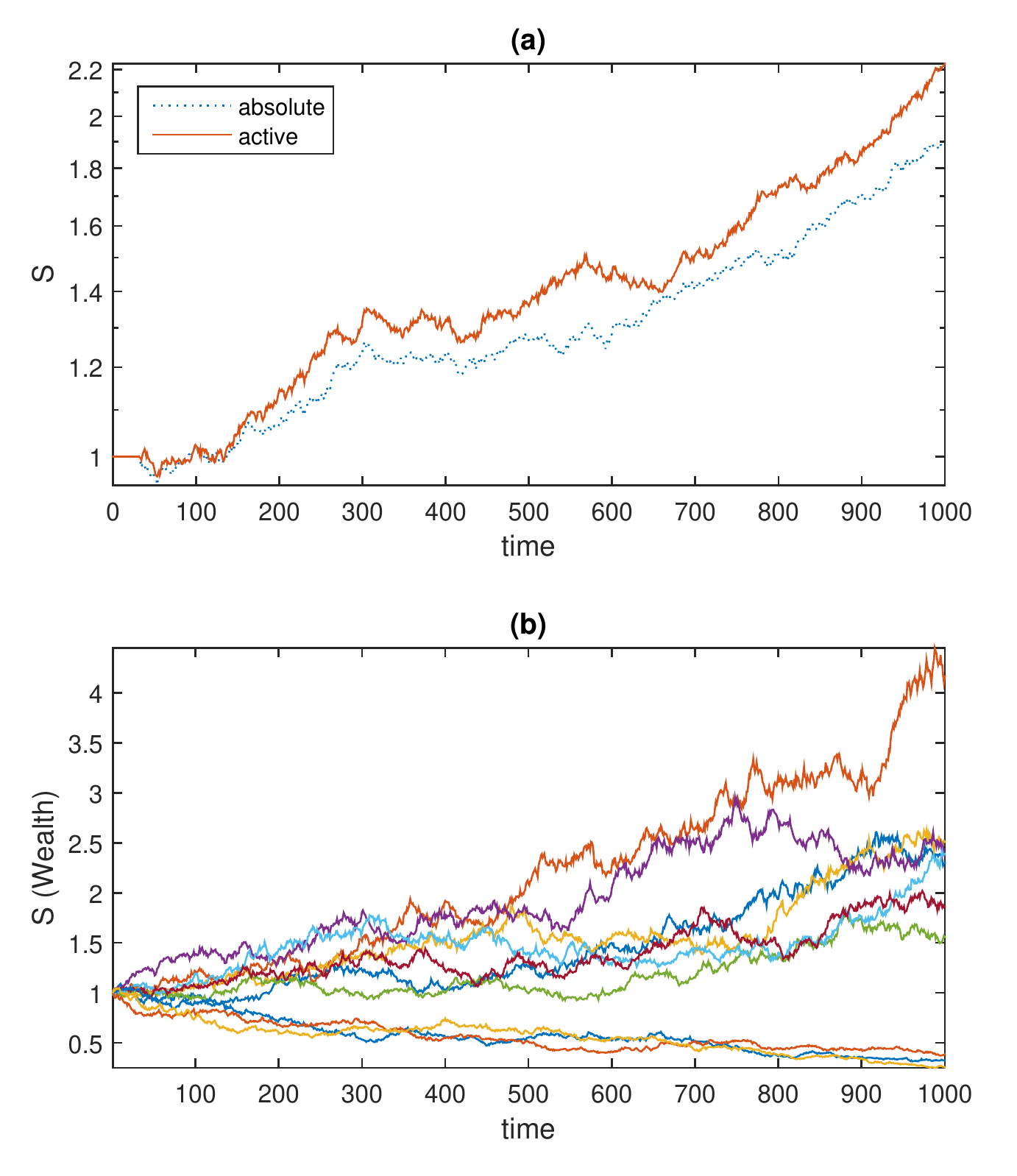}
\caption{(a) The wealth achieved by the active and absolute portfolios on SDC 4 that consists of a time period of 1000 and 10 stocks. (b) The wealth of each randomly generated stock.} 
\label{fig:SDC4_plot} 
\end{figure}

On all of the datasets the algorithm, when using absolute portfolio, eventually learns the stocks that are increasing in value over time as observed for SDC 2, 3 and 4. Similarly the algorithm, when using active portfolio, eventually learns to hold a long position on the stocks that are increasing in value over time and hold a short position on the stocks that are decreasing in value over time; as observed for SDC 3 and 4. Figures \ref{fig:SDC1_plot}, \ref{fig:SDC2_plot}, \ref{fig:SDC3_plot} and \ref{fig:SDC4_plot} shows the wealth achieved by the active and absolute portfolios, as well as the wealth achieved by each synthetic stock when randomly generated using an initial seed value of 7.

\begin{table}
\centering
\figuretitle{Wealth ($S$) from Investing in Synthetic Data}
\begin{tabular}{ | c | c | p{1.0cm} | p{1.0cm} | p{1.0cm} | p{1.0cm}|}  
\hline \multirow{2}{*}{ \textbf{Data}} & \multirow{2}{*}{\textbf{Port.}} & \multicolumn{2}{c|}{Wealth} & \multicolumn{2}{|c|}{Best Agent}  \\   
\cline{3-6}  &  & Best & Avg. & Best & Avg. \\   
 \hline SDC 1 & \begin{tabular}{l}Abs. \\ Act. \end{tabular} & \begin{tabular}{l}   1.231 \\    1.451 \end{tabular} & \begin{tabular}{l}   0.992 \\    1.052 \end{tabular} & \begin{tabular}{l}   1.806 \\    1.753 \end{tabular} & \begin{tabular}{l}   1.250 \\    1.358 \end{tabular}  \\ 
 \hline SDC 2 & \begin{tabular}{l}Abs. \\ Act. \end{tabular} & \begin{tabular}{l}   3.241 \\    1.451 \end{tabular} & \begin{tabular}{l}   2.612 \\    1.052 \end{tabular} & \begin{tabular}{l}   4.654 \\    1.753 \end{tabular} & \begin{tabular}{l}   3.270 \\    1.358 \end{tabular}  \\ 
 \hline SDC 3 & \begin{tabular}{l}Abs. \\ Act. \end{tabular} & \begin{tabular}{l}   2.320 \\    1.490 \end{tabular} & \begin{tabular}{l}   1.685 \\    1.171 \end{tabular} & \begin{tabular}{l}   3.091 \\    1.782 \end{tabular} & \begin{tabular}{l}   2.090 \\    1.451 \end{tabular}  \\ 
 \hline SDC 4 & \begin{tabular}{l}Abs. \\ Act. \end{tabular} & \begin{tabular}{l}   2.455 \\    2.927 \end{tabular} & \begin{tabular}{l}   1.896 \\    2.055 \end{tabular} & \begin{tabular}{l}   2.931 \\    3.270 \end{tabular} & \begin{tabular}{l}   2.250 \\    2.297 \end{tabular}  \\ 
\hline 
\end{tabular}
\caption{Wealth achieved by the active and absolute portfolios for  30 runs of each synthetic data case.}
\label{tab: syn data wealth}
\end{table}

Tables \ref{tab:p values act rand data} and \ref{tab:p values abs rand data} displays average $p$ values from the two-sample Kolmogorov-Smirnov tests when comparing the following combinations of the total wealth gained from the portfolio ($S_1$), the wealth gained from the best agent of the portfolio ($S_2$) and the wealth gained from the best stock ($S_3$):

\begin{enumerate}
\item $S_2>S_1$ : The alternative hypothesis that the cumulative distribution function (CDF) of the wealth gained from the best agent of the portfolio, $S_2$, is larger than the CDF of the total wealth gained from the portfolio, $S_1$, at the 5\% significance level.
\item $S_2>S_3$ : The alternative hypothesis that the CDF of the wealth gained from the best agent of the portfolio, $S_2$, is larger than the CDF of the wealth gained from the best stock, $S_3$, at the 5\% significance level.
\item $S_3>S_1$ : The alternative hypothesis that the CDF of the wealth gained from the best stock, $S_3$, is larger than the CDF of the total wealth gained from the portfolio, $S_1$, at the 5\% significance level.
\end{enumerate} 

The two-sample Kolmogorov-Smirnov test was chosen because it is a non-parametric test and makes no assumption about the distribution of the datasets.

\begin{table}
\centering
\figuretitle{Average $p$ values of Wealth ($S$) for Active Portfolios}
\begin{tabular}{ |c | c | c | c | c| c| c|}  
\hline   & \multicolumn{2}{p{2cm}}{{\begin{tabular}{c} Best Agent\\ vs.\\ Tot. Wealth \end{tabular} }} & \multicolumn{2}{|p{2cm}|}{{\begin{tabular}{c} Best Agent\\ vs. \\Best Stock \end{tabular} }} & \multicolumn{2}{p{2cm}|}{{\begin{tabular}{c} Best Stock\\ vs. \\Tot. Wealth \end{tabular}}} \\ \hline  
\textbf{Hyp.}  & \multicolumn{2}{c|}{$S_2>S_1$} & \multicolumn{2}{c|}{$S_2>S_3$} & \multicolumn{2}{c|}{$S_3>S_1$}  \\  
\hline  & $ \bar{p}$ & $p>\bar{p}$ & $\bar{p}$ & $p>\bar{p}$ & $\bar{p}$ & $p>\bar{p}$  \\ 
\hline 
 \hline SDC 1 &  0.809 &  0.172 &  0.031 &  0.000 &  0.654 &  0.013 \\  
 \hline SDC 2 &  0.809 &  0.172 &  0.000 &  0.000 &  0.873 &  0.407 \\  
 \hline SDC 3 &  0.830 &  0.172 &  0.000 &  0.000 &  0.904 &  0.563 \\  
 \hline SDC 4 &  0.725 &  0.013 &  0.000 &  0.000 &  0.622 &  0.006 \\  
\hline 
\end{tabular}
 
\caption{Comparisons of the average $p$ values of the wealth gained from the active portfolio. The first $p$ value in each column is average $p$ value, of the 30 data sets for each case, using two-sample Kolmogorov-Smirnov tests for the alternative hypotheses (Hyp.). The second $p$ value is obtained from the two-sample Kolmogorov-Smirnov tests for the alternative hypothesis that the cumulative distribution function (CDF) of the $p$ values for the 30 data sets for each case is larger than the CDF of the average $p$ value at the 5\% significance level.}
\label{tab:p values act rand data}
\end{table}

\begin{table}
\centering
\figuretitle{Average $p$ Values of Wealth Gained ($S$) from the Absolute Portfolio}
\begin{tabular}{ | c | c | c | c| c | c | c|}  
\hline   & \multicolumn{2}{|p{2cm}|}{{\begin{tabular}{c} Best Agent \\ vs. \\ Wealth \end{tabular} }} & \multicolumn{2}{p{2cm}|}{{\begin{tabular}{c} Best Agent \\ vs. \\ Best Stock \end{tabular} }} & \multicolumn{2}{p{2cm}|}{{\begin{tabular}{c} Best Stock \\ vs. \\ Wealth \end{tabular} }} \\ \hline  
{Hyp.}  & \multicolumn{2}{|c|}{$S_2>S_1$} & \multicolumn{2}{c|}{$S_2>S_3$} & \multicolumn{2}{c|}{$S_3>S_1$}  \\  
\hline  & $ \bar{p}$ & $p > \bar{p}$ & $\bar{p}$ & $p > \bar{p}$ & $\bar{p}$ & $p > \bar{p}$  \\
\hline 
 \hline SDC 1 &  0.893 &  0.407 &  0.000 &  0.000 &  0.647 &  0.013 \\  
 \hline SDC 2 &  0.642 &  0.013 &  0.000 &  0.000 &  0.567 &  0.001 \\  
 \hline SDC 3 &  0.593 &  0.002 &  0.000 &  0.000 &  0.795 &  0.100 \\  
 \hline SDC 4 &  0.846 &  0.274 &  0.000 &  0.000 &  0.644 &  0.006 \\  
\hline 
\end{tabular} 
\caption{Comparisons of the average $p$ values of the wealth gained from the absolute portfolio. The first $p$ value in each column is the average $p$ value, of the 30 data sets for each case, using two-sample Kolmogorov-Smirnov tests for the alternative hypotheses (Hyp.). The second $p$ value is obtained from the two-sample Kolmogorov-Smirnov tests for the alternative hypothesis that the cumulative distribution function (CDF) of the $p$ values for the 30 data sets for each case is larger than the CDF of the average $p$ value at the 5\% significance level.}
\label{tab:p values abs rand data}
\end{table}

\begin{table}
\centering
\figuretitle{Comparison of $p$ Values of Wealth Gained ($S$) from the Active Portfolio}
\begin{tabular}{ | p{1.3cm} | p{1.4cm} | p{1.4cm} | p{1.4cm} | p{1.4cm} |}  
\hline   & \textbf{SDC 1} & \textbf{SDC 2} & \textbf{SDC 3} & \textbf{SDC 4} \\ \hline 

 \hline \textbf{SDC 1} & \hspace{0.65cm} - & 0.9765  &      0.0669 &      0 \\  
 \hline \textbf{SDC 2} & 0.9786  & \hspace{0.65cm} - &      0.0669 &      0 \\  
 \hline \textbf{SDC 3} & 0.7928 & 0.7916 & \hspace{0.65cm} - & 0 \\  
 \hline \textbf{SDC 4} & 0.9653 & 0.9649 & 0.9535 & \hspace{0.65cm} - \\  
\hline 
\end{tabular}  
\caption{Comparison of the average $p$ values from two-sample Kolmogorov-Smirnov tests for the alternative hypothesis that the cumulative distribution function (CDF) of wealth gained from the active portfolio on SDC $i$ is larger than the CDF of wealth gained from the active portfolio on SDC $j$ at the 5\% significance level, where $i$ represents the rows and $j$ represents the columns of the table. The $p$ values is the average of 30 comparisons, each comparison using a seed value of $1,2,\ldots,30$ respectively.}
\label{tab:p values act comp}
\end{table}

\begin{table}
\centering
\figuretitle{Comparison of $p$ Values of Wealth Gained ($S$) from the Absolute Portfolio}
\begin{tabular}{ | p{1.3cm} | p{1.4cm} | p{1.4cm} | p{1.4cm} | p{1.4cm} |}  
\hline   & \textbf{SDC 1} & \textbf{SDC 2} & \textbf{SDC 3} & \textbf{SDC 4} \\ \hline 

 \hline \textbf{SDC 1} & \hspace{0.65cm} - & 0 &      0 &      0 \\  
 \hline \textbf{SDC 2} & 1.0000 & \hspace{0.65cm} - &  1.0000 &  0.9997 \\  
 \hline \textbf{SDC 3} & 1.0000 & 0 & \hspace{0.65cm} - & 0.1289 \\  
 \hline \textbf{SDC 4} & 1.0000 & 0 & 0.5055 & \hspace{0.65cm} - \\  
\hline 
\end{tabular}
\caption{Comparison of the average $p$ values from two-sample Kolmogorov-Smirnov tests for the alternative hypothesis that the cumulative distribution function (CDF) of wealth gained from the absolute portfolio on SDC $i$ is larger than the CDF of wealth gained from the absolute portfolio on SDC $j$ at the 5\% significance level, where $i$ represents the rows and $j$ represents the columns of the table. The $p$ values is the average of 30 comparisons, each comparison using a seed value of $1,2,\ldots,30$ respectively.}
\label{tab:p values abs comp}
\end{table}

\subsection{NYSE Data}\label{ssec:NYSE}

The algorithm was run on the NYSE data set for both absolute and active portfolios on the same pairs of stocks used by Cover in \cite{C1991} and by Gy{\"o}rfi \textit{et al} in \cite{GUV2007,GUW2008}. Table \ref{table:NYSE} shows the wealth achieved by the active and absolute portfolios and is compared to reference results from the literature  when using the nearest neighbour strategy ($G_{NN}$) by Gy{\"o}rfi \textit{et al} \cite{GUV2007,GUW2008} and the universal portfolio strategy (UP) by Cover \cite{C1991}.

$G^{*}_{NN}$ denotes our best recovery of the results of the nearest neighbour strategy \cite{GUV2007,GUW2008}. The results achieved by from the universal portfolio strategy was identically recovered \cite{C1991}. The last row of table \ref{table:NYSE} shows the results of the strategies when running on all 36 NYSE stocks.

\begin{table}
\centering
\figuretitle{Wealth Achieved by Investing in Various Combinations of Stocks From NYSE Dataset}
\begin{tabular}{ | p{1.5cm} | p{1.2cm}| p{1.7cm} | p{1.7cm} |} 
\hline \textbf{Stocks} & \textbf{Strat.} & \textbf{Wealth} & \textbf{Best Agent}  \\ \hline 
\hline \begin{tabular}{@{}l@{}} \textbf{IROQU} \\ \textbf{KINAR} \end{tabular} & \begin{tabular}{@{}l@{}}Abs. \\ Act. \\ $G_{NN}$ \\ $G^{*}_{NN}$ \\ UP \\ Best \end{tabular} & \begin{tabular}{@{}l@{}} 1.02e+12 \\ 1.00e+11 \\ 1.16e+12 \\ 1.01e+12 \\ 38.67 \\ 8.92 \end{tabular} & \begin{tabular}{@{}l@{}} 1.76e+13 \\ 4.97e+11 \\ 1.44e+13 \\ 1.63e+13 \\ \\ \\ \end{tabular} \\ 
\hline \begin{tabular}{@{}l@{}} \textbf{COMME} \\ \textbf{MEICO} \end{tabular} & \begin{tabular}{@{}l@{}}Abs. \\ Act. \\ $G_{NN}$ \\ $G^{*}_{NN}$ \\ UP \\ Best \end{tabular} & \begin{tabular}{@{}l@{}} 3.56e+03 \\ 4.28e+01 \\ 3.51e+3 \\ 3.58e+03 \\ 72.63 \\ 52.02 \end{tabular} & \begin{tabular}{@{}l@{}} 2.61e+04 \\ 2.49e+02 \\ 3.15e+4 \\ 2.60e+04 \\ \\ \\ \end{tabular} \\ 
\hline \begin{tabular}{@{}l@{}} \textbf{COMME} \\ \textbf{KINAR} \end{tabular} & \begin{tabular}{@{}l@{}}Abs. \\ Act. \\ $G_{NN}$ \\ $G^{*}_{NN}$ \\ UP \\ Best \end{tabular} & \begin{tabular}{@{}l@{}} 2.99e+12 \\ 1.05e+11 \\ 4.78e+12 \\ 3.09e+12 \\ 78.47 \\ 52.02 \end{tabular} & \begin{tabular}{@{}l@{}} 3.46e+13 \\ 3.75e+11 \\ 8.26e+13 \\ 3.75e+13 \\ \\ \\ \end{tabular} \\ 
\hline \begin{tabular}{@{}l@{}} \textbf{IBM} \\ \textbf{COKE} \end{tabular} & \begin{tabular}{@{}l@{}}Abs. \\ Act. \\ $G_{NN}$ \\ $G^{*}_{NN}$ \\ UP \\ Best \end{tabular} & \begin{tabular}{@{}l@{}} 7.84e+01 \\ 9.70e+00 \\ 74.37 \\ 7.83e+01 \\ 14.18 \\ 13.36 \end{tabular} & \begin{tabular}{@{}l@{}} 2.74e+02 \\ 1.98e+01 \\ 296.3 \\ 2.67e+02 \\ \\ \\ \end{tabular} \\ 
\hline \begin{tabular}{@{}l@{}} \textbf{36} \\ \textbf{STOCKS} \end{tabular} & \begin{tabular}{@{}l@{}}Abs. \\ Act. \\ $G_{NN}$ \\ $G^{*}_{NN}$ \\ Best \end{tabular} & \begin{tabular}{@{}l@{}} 5.42e+01 \\ 5.29e+01 \\ 3.3e+11 \\ 3.43e+11 \\ 54.14 \end{tabular} & \begin{tabular}{@{}l@{}} 1.36e+02 \\ 7.13e+01 \\ 7.7e+12 \\ 7.45e+12 \\ \\ \end{tabular} \\ 
\hline 
\end{tabular}
\caption{Comparison of the total wealth achieved from the active (Act.) and absolute (Abs.) portfolios to the wealth achieved from the Gy{\"o}rfi \textit{et al} nearest neighbour ($G_{NN}$), attempted recovery of the Gy{\"o}rfi \textit{et al} nearest neighbour ($G^{*}_{NN}$), the universal portfolio (UP) and a buy-and-hold of the best stock strategies. }
\label{table:NYSE}
\end{table}
  
The algorithm compares well to the two stocks combinations used by Cover in \cite{C1991} and by Gy{\"o}rfi \textit{et al} in \cite{GUV2007,GUW2008}.  A surprising result is how the wealth achieved by the portfolio when run over all 36 stocks compares to results by Gy{\"o}rfi \textit{et al} in \cite{GUV2007,GUW2008}, this may be due to a loss of accuracy in the quadratic approximation step of the algorithm as the number of stocks increase.

\begin{figure}
    \centering
    \figuretitle{Wealth Gained in NYSE Data Experiments}
    \begin{subfigure}[b]{0.45\textwidth}
        \includegraphics[width=\textwidth]{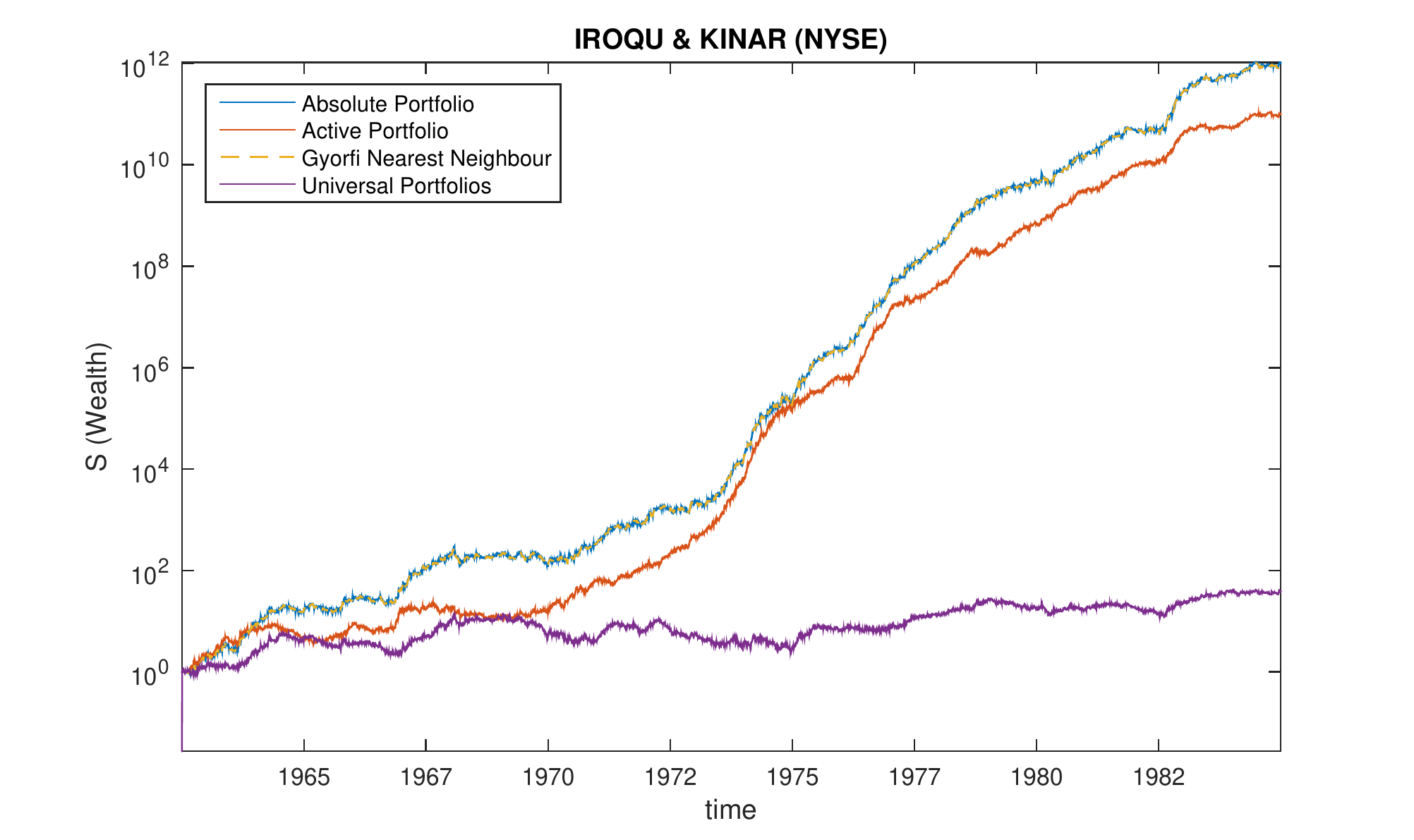}
        \caption{•}
        \label{fig:nyseold_iroqu_kinar_plot}
    \end{subfigure}
    \begin{subfigure}[b]{0.45\textwidth}
        \includegraphics[width=\textwidth]{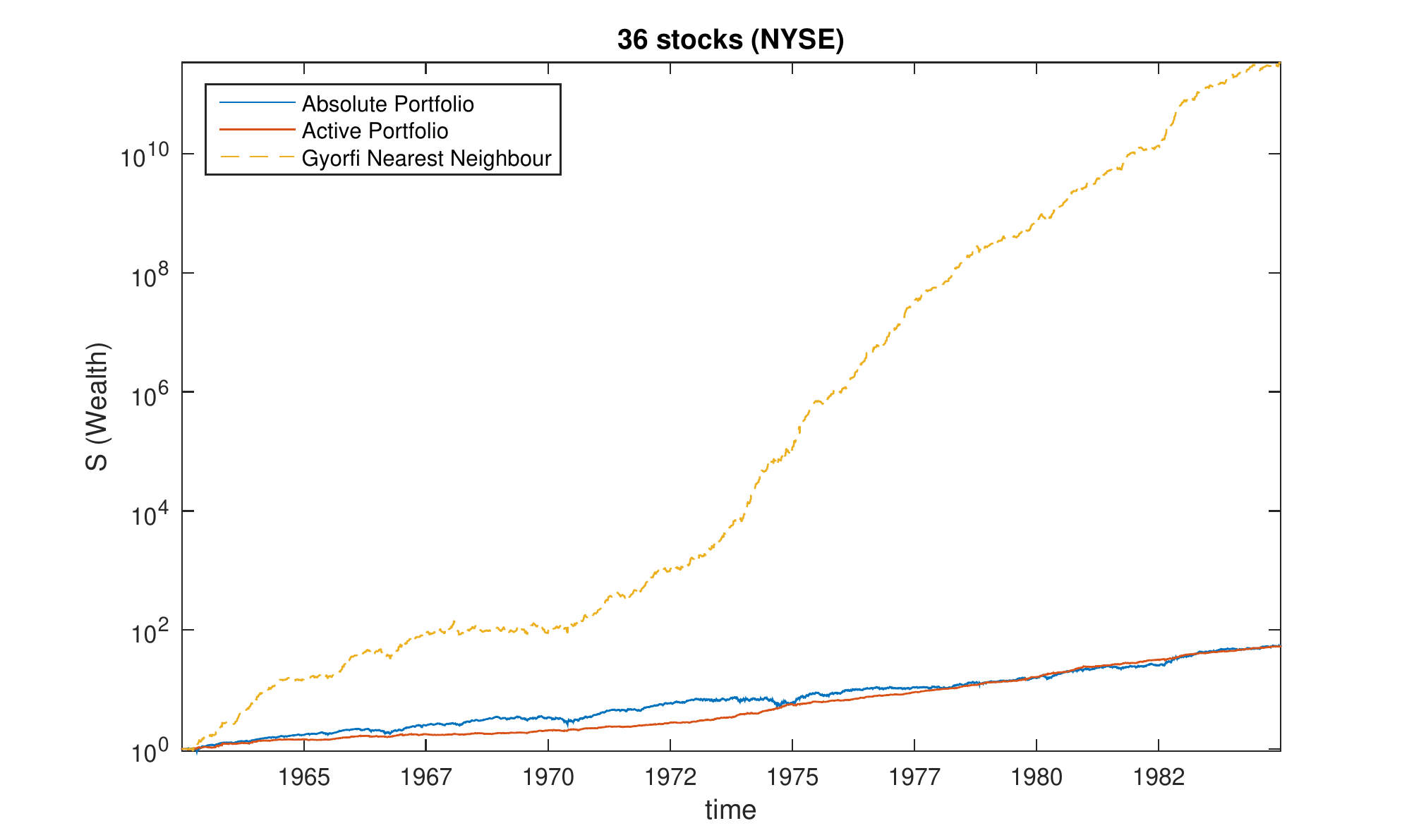}
        \caption{•}
        \label{fig:nyseold_36_stocks_plot}
    \end{subfigure}
    \caption{ Comparison of the wealth gained from different methods when investing in (a) iroqu and kinar (b) 36 stocks from  the NYSE dataset.}\label{fig:NYSE_wealthplot}
\end{figure}

\subsection{NYSE Merged Data} \label{ssec:mergedNYSE}

The algorithm was run for both absolute and active portfolios on the NYSE Merged dataset on two stock combinations and on all of the 23 stocks in the dataset. The two stock combinations chosen were stocks Commercial Metals and Kin Ark Corp. and stocks IBM and Coca-Cola.

\begin{table}
\centering
\figuretitle{Wealth Achieved by Investing in Various Combinations of Stocks From NYSE Merged Dataset}
\begin{tabular}{ | p{1.5cm} | p{1.2cm}| p{1.7cm} | p{1.7cm} |} 
\hline \textbf{Stocks} & \textbf{Strat.} & \textbf{Wealth} & \textbf{Best Agent}  \\ \hline 
\hline \begin{tabular}{@{}l@{}} \textbf{COMME} \\ \textbf{KINAR} \end{tabular} & \begin{tabular}{@{}l@{}}Abs. \\ Act. \\ $G^{*}_{NN}$ \\ UP \\ Best \end{tabular} & \begin{tabular}{@{}l@{}} 3.07e+19 \\ 7.36e+16 \\ 3.19e+19 \\ 2192.43 \\ 1344.3 \end{tabular} & \begin{tabular}{@{}l@{}} 4.37e+20 \\ 7.93e+16 \\ 4.73e+20 \\ \\ \\ \end{tabular} \\ 
\hline \begin{tabular}{@{}l@{}} \textbf{IBM} \\ \textbf{COKE} \end{tabular} & \begin{tabular}{@{}l@{}}Abs. \\ Act. \\ $G^{*}_{NN}$ \\ UP \\ Best \end{tabular} & \begin{tabular}{@{}l@{}} 1.79e+03 \\ 7.79e+00 \\ 1.79e+03 \\ 229.13 \\ 365.92 \end{tabular} & \begin{tabular}{@{}l@{}} 5.12e+03 \\ 2.63e+01 \\ 4.75e+03 \\ \\ \\ \end{tabular} \\ 
\hline \begin{tabular}{@{}l@{}} \textbf{23} \\ \textbf{STOCKS} \end{tabular} & \begin{tabular}{@{}l@{}}Abs. \\ Act. \\ $G^{*}_{NN}$ \\ Best \end{tabular} & \begin{tabular}{@{}l@{}} 8.05e+04 \\ 1.45e+06 \\ 3.68e+17 \\ 3496.7 \end{tabular} & \begin{tabular}{@{}l@{}} 3.34e+05 \\ 5.42e+05 \\ 5.60e+18 \\ \\ \end{tabular} \\ 
\hline 
\end{tabular}
\caption{Comparison of the total wealth achieved from the active (Act.) and absolute (Abs.) portfolios to the wealth achieved from the attempted recovery of the Gy{\"o}rfi \textit{et al} nearest neighbour strategy ($G^{*}_{NN}$), the attempted recovery of the universal portfolio strategy ($UP^*$) and a buy-and-hold strategy of the best stock (Best). }
 \label{table: NYSEmerged}
\end{table}

\begin{figure}
    \centering
    \figuretitle{Wealth Gained in NYSE Merged Data Experiments}
    \begin{subfigure}[b]{0.45\textwidth}
        \includegraphics[width=\textwidth]{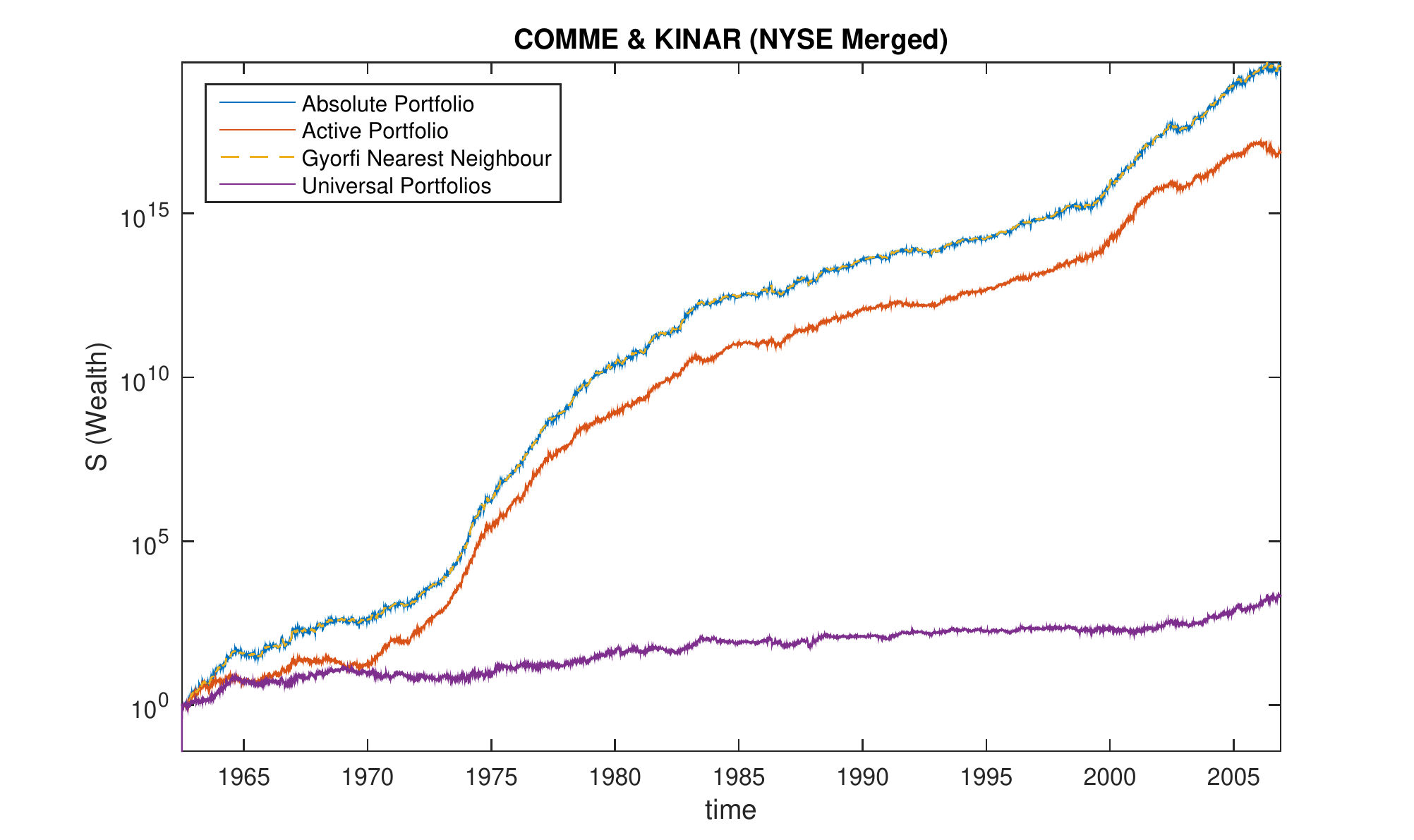}
        \caption{•}
        \label{fig:nysemerged_comme_kinar_plot}
    \end{subfigure}
    \begin{subfigure}[b]{0.45\textwidth}
        \includegraphics[width=\textwidth]{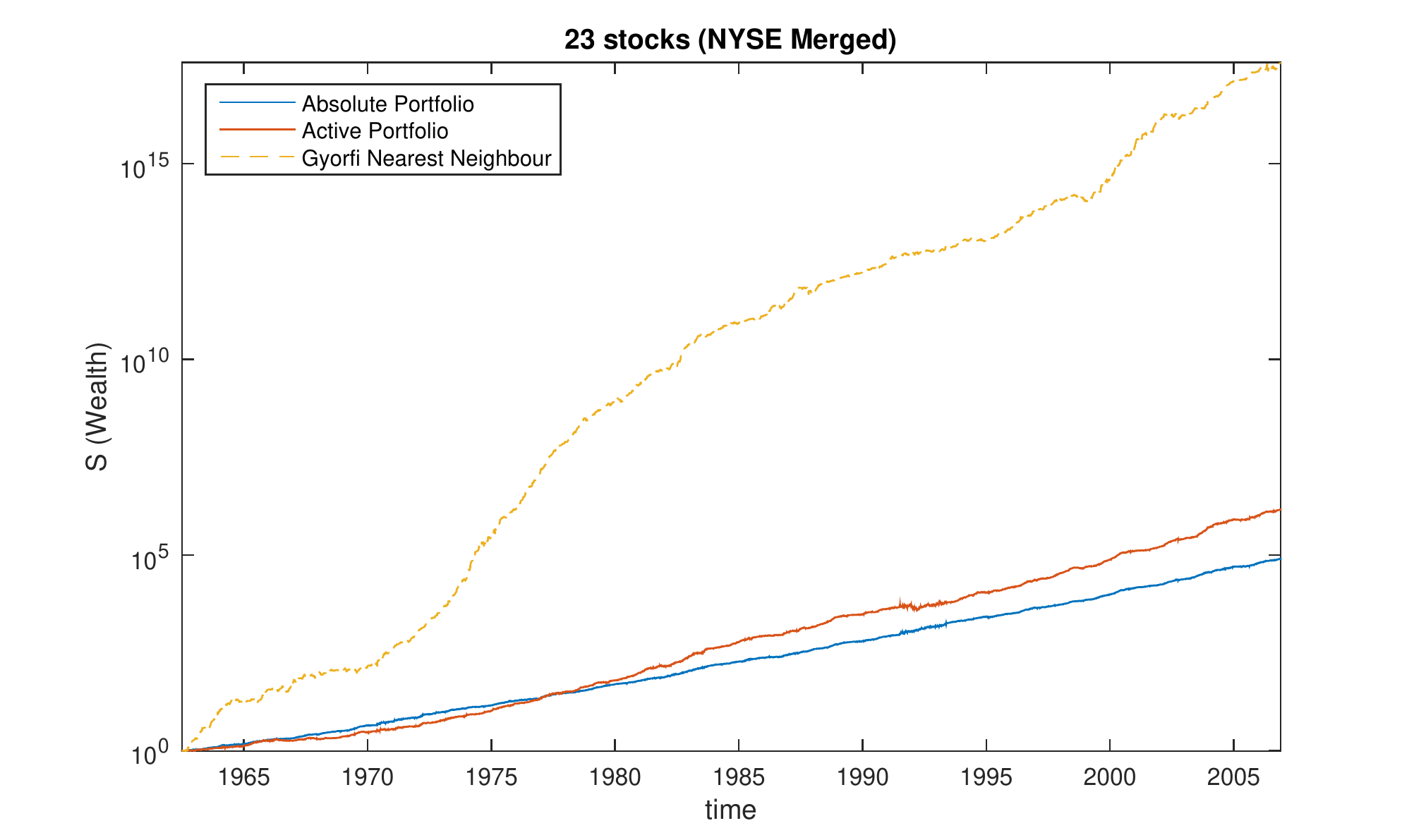}
        \caption{•}
        \label{fig:nysemerged_23_stocks_plot}
    \end{subfigure}
    \caption{ Comparison of the wealth gained from different methods when investing in (a) comme and kinar (b) 23 stocks from  the NYSE Merged dataset.}\label{fig:NYSEMerged_wealthplot}
\end{figure}

\subsection{Daily sampled JSE data} \label{ssec:dailyJSE}

The algorithm was run for both absolute and active portfolios on various sets of two stock combinations, namely stocks AngloGold Ashanti Ltd and Anglo American PLC, stocks Standard Bank Group Ltd and FirstRand Ltd,  and stocks Tiger Brands Ltd and Woolworths Holdings Ltd. The algorithm was also run for both absolute and active portfolios on a combination of 10 stocks, 20 stocks and 30 stocks. In each case the date for which the data of a stock starts may be different, the time period for the algorithm therefore starts with the stock that has a later starting time. The JSE OHLC dataset was processed into four datasets containing \textit{close-to-close}, \textit{open-to-close}, \textit{close-to-open} and \textit{open-to-open} price relatives, the algorithm was run on each of these datasets.

\begin{table}
\centering
\figuretitle{Wealth Achieved by Investing in Various Combinations of Stocks From JSE OHLC Dataset (close-to-close)}
\begin{tabular}{ | p{1.5cm} | p{1.2cm}| p{1.7cm} | p{1.7cm} |} 
\hline \textbf{Stocks} & \textbf{Strat.} & \textbf{Wealth} & \textbf{Best Agent}  \\ \hline 
\hline \begin{tabular}{@{}l@{}} \textbf{ANGJ} \\ \textbf{AGLJ} \end{tabular} & \begin{tabular}{@{}l@{}}Abs. \\ Act. \\ $G^{*}_{NN}$ \\ UP \\ Best \end{tabular} & \begin{tabular}{@{}l@{}}  4.05 \\  1.28 \\  4.02 \\  2.52 \\  3.61 \end{tabular} & \begin{tabular}{@{}l@{}} 13.10 \\  3.40 \\ 13.27 \\ \\ \\ \end{tabular} \\ 
\hline \begin{tabular}{@{}l@{}} \textbf{SBKJ} \\ \textbf{FSRJ} \end{tabular} & \begin{tabular}{@{}l@{}}Abs. \\ Act. \\ $G^{*}_{NN}$ \\ UP \\ Best \end{tabular} & \begin{tabular}{@{}l@{}} 55.53 \\  7.77 \\ 55.60 \\ 18.17 \\ 21.09 \end{tabular} & \begin{tabular}{@{}l@{}} 187.62 \\ 13.40 \\ 189.08 \\ \\ \\ \end{tabular} \\ 
\hline \begin{tabular}{@{}l@{}} \textbf{TBSJ} \\ \textbf{WHLJ} \end{tabular} & \begin{tabular}{@{}l@{}}Abs. \\ Act. \\ $G^{*}_{NN}$ \\ UP \\ Best \end{tabular} & \begin{tabular}{@{}l@{}}  7.07 \\  0.49 \\  7.07 \\  8.24 \\  8.97 \end{tabular} & \begin{tabular}{@{}l@{}} 19.97 \\  1.80 \\ 19.49 \\ \\ \\ \end{tabular} \\ 
\hline \begin{tabular}{@{}l@{}} \textbf{10} \\ \textbf{STOCKS} \end{tabular} & \begin{tabular}{@{}l@{}}Abs. \\ Act. \\ $G^{*}_{NN}$ \\ Best \end{tabular} & \begin{tabular}{@{}l@{}} 68.49 \\  9.28 \\ 194.76 \\ 89.72 \end{tabular} & \begin{tabular}{@{}l@{}} 135.03 \\ 12.84 \\ 854.62 \\ \\ \end{tabular} \\ 
\hline \begin{tabular}{@{}l@{}} \textbf{20} \\ \textbf{STOCKS} \end{tabular} & \begin{tabular}{@{}l@{}}Abs. \\ Act. \\ $G^{*}_{NN}$ \\ Best \end{tabular} & \begin{tabular}{@{}l@{}} 16.20 \\  9.52 \\ 98.84 \\ 89.72 \end{tabular} & \begin{tabular}{@{}l@{}} 63.18 \\ 12.11 \\ 330.37 \\ \\ \end{tabular} \\ 
\hline \begin{tabular}{@{}l@{}} \textbf{30} \\ \textbf{STOCKS} \end{tabular} & \begin{tabular}{@{}l@{}}Abs. \\ Act. \\ $G^{*}_{NN}$ \\ Best \end{tabular} & \begin{tabular}{@{}l@{}} 20.48 \\  7.34 \\ 124.91 \\ 85.03 \end{tabular} & \begin{tabular}{@{}l@{}} 58.27 \\  6.24 \\ 590.51 \\ \\ \end{tabular} \\ 
\hline 
\end{tabular}
\caption{The total wealth achieved by the active (Act.) and absolute (Abs.) portfolios compared to the wealth achieved from the attempted recovery of the Gy{\"o}rfi \textit{et al} nearest neighbour strategy ($G^{*}_{NN}$), the attempted recovery of the universal portfolio strategy ($UP^*$) and a buy-and-hold strategy of the best stock (Best) on the \textit{close-to-close} dataset.}
\label{table:JSEc2c}
\end{table}

\begin{table}
\centering
\figuretitle{Wealth Achieved by Investing in Various Combinations of Stocks From JSE OHLC Dataset (close-to-open)}
\begin{tabular}{ | p{1.5cm} | p{1.2cm}| p{1.7cm} | p{1.7cm} |} 
\hline \textbf{Stocks} & \textbf{Strat.} & \textbf{Wealth} & \textbf{Best Agent}  \\ \hline 
\hline \begin{tabular}{@{}l@{}} \textbf{ANGJ} \\ \textbf{AGLJ} \end{tabular} & \begin{tabular}{@{}l@{}}Abs. \\ Act. \\ $G^{*}_{NN}$ \\ UP \\ Best \end{tabular} & \begin{tabular}{@{}l@{}}  1.24 \\  1.04 \\  1.24 \\  0.97 \\  2.66 \end{tabular} & \begin{tabular}{@{}l@{}}  2.60 \\  2.82 \\  2.56 \\ \\ \\ \end{tabular} \\ 
\hline \begin{tabular}{@{}l@{}} \textbf{SBKJ} \\ \textbf{FSRJ} \end{tabular} & \begin{tabular}{@{}l@{}}Abs. \\ Act. \\ $G^{*}_{NN}$ \\ UP \\ Best \end{tabular} & \begin{tabular}{@{}l@{}} 12.35 \\  1.57 \\ 12.31 \\  7.89 \\ 12.47 \end{tabular} & \begin{tabular}{@{}l@{}} 25.13 \\  3.06 \\ 25.14 \\ \\ \\ \end{tabular} \\ 
\hline \begin{tabular}{@{}l@{}} \textbf{TBSJ} \\ \textbf{WHLJ} \end{tabular} & \begin{tabular}{@{}l@{}}Abs. \\ Act. \\ $G^{*}_{NN}$ \\ UP \\ Best \end{tabular} & \begin{tabular}{@{}l@{}}  4.38 \\  0.79 \\  4.46 \\  4.89 \\  4.98 \end{tabular} & \begin{tabular}{@{}l@{}}  9.00 \\  1.88 \\  8.83 \\ \\ \\ \end{tabular} \\ 
\hline \begin{tabular}{@{}l@{}} \textbf{10} \\ \textbf{STOCKS} \end{tabular} & \begin{tabular}{@{}l@{}}Abs. \\ Act. \\ $G^{*}_{NN}$ \\ Best \end{tabular} & \begin{tabular}{@{}l@{}}  7.11 \\  2.68 \\ 13.45 \\ 56.36 \end{tabular} & \begin{tabular}{@{}l@{}} 11.43 \\  3.39 \\ 30.53 \\ \\ \end{tabular} \\ 
\hline \begin{tabular}{@{}l@{}} \textbf{20} \\ \textbf{STOCKS} \end{tabular} & \begin{tabular}{@{}l@{}}Abs. \\ Act. \\ $G^{*}_{NN}$ \\ Best \end{tabular} & \begin{tabular}{@{}l@{}}  8.82 \\  5.28 \\ 11.00 \\ 56.36 \end{tabular} & \begin{tabular}{@{}l@{}} 16.87 \\  5.24 \\ 27.01 \\ \\ \end{tabular} \\ 
\hline \begin{tabular}{@{}l@{}} \textbf{30} \\ \textbf{STOCKS} \end{tabular} & \begin{tabular}{@{}l@{}}Abs. \\ Act. \\ $G^{*}_{NN}$ \\ Best \end{tabular} & \begin{tabular}{@{}l@{}}  8.07 \\  5.66 \\ 22.53 \\ 51.02 \end{tabular} & \begin{tabular}{@{}l@{}} 15.80 \\  5.50 \\ 54.80 \\ \\ \end{tabular} \\ 
\hline 
\end{tabular}
\caption{The total wealth achieved by the active (Act.) and absolute (Abs.) portfolios compared to the wealth achieved from the attempted recovery of the Gy{\"o}rfi \textit{et al} nearest neighbour strategy ($G^{*}_{NN}$), the attempted recovery of the universal portfolio strategy ($UP^*$) and a buy-and-hold strategy of the best stock (Best) on the \textit{close-to-open} dataset.}
\label{table:JSEc2o}
\end{table}

\begin{table}
\centering
\figuretitle{Wealth Achieved by Investing in Various Combinations of Stocks From JSE OHLC Dataset (open-to-close)}
\begin{tabular}{ | p{1.5cm} | p{1.2cm}| p{1.7cm} | p{1.7cm} |} 
\hline \textbf{Stocks} & \textbf{Strat.} & \textbf{Wealth} & \textbf{Best Agent}  \\ \hline 
\hline \begin{tabular}{@{}l@{}} \textbf{ANGJ} \\ \textbf{AGLJ} \end{tabular} & \begin{tabular}{@{}l@{}}Abs. \\ Act. \\ $G^{*}_{NN}$ \\ UP \\ Best \end{tabular} & \begin{tabular}{@{}l@{}} 97.75 \\ 50.21 \\ 97.53 \\  3.41 \\  4.68 \end{tabular} & \begin{tabular}{@{}l@{}} 395.58 \\ 91.82 \\ 398.74 \\ \\ \\ \end{tabular} \\ 
\hline \begin{tabular}{@{}l@{}} \textbf{SBKJ} \\ \textbf{FSRJ} \end{tabular} & \begin{tabular}{@{}l@{}}Abs. \\ Act. \\ $G^{*}_{NN}$ \\ UP \\ Best \end{tabular} & \begin{tabular}{@{}l@{}}  2.53 \\  1.04 \\  2.53 \\  2.66 \\  2.97 \end{tabular} & \begin{tabular}{@{}l@{}}  5.12 \\  1.80 \\  5.00 \\ \\ \\ \end{tabular} \\ 
\hline \begin{tabular}{@{}l@{}} \textbf{TBSJ} \\ \textbf{WHLJ} \end{tabular} & \begin{tabular}{@{}l@{}}Abs. \\ Act. \\ $G^{*}_{NN}$ \\ UP \\ Best \end{tabular} & \begin{tabular}{@{}l@{}}  5.59 \\  5.00 \\  5.59 \\  1.94 \\  1.89 \end{tabular} & \begin{tabular}{@{}l@{}} 12.73 \\  7.19 \\ 12.75 \\ \\ \\ \end{tabular} \\ 
\hline \begin{tabular}{@{}l@{}} \textbf{10} \\ \textbf{STOCKS} \end{tabular} & \begin{tabular}{@{}l@{}}Abs. \\ Act. \\ $G^{*}_{NN}$ \\ Best \end{tabular} & \begin{tabular}{@{}l@{}} 15.90 \\ 15.21 \\ 101.80 \\ 169.83 \end{tabular} & \begin{tabular}{@{}l@{}} 33.40 \\ 19.86 \\ 415.97 \\ \\ \end{tabular} \\ 
\hline \begin{tabular}{@{}l@{}} \textbf{20} \\ \textbf{STOCKS} \end{tabular} & \begin{tabular}{@{}l@{}}Abs. \\ Act. \\ $G^{*}_{NN}$ \\ Best \end{tabular} & \begin{tabular}{@{}l@{}}  4.80 \\  6.80 \\ 30.26 \\ 169.83 \end{tabular} & \begin{tabular}{@{}l@{}}  8.88 \\  6.07 \\ 104.46 \\ \\ \end{tabular} \\ 
\hline \begin{tabular}{@{}l@{}} \textbf{30} \\ \textbf{STOCKS} \end{tabular} & \begin{tabular}{@{}l@{}}Abs. \\ Act. \\ $G^{*}_{NN}$ \\ Best \end{tabular} & \begin{tabular}{@{}l@{}}  5.16 \\  4.58 \\ 50.17 \\ 1278.83 \end{tabular} & \begin{tabular}{@{}l@{}} 11.98 \\  5.57 \\ 317.92 \\ \\ \end{tabular} \\ 
\hline 
\end{tabular}
\caption{The total wealth achieved by the active (Act.) and absolute (Abs.) portfolios compared to the wealth achieved from the attempted recovery of the Gy{\"o}rfi \textit{et al} nearest neighbour strategy ($G^{*}_{NN}$), the attempted recovery of the universal portfolio strategy ($UP^*$) and a buy-and-hold strategy of the best stock (Best) on the \textit{open-to-close} dataset.}
\label{table:JSEo2c}
\end{table}

\begin{table}
\centering
\figuretitle{Wealth Achieved by Investing in Various Combinations of Stocks From JSE OHLC Dataset (open-to-open)}
\begin{tabular}{ | p{1.5cm} | p{1.2cm}| p{1.7cm} | p{1.7cm} |} 
\hline \textbf{Stocks} & \textbf{Strat.} & \textbf{Wealth} & \textbf{Best Agent}  \\ \hline 
\hline \begin{tabular}{@{}l@{}} \textbf{ANGJ} \\ \textbf{AGLJ} \end{tabular} & \begin{tabular}{@{}l@{}}Abs. \\ Act. \\ $G^{*}_{NN}$ \\ UP \\ Best \end{tabular} & \begin{tabular}{@{}l@{}}  5.11 \\  1.43 \\  5.07 \\  2.28 \\  3.18 \end{tabular} & \begin{tabular}{@{}l@{}} 26.65 \\  9.10 \\ 25.66 \\ \\ \\ \end{tabular} \\ 
\hline \begin{tabular}{@{}l@{}} \textbf{SBKJ} \\ \textbf{FSRJ} \end{tabular} & \begin{tabular}{@{}l@{}}Abs. \\ Act. \\ $G^{*}_{NN}$ \\ UP \\ Best \end{tabular} & \begin{tabular}{@{}l@{}} 113.43 \\ 13.97 \\ 113.02 \\ 20.29 \\ 20.55 \end{tabular} & \begin{tabular}{@{}l@{}} 412.75 \\ 18.87 \\ 415.45 \\ \\ \\ \end{tabular} \\ 
\hline \begin{tabular}{@{}l@{}} \textbf{TBSJ} \\ \textbf{WHLJ} \end{tabular} & \begin{tabular}{@{}l@{}}Abs. \\ Act. \\ $G^{*}_{NN}$ \\ UP \\ Best \end{tabular} & \begin{tabular}{@{}l@{}} 45.30 \\  5.52 \\ 45.80 \\  8.81 \\  9.24 \end{tabular} & \begin{tabular}{@{}l@{}} 266.17 \\ 28.54 \\ 273.44 \\ \\ \\ \end{tabular} \\ 
\hline \begin{tabular}{@{}l@{}} \textbf{10} \\ \textbf{STOCKS} \end{tabular} & \begin{tabular}{@{}l@{}}Abs. \\ Act. \\ $G^{*}_{NN}$ \\ Best \end{tabular} & \begin{tabular}{@{}l@{}} 60.62 \\ 12.03 \\ 254.00 \\ 69.49 \end{tabular} & \begin{tabular}{@{}l@{}} 342.40 \\ 20.57 \\ 2292.47 \\ \\ \end{tabular} \\ 
\hline \begin{tabular}{@{}l@{}} \textbf{20} \\ \textbf{STOCKS} \end{tabular} & \begin{tabular}{@{}l@{}}Abs. \\ Act. \\ $G^{*}_{NN}$ \\ Best \end{tabular} & \begin{tabular}{@{}l@{}} 14.73 \\  9.13 \\ 44.88 \\ 69.49 \end{tabular} & \begin{tabular}{@{}l@{}} 73.61 \\ 14.76 \\ 192.99 \\ \\ \end{tabular} \\ 
\hline \begin{tabular}{@{}l@{}} \textbf{30} \\ \textbf{STOCKS} \end{tabular} & \begin{tabular}{@{}l@{}}Abs. \\ Act. \\ $G^{*}_{NN}$ \\ Best \end{tabular} & \begin{tabular}{@{}l@{}} 21.54 \\ 11.48 \\ 161.16 \\ 131.89 \end{tabular} & \begin{tabular}{@{}l@{}} 73.88 \\ 18.89 \\ 1069.81 \\ \\ \end{tabular} \\ 
\hline 
\end{tabular}
\caption{The total wealth achieved by the active (Act.) and absolute (Abs.) portfolios compared to the wealth achieved from the attempted recovery of the Gy{\"o}rfi \textit{et al} nearest neighbour strategy ($G^{*}_{NN}$), the attempted recovery of the universal portfolio strategy ($UP^*$) and a buy-and-hold strategy of the best stock (Best) on the \textit{open-to-open} dataset.}
\label{table:JSEo2o}
\end{table}

\begin{figure}
    \centering
    \figuretitle{Wealth Gained close-to-close JSE OHLC Data}
    \begin{subfigure}[b]{0.45\textwidth}
        \includegraphics[width=\textwidth]{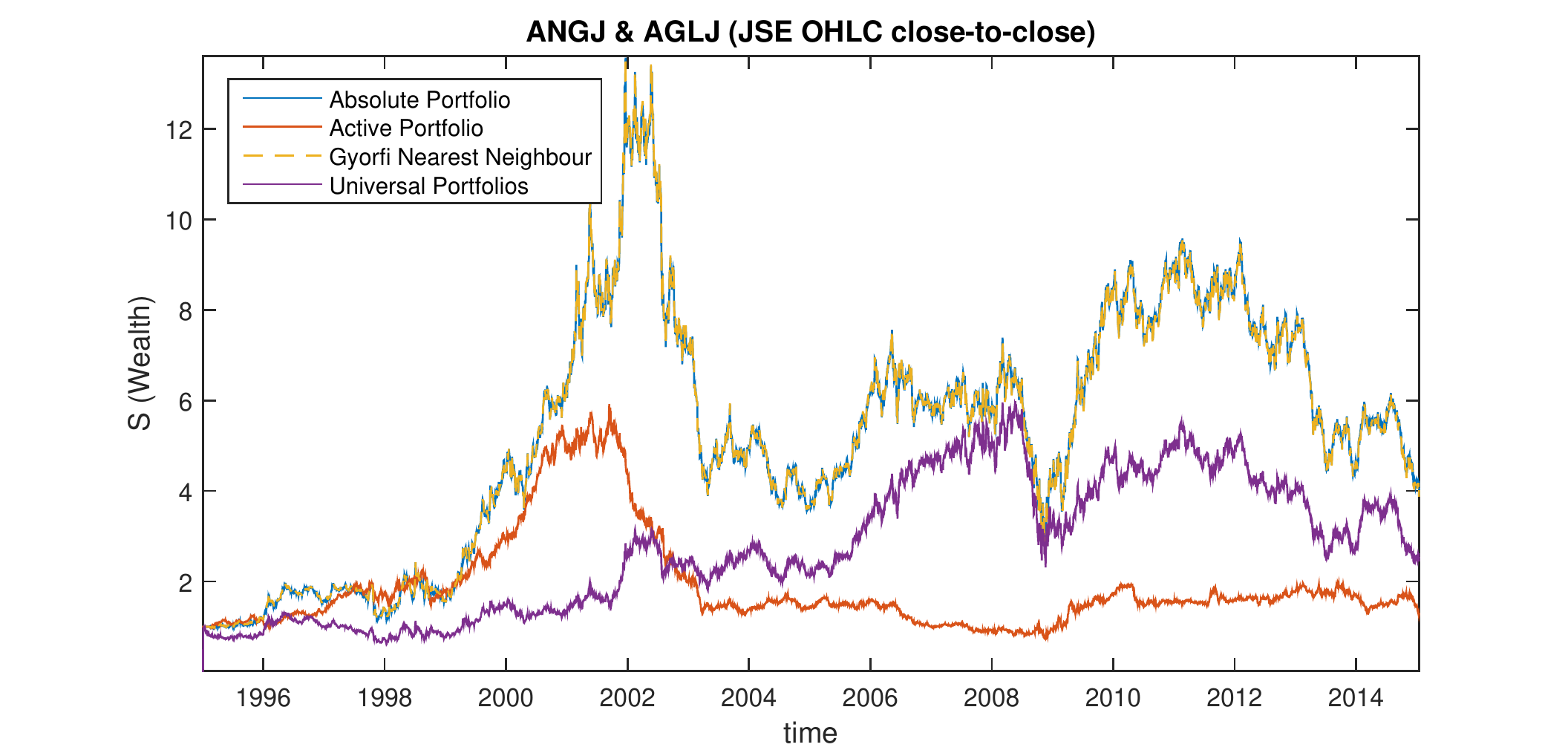}
        \caption{•}
        \label{fig:JSE_ANGAGL_Wealth}
    \end{subfigure}
    \begin{subfigure}[b]{0.45\textwidth}
        \includegraphics[width=\textwidth]{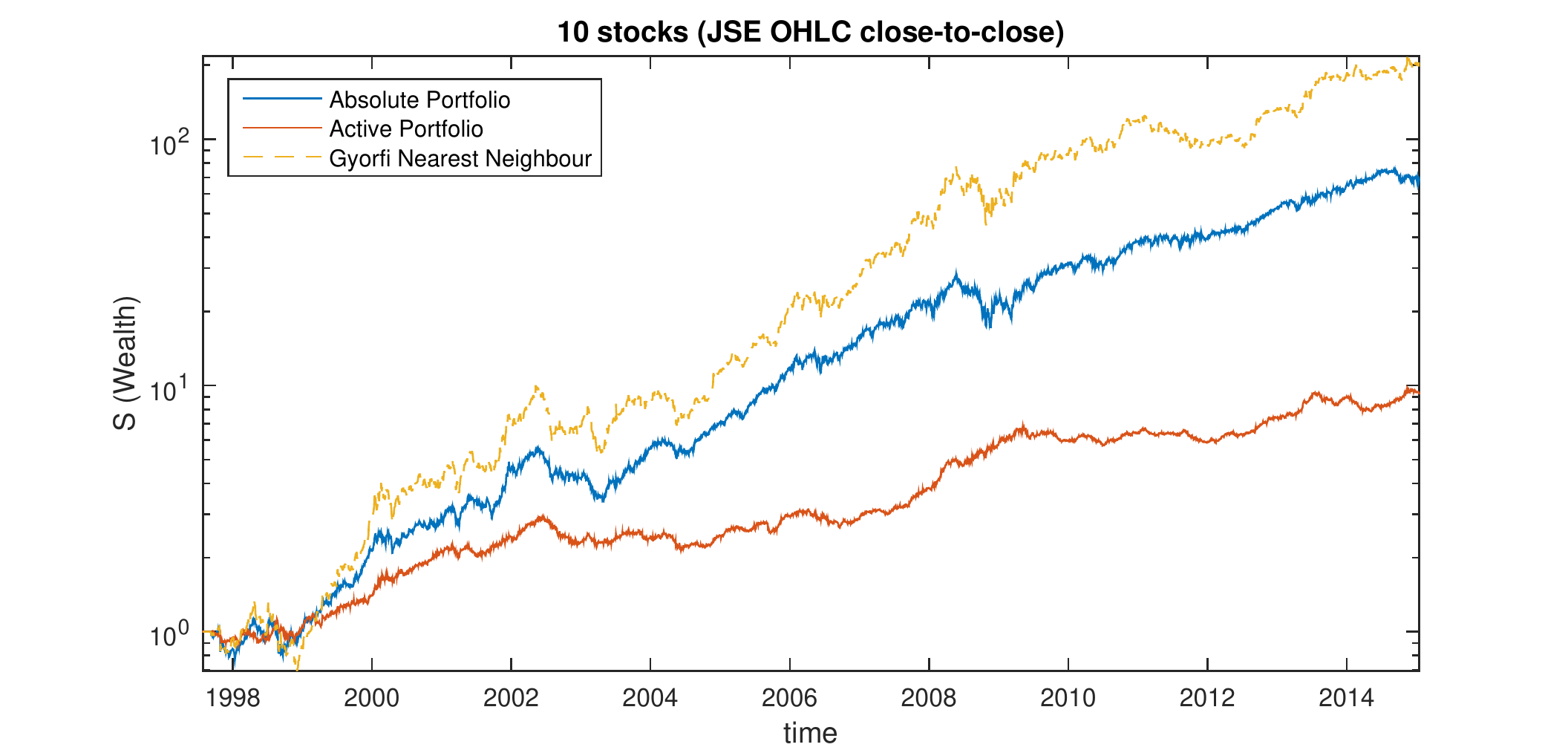}
        \caption{•}
        \label{fig:JSE_SBKFSR_Wealth}
    \end{subfigure}
    \begin{subfigure}[b]{0.45\textwidth}
        \includegraphics[width=\textwidth]{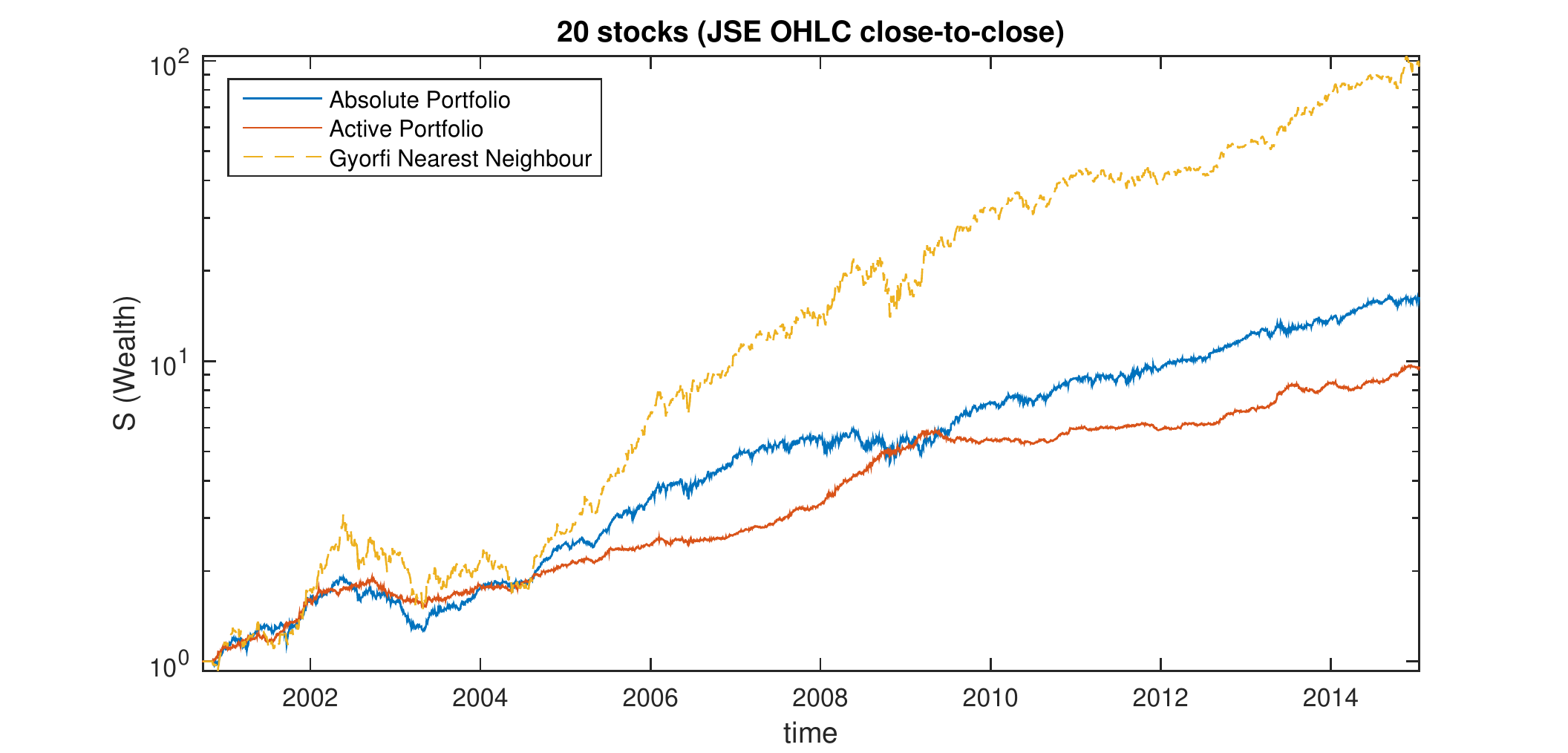}
        \caption{•}
        \label{fig:JSE_TBSWHL_Wealth}
    \end{subfigure}
    \begin{subfigure}[b]{0.45\textwidth}
        \includegraphics[width=\textwidth]{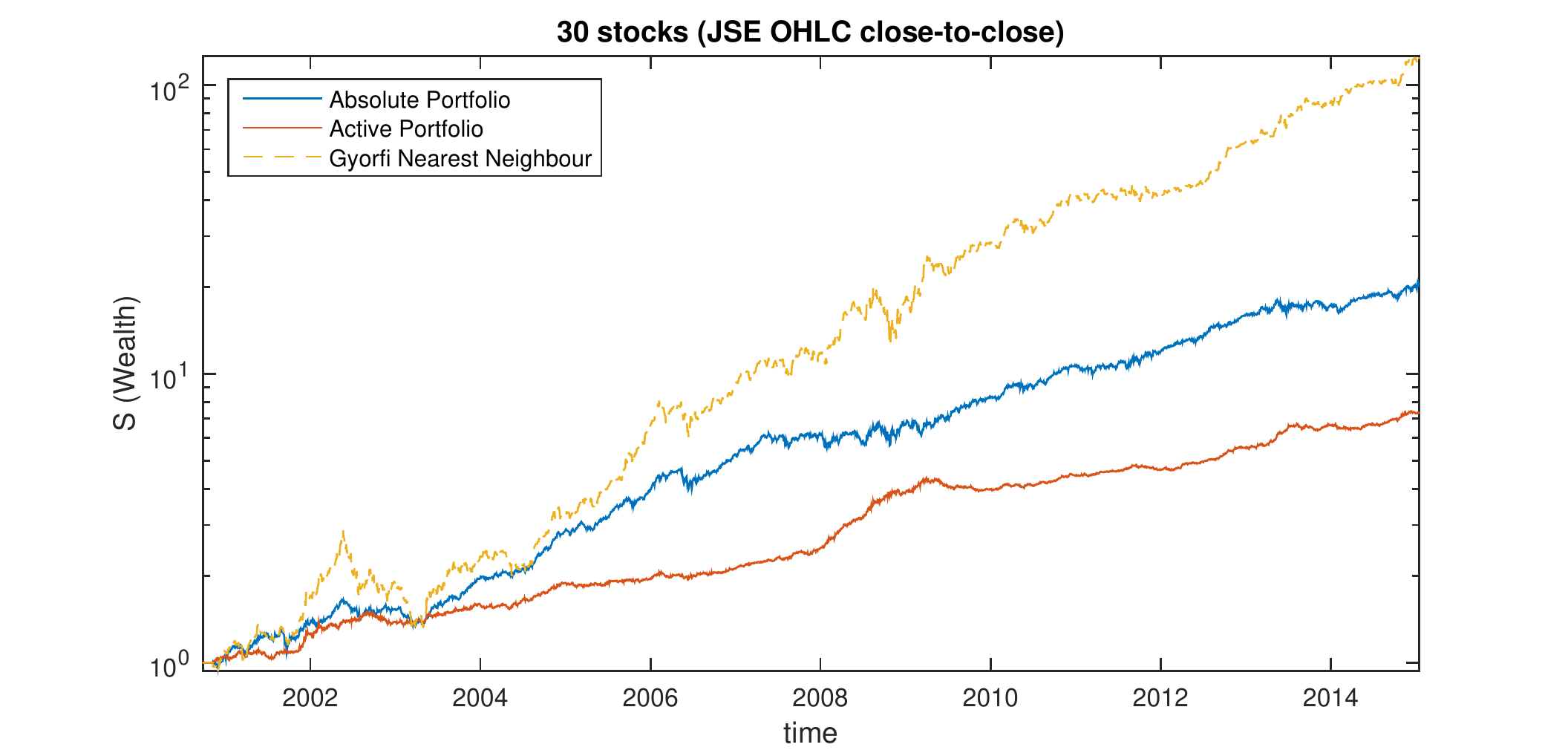}
        \caption{•}
        \label{fig:JSE_10Stocks_Wealth}
    \end{subfigure}
    \caption{ Comparison of the wealth gained from different methods when investing in (a) ANGJ and AGLJ (b) 10 stocks (c) 20 stocks (d) 30 stocks from  the JSE OHLC \textit{close-close} dataset. This does not account for price-impacts and frictions, nor for the need to approximate an expected close price just prior to market close as one solves for the portfolio controls, there will always be a difference between the controls solved for just prior to market close and those required once the market has closed and the official closing prices printed. }\label{fig:JSEOHLC_wealthplot}
\end{figure}

\begin{figure}
    \centering
    \figuretitle{Wealth from Different JSE OHLC Data Sets}
    \begin{subfigure}[b]{0.45\textwidth}
        \includegraphics[width=\textwidth]{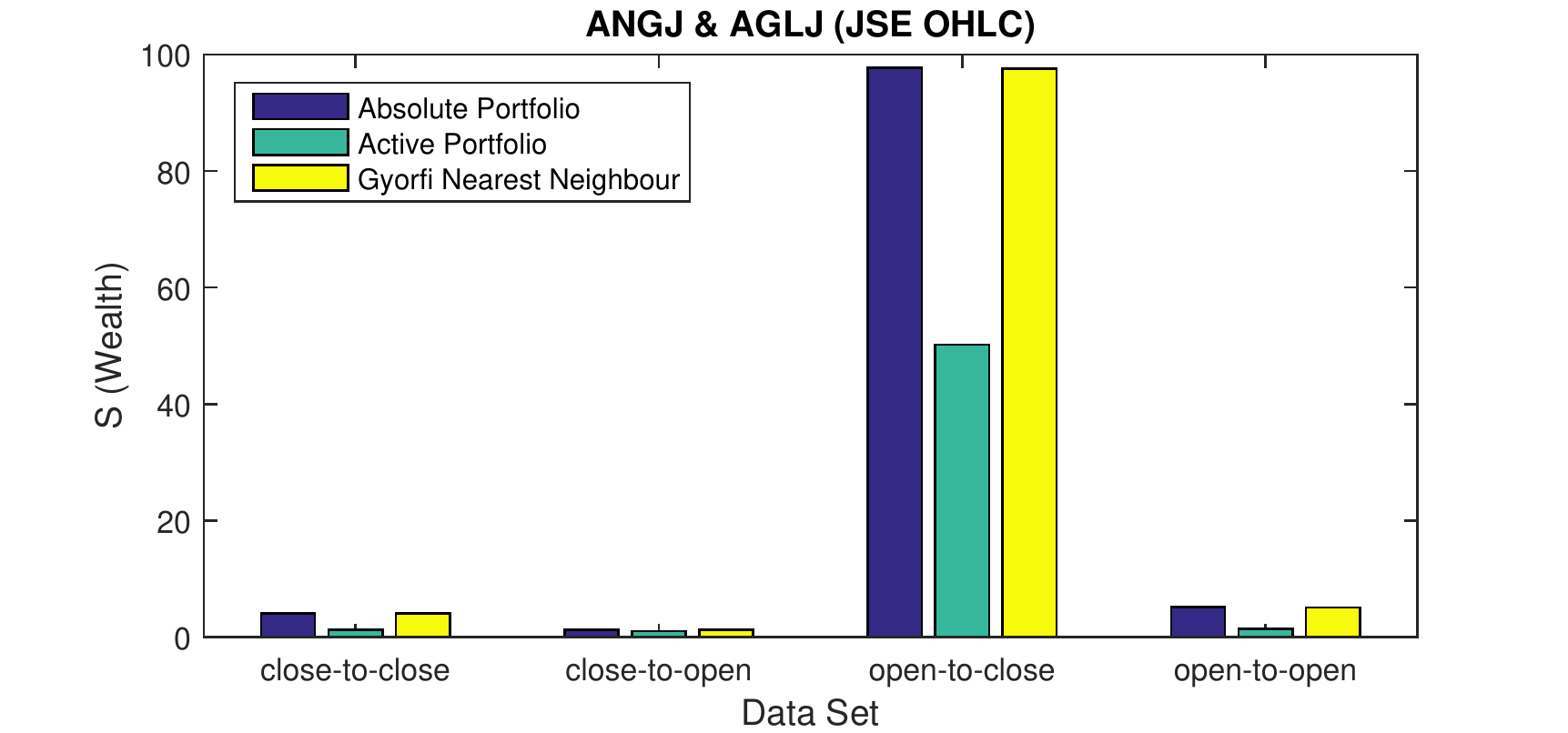}
        \caption{•}
        \label{fig:JSE_ANGAGL_Bar}
    \end{subfigure}
    \begin{subfigure}[b]{0.45\textwidth}
        \includegraphics[width=\textwidth]{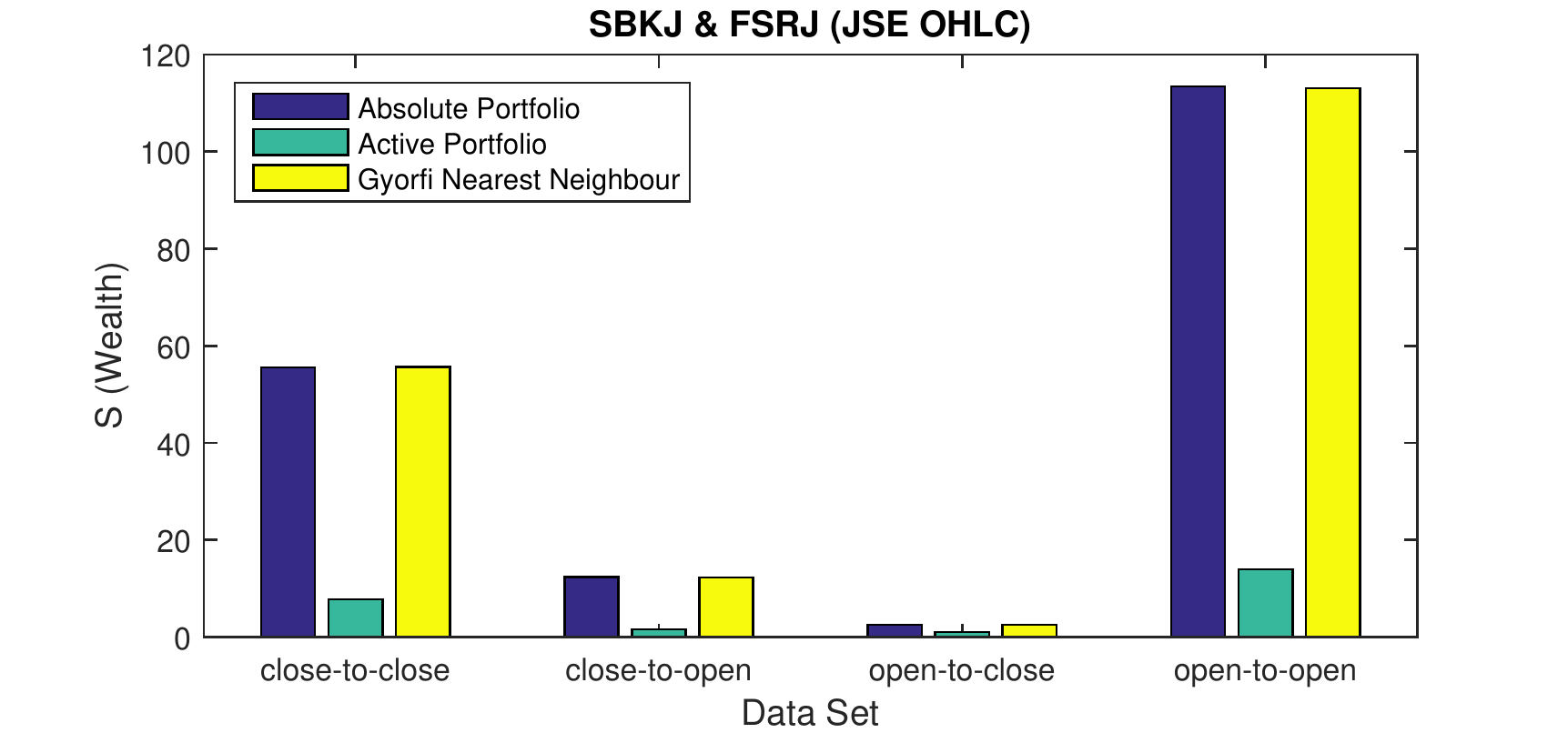}
        \caption{•}
        \label{fig:JSE_SBKFSR_Bar}
    \end{subfigure}
    \begin{subfigure}[b]{0.45\textwidth}
        \includegraphics[width=\textwidth]{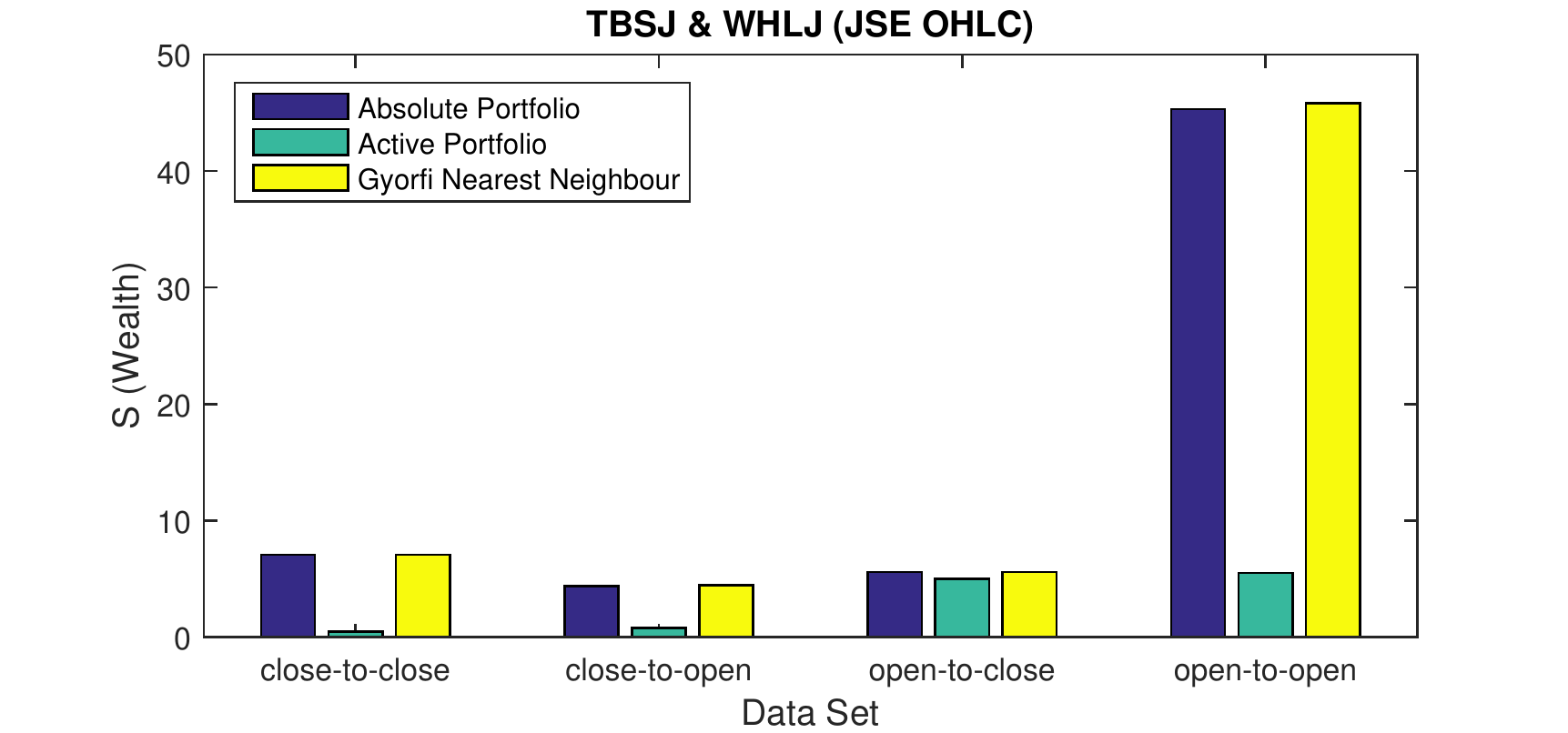}
        \caption{•}
        \label{fig:JSE_TBSWHL_Bar}
    \end{subfigure}
    \begin{subfigure}[b]{0.45\textwidth}
        \includegraphics[width=\textwidth]{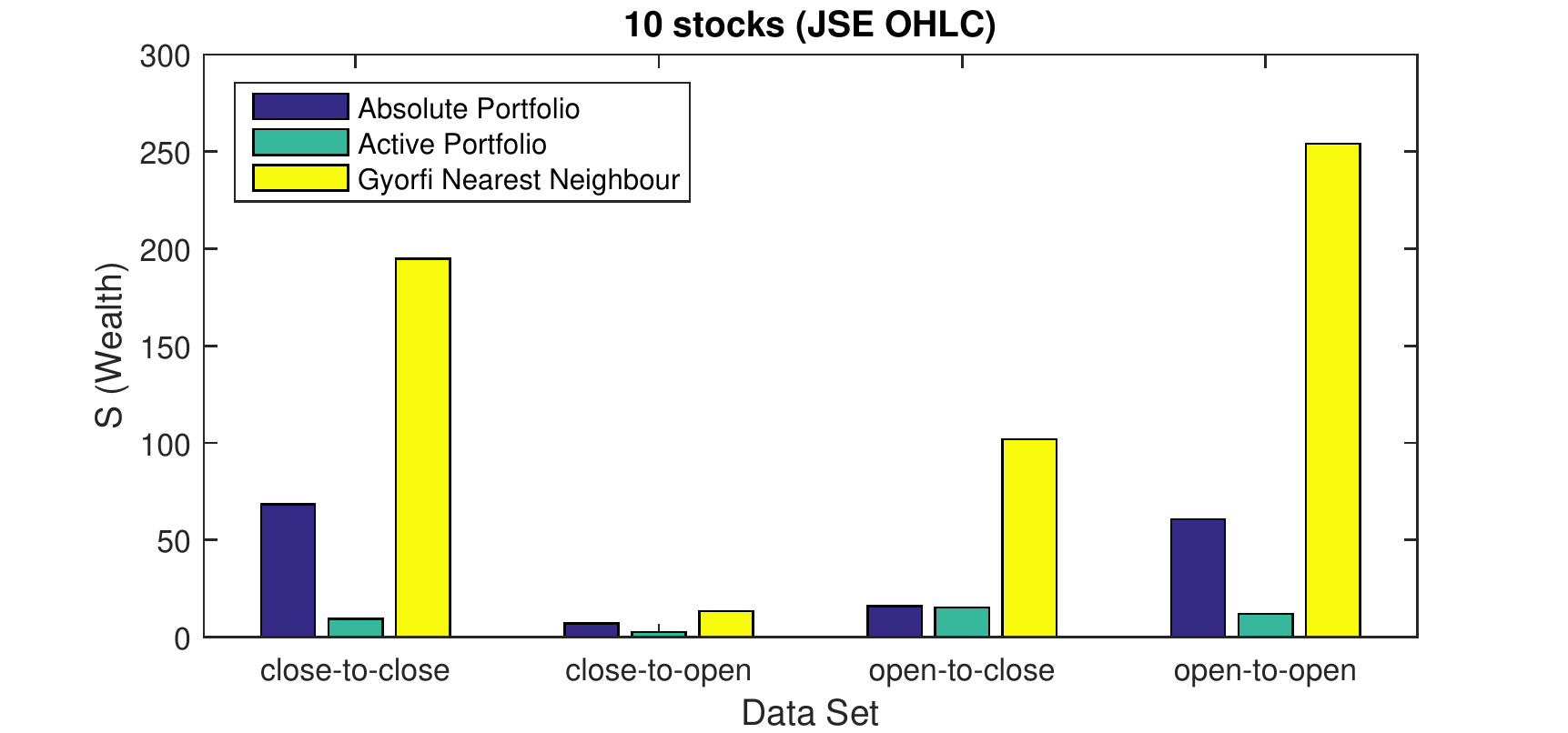}
        \caption{•}
        \label{fig:JSE_10Stocks_Bar}
    \end{subfigure}
    \caption{ Comparison of wealth achieved from the absolute portfolio, active portfolio and Gy{\"o}rfi nearest neighbour porfolio on the(a)\textit{close-to-close} (b) \textit{open-to-close} (c) \textit{close-to-open} and (d) \textit{open-to-open} JSE OHLC datasets. Here we find that there is no particular combination of OHLC data for which there is a systematic preference, e.g. \textit{close-to-close}, the case of considering the close price change from one day end to another is not systematically more profitable than other combinations of data times. These tests do consider the reality of trading prior to a time point, for example market close, one cannot {\it a-priori} know what the close price will be, this has to be approximated. This excludes price-impact effects. }\label{fig:JSEOHLC_Barplot}
\end{figure}

\subsection{Intraday JSE data} \label{ssec:intradayJSE}

The algorithm was run for both absolute and active portfolios on various sets of two stock combinations, namely stocks AngloGold Ashanti Ltd and Anglo American PLC, stocks Standard Bank Group Ltd and FirstRand Ltd, stocks Tiger Brands Ltd and Woolworths Holdings Ltd, and stocks MTN Group Ltd and Vodacom Group Ltd. The algorithm was also run for both absolute and active portfolios on the same combination of 10 stocks used on the JSE OHLC Dataset.

\begin{table}
\centering
\figuretitle{Wealth from Stock-Pairs of JSE Intraday Data}
\begin{tabular}{ | p{1.5cm} | p{1.2cm}| p{1.7cm} | p{1.7cm} |} 
\hline \textbf{Stocks} & \textbf{Strat.} & \textbf{Wealth} & \textbf{Best Agent}  \\ \hline 
\hline \begin{tabular}{@{}l@{}} \textbf{ANGJ} \\ \textbf{AGLJ} \end{tabular} & \begin{tabular}{@{}l@{}}Abs. \\ Act. \\ $G^{*}_{NN}$ \\ UP \\ Best \end{tabular} & \begin{tabular}{@{}l@{}}  1.38 \\  2.16 \\  1.36 \\  0.66 \\  0.87 \end{tabular} & \begin{tabular}{@{}l@{}}  3.33 \\  5.01 \\  3.21 \\ \\ \\ \end{tabular} \\ 
\hline \begin{tabular}{@{}l@{}} \textbf{SBKJ} \\ \textbf{FSRJ} \end{tabular} & \begin{tabular}{@{}l@{}}Abs. \\ Act. \\ $G^{*}_{NN}$ \\ UP \\ Best \end{tabular} & \begin{tabular}{@{}l@{}}  1.82 \\  2.02 \\  1.82 \\  1.08 \\  1.08 \end{tabular} & \begin{tabular}{@{}l@{}}  2.19 \\  2.01 \\  2.15 \\ \\ \\ \end{tabular} \\ 
\hline \begin{tabular}{@{}l@{}} \textbf{TBSJ} \\ \textbf{WHLJ} \end{tabular} & \begin{tabular}{@{}l@{}}Abs. \\ Act. \\ $G^{*}_{NN}$ \\ UP \\ Best \end{tabular} & \begin{tabular}{@{}l@{}}  1.95 \\  2.24 \\  1.95 \\  0.91 \\  0.99 \end{tabular} & \begin{tabular}{@{}l@{}}  3.06 \\  3.29 \\  2.99 \\ \\ \\ \end{tabular} \\ 
\hline \begin{tabular}{@{}l@{}} \textbf{MTNJ} \\ \textbf{VODJ} \end{tabular} & \begin{tabular}{@{}l@{}}Abs. \\ Act. \\ $G^{*}_{NN}$ \\ UP \\ Best \end{tabular} & \begin{tabular}{@{}l@{}}  2.13 \\  2.13 \\  2.13 \\  1.12 \\  1.18 \end{tabular} & \begin{tabular}{@{}l@{}}  3.55 \\  3.25 \\  3.57 \\ \\ \\ \end{tabular} \\ 
\hline \begin{tabular}{@{}l@{}} \textbf{10} \\ \textbf{STOCKS} \end{tabular} & \begin{tabular}{@{}l@{}}Abs. \\ Act. \\ $G^{*}_{NN}$ \\ Best \end{tabular} & \begin{tabular}{@{}l@{}}  1.89 \\  3.86 \\  3.95 \\  1.93 \end{tabular} & \begin{tabular}{@{}l@{}}  2.79 \\  5.40 \\ 14.05 \\ \\ \end{tabular} \\ 
\hline \begin{tabular}{@{}l@{}} \textbf{20} \\ \textbf{STOCKS} \end{tabular} & \begin{tabular}{@{}l@{}}Abs. \\ Act. \\ $G^{*}_{NN}$ \\ Best \end{tabular} & \begin{tabular}{@{}l@{}}  1.74 \\  3.42 \\  5.68 \\  1.93 \end{tabular} & \begin{tabular}{@{}l@{}}  2.35 \\  4.29 \\ 21.62 \\ \\ \end{tabular} \\ 
\hline \begin{tabular}{@{}l@{}} \textbf{30} \\ \textbf{STOCKS} \end{tabular} & \begin{tabular}{@{}l@{}}Abs. \\ Act. \\ $G^{*}_{NN}$ \\ Best \end{tabular} & \begin{tabular}{@{}l@{}}  1.67 \\  3.07 \\  6.03 \\  1.93 \end{tabular} & \begin{tabular}{@{}l@{}}  2.16 \\  3.76 \\ 12.61 \\ \\ \end{tabular} \\ 
\hline 
\end{tabular}
\caption{The total wealth achieved by the active (Act.) and absolute (Abs.) portfolios compared to the wealth achieved from the attempted recovery of the Gy{\"o}rfi \textit{et al} nearest neighbour strategy ($G^{*}_{NN}$), the attempted recovery of the universal portfolio strategy ($UP^*$) and a buy-and-hold strategy of the best stock (Best) on the JSE Intraday Dataset.}
\label{table:JSEIntraday}
\end{table}

\begin{figure}
    \centering
    \figuretitle{Wealth from JSE Intraday Data Experiments}
    \begin{subfigure}[b]{0.45\textwidth}
        \includegraphics[width=\textwidth]{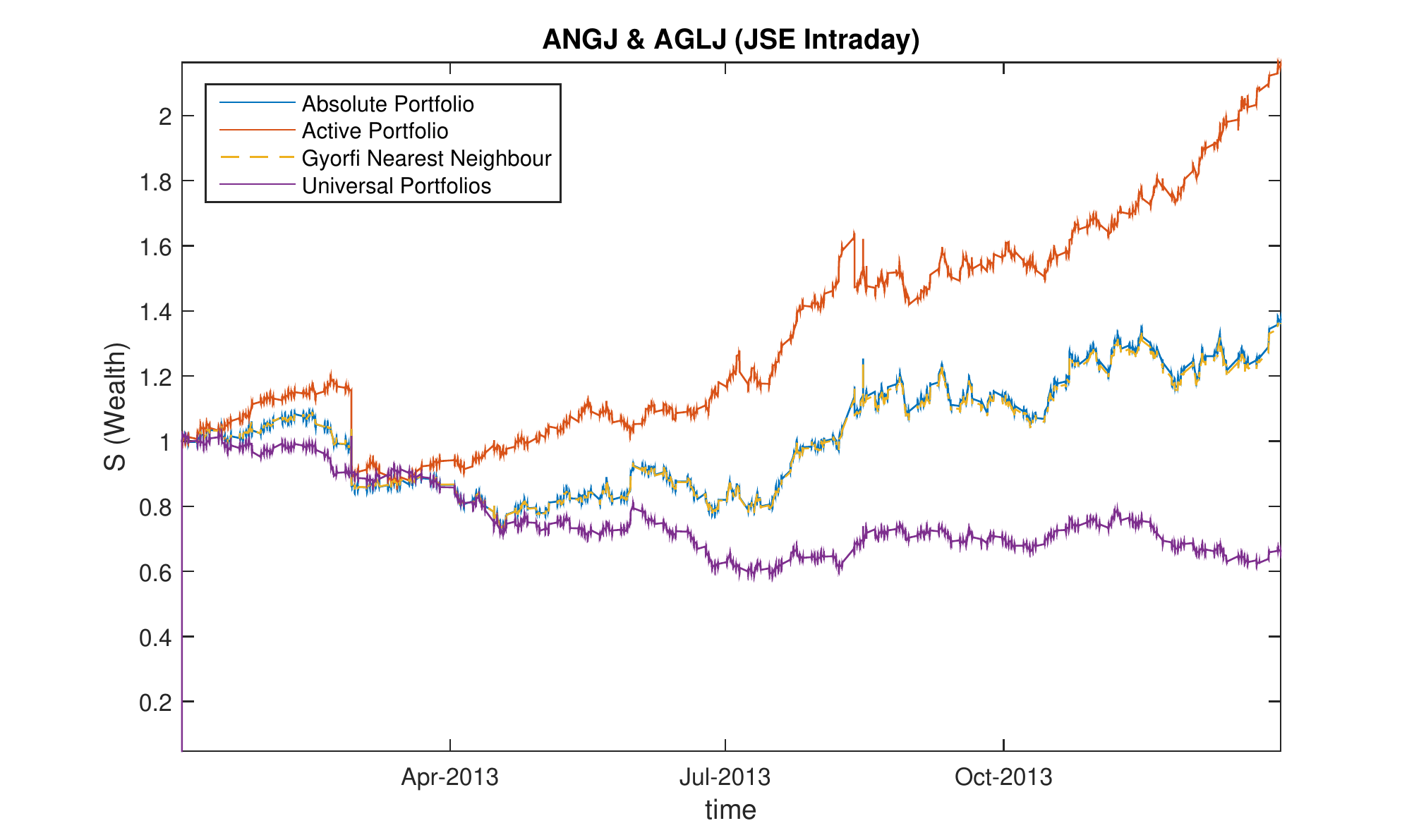}
        \caption{•}
        \label{fig:ANGJ_AGLJ_plot_JSEIntraday}
    \end{subfigure}
    \begin{subfigure}[b]{0.45\textwidth}
        \includegraphics[width=\textwidth]{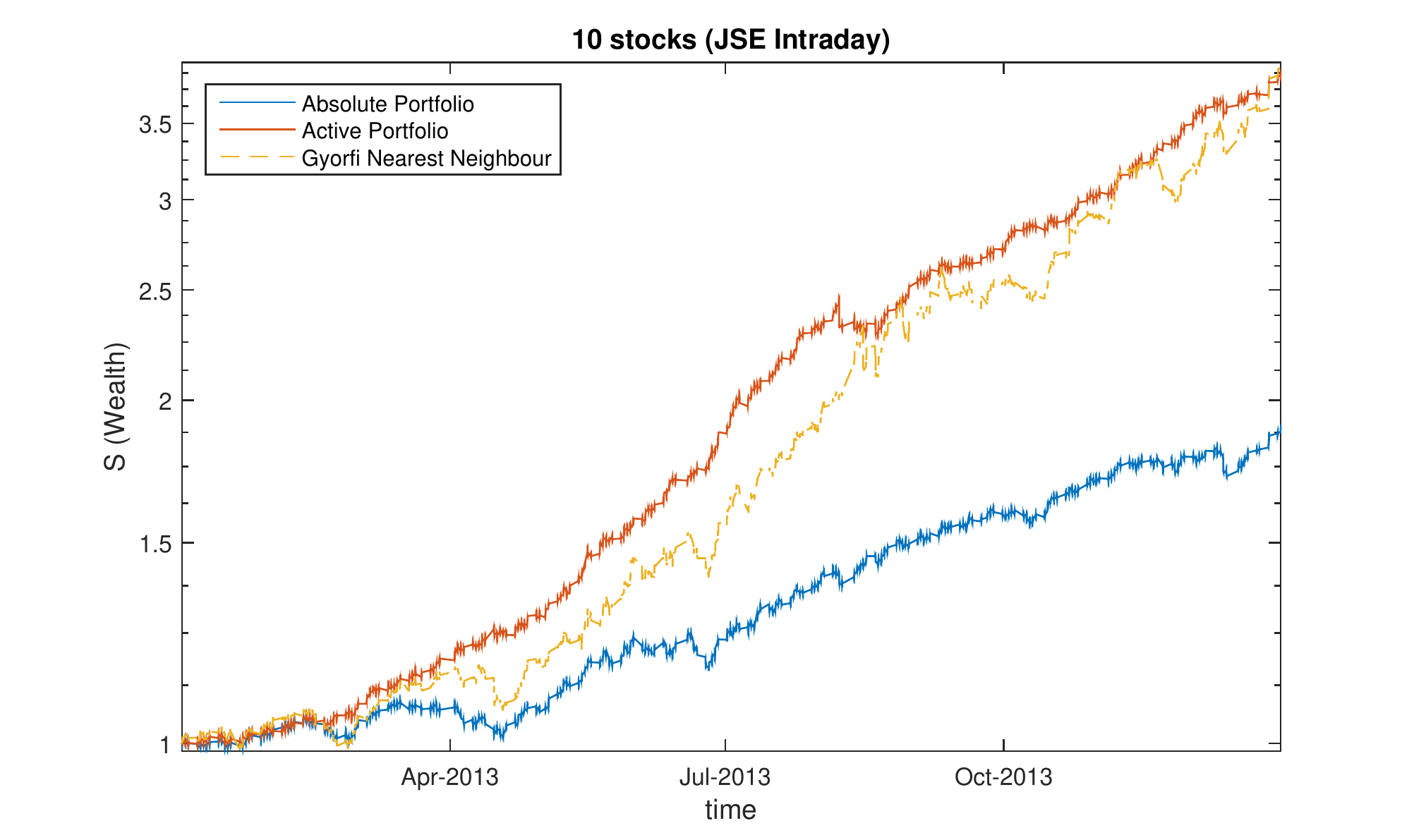}
        \caption{•}
        \label{fig:10_stocks_plot_JSEIntraday}
    \end{subfigure}
    \caption{ Comparison of the wealth gained from different methods when investing in (a) ANGJ and AGLJ (b) 10 stocks from  the JSE Intraday dataset.}\label{fig:JSEIntraday_wealthplot}
\end{figure}

\begin{table} 
\centering
\figuretitle{JSE Intraday Data with cluster defined agents}
\begin{tabular}{ | p{0.22cm} | p{2.5cm} | p{1.2cm} | p{1.2cm} | p{1.2cm} |}
\hline & \textbf{} & \textbf{RESI} & \textbf{INDI} & \textbf{FINI}\\ \hline
\hline \parbox[t]{2mm}{\multirow{2}{*}{\rotatebox[origin=c]{90}{Act.}}} & Best Agent  & 9.1018 & 3.4470 & 3.6698 \\
\cline{2-5} & Total Wealth & \multicolumn{3}{c|}{4.6368} \\
\hline \parbox[t]{2mm}{\multirow{2}{*}{\rotatebox[origin=c]{90}{Abs.}}} & Best Agent  & 5.6655 & 2.6341 & 2.6059  \\
\cline{2-5} & Total Wealth & \multicolumn{3}{c|}{2.2093}    \\
\hline
\end{tabular} 
\caption{Wealth achieved by active and absolute portfolios when using economic sectors as clusters. Using three clusters increases the number of competing agents by a factor of 3, from 50 to 150. The inclusion of a larger set of agents increase the out-of-sample wealth performance of the two strategies. See Tables \ref{tab:resi}, \ref{tab:fini}, and \ref{tab:indi} for the particular stocks in each sectors, Resources (RESI), Financials (FINI) and Industrials (INDI), respectively. The total wealth includes the relative competition between agents defined by the three economic sectors. The sector specific wealth is given the best performing agents from each cluster groups of stocks.}
\label{tab: JSE IntradayClusters}
\end{table}

\begin{figure}
\centering
\figuretitle{20 JSE Stocks}
\includegraphics[width=0.5\textwidth]{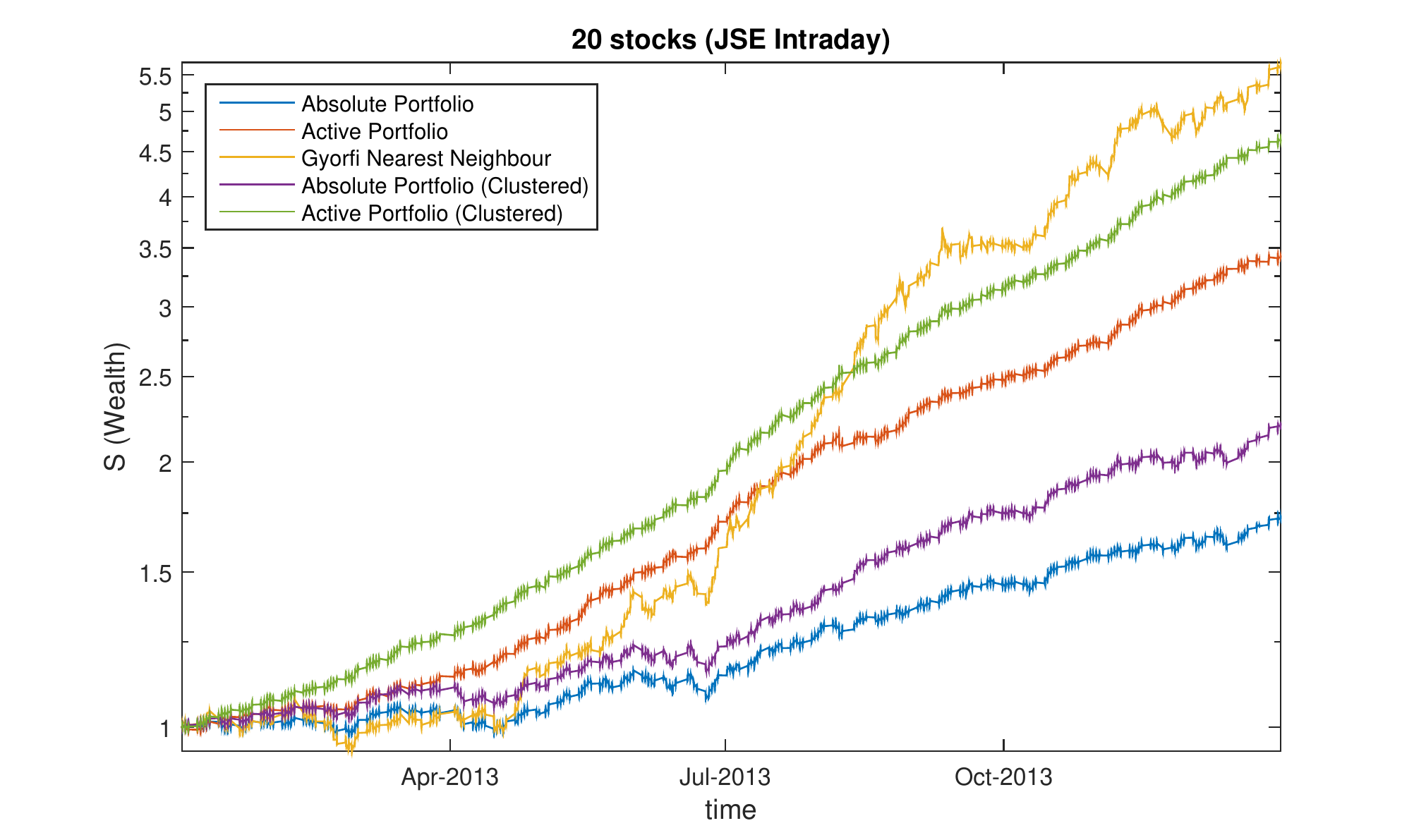}
\caption{Comparison of the wealth gained from different methods when investing in 20 stocks from  the JSE Intraday dataset, the plot includes the results of using clusters on the stocks. It is important to note that the clustered portfolios have 150 agents and the portfolios without clusters have 50 agents.}
\label{fig:20_stocks_clustered_plot_JSEIntraday} 
\end{figure}

\begin{figure}
\centering
\figuretitle{JSE Intraday Data Running Times}
\includegraphics[width=0.5\textwidth]{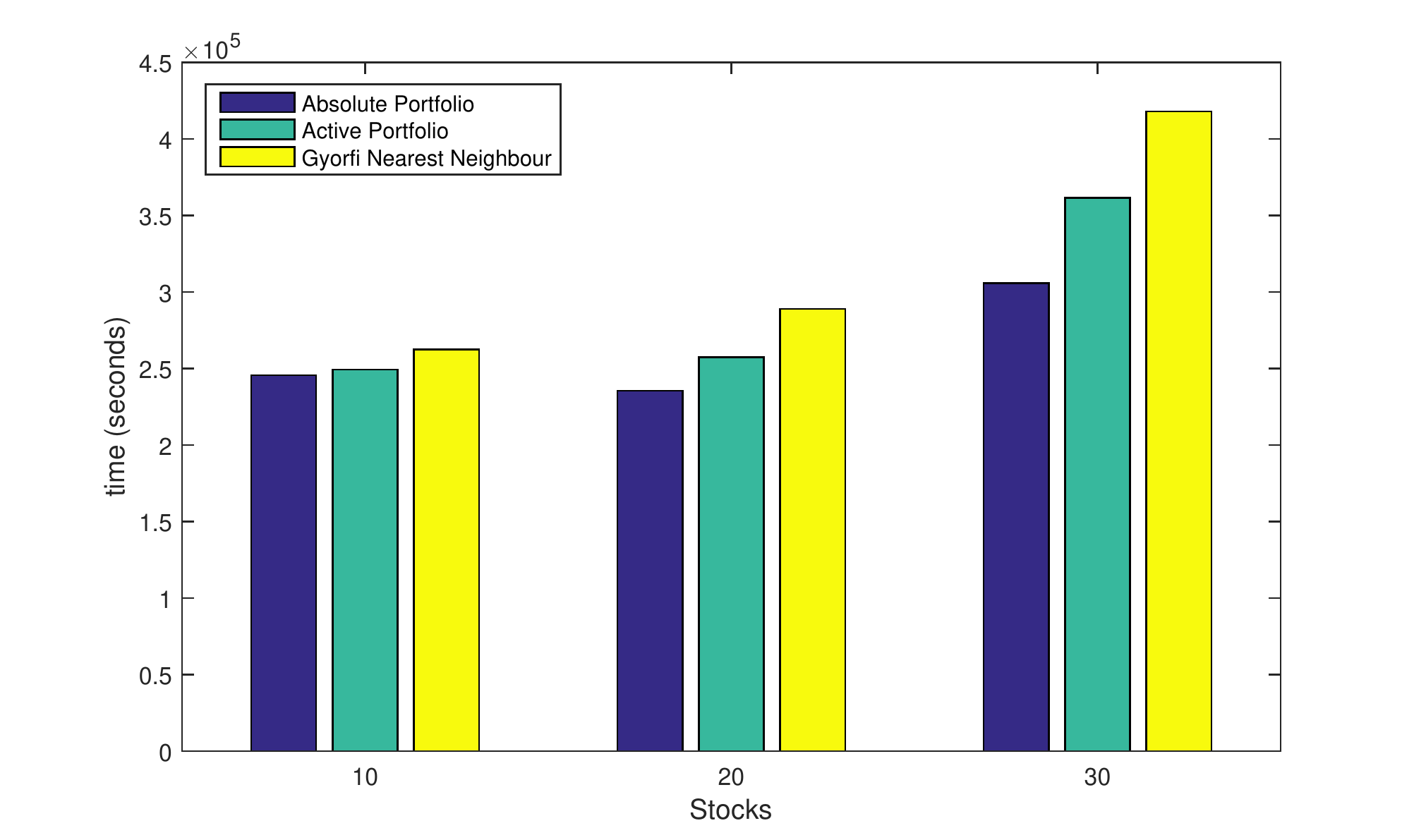}
\caption{Running time of the portfolios in seconds of the different strategies. This demonstrates the speed advantage of using the analytic quadratic approximation as compared to numerically solving the log-optimal constrained optimization at each time-step for each agent combination. As expected the fully invested analytic solution is fastest, the zero-cost portfolio next, because of the additional leverage constraint, and the slowest the algorithm that required the numerical solution of the optimization. }
\label{fig: JSE Intraday Timings}
\end{figure}

\subsection{The Impact of Market Frictions} \label{ssec:frictions}

An important criticism of any strategy simulation relates to the need to account for the impact of market frictions, this includes: transaction costs, the cost of the capital for trading, the cost of market access, the cost of regulatory capital for taking risky trading positions, and market impact. These are all required to be included in any estimate of performance slippage for any realistic assessment of the viability of trading activity.

\subsubsection{Daily strategy trading frictions} 

The argument that the zero-cost low frequency (daily traded) strategies are viable, even when unleveraged, is based on the 7 years of history \footnote{Although this may be considered short in the context of many academic studies, this is of the order of the time-scale of the business cycle so we considered this realistic.}. Consider Table \ref{table:JSEc2c} for the active case for the Top 10 JSE stocks (see appendix \ref{app:ticker_lists}). Here we would argue for 15bps of daily profit before costs (from Table \ref{table:JSEc2c} using the accumulated daily wealth of 9.53). We consider the strategy that trades close of the one day to the close of the next day (close-to-close). 

This is considered in order to take into account liquidity effects. The closing auction is the most liquid time to trade on the JSE. It is unlikely that one would be able to achieve low slippage trading near the daily market opening. We consider the combination of cost of capital (the borrowing costs required to source trading capital and cost of regulatory capital) and a small penalty for slippage due to the differences between the realized closing prices and the estimated closing prices that the algorithm would require in order to estimate the portfolio controls \footnote{This can in practice be carried out during the closing auction, just prior to the market close, by estimating online, the equilibrium price that could be the result of maximizing execution volume for the lowest surplus when the market is cleared at market close, bearing in mind that there is a short randomization period at the end of the auction that needs to be accounted for. It is fairly straight-forward to estimate sufficiently reasonable market clearing prices. If this is not considered realistic, one can then merely consider the trading to have occurred during a post market close period (such as that found on the LSE and JSE) where one can transact at the market close price – but at lower volumes.} as 10bps per day. In practise it should also be noted that such a trading strategy can be converted to one that trades in equity swaps, so-called contracts-for-difference (CFDs), this would convert the uncertainty about slippage into an up-front fee and allow for excellent implementation of the required model positions with a known cost and no meaningful liquidity concerns. If the daily strategy was implemented with these types of delta-one instruments our estimates of slippage can then be considered conservative.

We argue that we can realistically earn a modest 5bps of unleveraged self-funded trading profit per day, or an annual return of 12\% of unleveraged profit-and-loss.

\subsubsection{Intraday strategy trading frictions}

We assume: 1.) a daily slippage of 50bps for the (self-financing) zero-cost statistical arbitrage strategy, 2.) borrowing costs on the capital required for trading over the year to be 10\%, and 3.) that the strategy we denote as the active strategy generated a 4.63 wealth gained over a year of trading (see Table \ref{tab: JSE IntradayClusters}). Putting these together we argue for an upper limit on the profit, even when unleveraged, to be a return of 20\% for 250 days of trading\footnote{The strategy generates 62bps per day, the slippage is 50bps, leaving 12bps, less the 5bps for the cost of capital, then leaves 7bps to accumulate as profit-and-loss per trading day.}. 

The strategy turn-over is important in realistic assessments of profitability for intraday trading. We try to account for this in our indicative costing of the slippage by assuming that we have 100\% turn-over of inventory at each trading period with a consistent cost of 0.55bps (0.0055\%),  per trading period\footnote{If we considered trading to be for a 7 hour period starting a half hour after market opening, and stopping 15 minutes before the market closing auction (using the JSE market times), this then leads to $84 =12 \times 7$ 5-minute trading periods across the day, this would give the worst case scenario of 8400\% turn-over per day, at a transaction cost of 0.55bps per trading period, we argue that this then leads to a transaction cost of 46bps per day, additional frictions of 4bps are added to this to get the over-all daily slippage estimate of 50bps per day.} with an additional 4bps to give an indicative slippage of 50bps for the intraday trading per day. 

\section{Conclusion} \label{sec:conclusions}

In prior work it has been shown that in South African financial markets persistence and long-memory are generic \cite{WG2008}. This paper adds to our knowledge of the South African market by showing that in addition to evidence supporting long-memory processes, price processes have patterns that are exploitable in a straight-forward manner. 

We provide a simple portfolio value based learning algorithm, a multi-manager in the language of asset management, that selects an over-all portfolio with weights $\ve b$ by considering a selection of N different strategies $\ve H_{n}$ with their underlying portfolio weights being constructed for underlying strategies that are enumerated over a variety of combinations of time-series patterns, time-scales, clusters and partitions. This is considered in the context of universally consistent strategies \cite{C1991,GUW2008} but with an extension to directly consider self-financing zero-cost quantitative trading strategies - what we call the active portfolio.  

When applying the algorithms to real daily test data, it compares well to results from our implementation of algorithms from the literature \cite{C1991,GUW2008} and actual results from the literature from the New York Stock Exchange (NYSE) dataset (see Tables \ref{table:NYSE} and \ref{table: NYSEmerged}).

The active version of the algorithm, when applied to intraday data from the Johannesburg Stock  Exchange (JSE), is shown to have performed well in comparison to the best stock, and compares favourably with methods from prior work \cite{C1991,GUW2008} (see \ref{table:JSEIntraday}). We show that on the Johannesburg Stock Exchange data the algorithms can learn trends and patterns and enhance out-of-sample wealth accumulation for both daily and intraday applications (see Tables \ref{table:JSEc2c} and \ref{table:JSEIntraday}). This is demonstrated on both low frequency data, daily sampled data, and higher frequency data, intraday uniformly sampled transaction data. 

We have shown that there is an advantage to include agents that are clustered on stock economic sector classifications (see Table \ref{tab: JSE IntradayClusters}); this increases the number of agents (or experts) considered by the learning algorithm through including sector membership into the resource, financial and industrial stocks sectors, which in turn boosts the out-of-sample performance. This suggests that combining more sophisticated clustering algorithms  \cite{HWG2016a} with machine learning can be advantageous in the domain of quantitative trading.

The pay-off between computational performance and wealth accumulation can be seen by considering the increased duration of the simulation as one increases from 10 stocks, to 20 stocks through to 30 stocks in Figure \ref{fig: JSE Intraday Timings}. The commensurate loss in performance can be seen in Table \ref{table:JSEIntraday}.  For example, the 20 stock simulation generated a wealth of 1.74 for the absolute portfolio and 5.68 for the Gyorfi {\it et al} nearest-neighbour strategy with the absolute portfolio being almost $5 \times 10^4$ seconds faster (or 18\% faster). For intraday statistical arbitrage problems for quantitative trading with many (50$>$) assets, computational delays can lead to lags between information arrival and order-execution that can negatively impact a strategies profit-and-loss performance. 

We have shown that in the daily dataset for the Johannesburg Stock Exchange, when considering open, high, low and close price data, there is an advantage when considering strategies that relate to the patterns arising across closing price to closing price data (see Table \ref{table:JSEc2c}).  It is difficult to profitable trade the market opening price to the market closing price as intraday dynamics seems to become important and one tends to incur significant market frictions associated with poor market liquidity near the market opening. This provides evidence that one can in principle beat the best stock (or the money market account in the case of the self-financing strategy) as pattern persistence is sufficiently robust in the markets considered. 

Towards addressing the key criticism related to correctly estimating the impact of market frictions, in Section \ref{ssec:frictions} we give our estimates for the impact of market frictions on both the intraday strategy, where an estimated annual return of upto 20\% for the unleveraged self-financing strategy, and the daily strategy, trading the closing price of the market from one day to the next, at an annual return of upto 12\% (see Section \ref{ssec:frictions}). Using this we argue that the self-financing zero-cost portfolio strategy can be considered tractable both intraday and across days, that after a reasonable estimates of costs, one is still able to learn how to exploit patterns the recur in the financial times-series data considered in this study. It should also be noted that for optimised intraday trading the event-time paradigm should be implemented rather than the calendar-time approach that was used for simplicity in the experiments in this paper. This is fairly straight-forward to implement using equal volume buckets and online down-sampling transaction data to a time-series of volume-weighted average prices for equal volume buckets \cite{ELO2012}.

The aim here is to show that there are repeated patterns that can be exploited on both daily samples and intraday time-scales.  We do not claim that being able to exploit such patterns is necessarily profitable as a commercial enterprise, what we are claiming is that structure does exist in the financial market time-series that is indicative of existence of repeated structures that emerge and change through time, but after reasonable costs can be considered a riskless profit, or at least be considered a signature of the ability to generate systematic profits from patterns in financial time-series data. 

We do not know whether there is a finite state representation of the system that could be used to generate the observed time-series dynamics. We have evidence for non-linear structure in the time-series data, by providing a simplistic algorithm that can exploit structure in time-series data, when it exists, and we know that the algorithm would behave quite differently for random data. To show that this is indicative of some finite state representation would require online state-detection, either via some type of cluster methodology \cite{HGW2016b}, or via some sort of state-space reconstruction algorithm following the methods of deterministic chaos \cite{PCFS1980,CEFG1991,SYC1991}. This paper makes no statement about the existence of a finite and sufficiently stable finite state representation.

The other criticisms could relate to both barriers to entry to reasonably cost effect market access, as well as the scalability of these types of strategies due to stock liquidity. In terms of the prior, many proprietary trading structures within hedge-funds and banks would have very low transaction costs due to bulk trading activities - hence we consider our daily transaction costs of 50bps as onerous but realistic. In terms of the liquidity concerns, we have limited ourselves, in the Johannesburg Stock Exchange data set, to collections of the 10 and 20 most liquid stocks. These stocks can be traded in meaningful volumes. 

We could speculate that it is the buying and selling patterns of large institutional mutual funds, or capital flows of large institutional participants in capital markets, that create key feed-backs which generate persistence in patterns of price dynamics. Realistically there are a variety of potential candidate sources of top-down and bottom-up feedbacks within a system as complex and adaptive as the financial market systems; these could provide various mechanisms that can balance disorder with order in a meta-stable configuration of states over various time-scales \cite{WGHC2014}. We argue that fairly naive computational learning agents can generate wealth within the system without special insights or understanding of the system itself. 

\section{Acknowledgements}

TG would like to thank AIMS South Africa for their support and hospitality at Muizenberg. The authors would like to thank Diane Wilcox for conceptual contributions, Turgay Celik for discussions and ideas relating to algorithm testing, Dieter Hendricks and Raphael Nkomo for various conversations relating the quantitative trading and machine learning for trading. This work was in part supported by NRF grant number 89250. The conclusions herein are due to the authors and the NRF
accepts no liability in this regard.

\appendix
\section{Ticker lists} \label{app:ticker_lists}
\subsection{NYSE}

\begin{tabular}{ | p{4.5cm} |p{1cm} |p{1cm} |}
\hline
 Name & NYSE & NYSE Merged \\
\hline
3M Company & \checkmark & \checkmark \\
Alcoa & \checkmark & \checkmark \\
Altria Group & \checkmark & \checkmark \\
Arco & \checkmark &  \\
Coca Cola & \checkmark & \checkmark \\
Commercial Metals & \checkmark & \checkmark \\
Dow Chemical Company & \checkmark & \checkmark \\
DuPont & \checkmark & \checkmark \\
Eastman Kodak & \checkmark & \checkmark \\
Espey & \checkmark &   \\
Exxon Mobil & \checkmark &  \\
Fischback & \checkmark &  \\
Ford Motor Company & \checkmark & \checkmark \\
Fortune Brands & \checkmark & \checkmark \\
General Electric & \checkmark & \checkmark \\
General Motors & \checkmark & \checkmark \\
Gran Tierra Energy Inc. & \checkmark &  \\
Gulf Oill & \checkmark &  \\
Hewlett-Packard & \checkmark & \checkmark \\
IBM & \checkmark & \checkmark \\
Ingersoll-Rand Plc & \checkmark & \checkmark \\
Iroquois Ltd. & \checkmark &  \\
Johnson \& Johnson & \checkmark & \checkmark \\
Kimberly-Clark Corp & \checkmark & \checkmark \\
Lukens & \checkmark &  \\
Mei Corp. & \checkmark &  \\
Merck \& Company & \checkmark & \checkmark \\
Mobil & \checkmark &  \\
Kin Ark Corp. & \checkmark & \checkmark \\
Pillsbury & \checkmark &  \\
Procter \& Gamble & \checkmark & \checkmark \\
Schlumberger N.V. & \checkmark & \checkmark \\
Sears & \checkmark &  \\
Sherwin-Williams & \checkmark & \checkmark \\
Texaco & \checkmark &  \\
Wyeth & \checkmark & \checkmark \\

\hline
\end{tabular}

%

\subsection{JSE intraday and daily data}
In the following tables $\checkmark^*$ represents stocks that are included in the JSE daily dataset grouping and not the JSE intraday dataset grouping, and $\checkmark^+$ represents stocks that are include in the JSE intraday dataset grouping and not the JSE daily dataset grouping.

\subsubsection{Financials / JSE-FINI (J212)}

\begin{tabular}{ | p{4cm} |c | c | c | c |}
\hline Name & RIC & \multicolumn{3}{c|}{Stock grouping}\\ \hline
Financial (JSE-FINI) & J212 & 10  & 20  & 30  \\
\hline 
Standard Bank Grp. Ltd & SBKJ.J & \checkmark & \checkmark & \checkmark \\
Firstrand Ltd & FSRJ.J & & \checkmark & \checkmark \\
Absa Group Ltd & ASAJ.J  & & $\checkmark^*$ & $\checkmark^*$ \\ 
Old Mutual Plc & OMLJ.J & & \checkmark & \checkmark \\ 
Nedbank Group Ltd & NEDJ.J & &\checkmark & \checkmark\\ 
Sanlam Ltd & SLMJ.J & &\checkmark &\checkmark \\
Investec Plc & INPJ.J & & & \\
RMB Holdings Ltd & RMHJ.J & & & \\ 
Growthpoint Prop Ltd & GRTJ.J & & & \checkmark \\ 
African Bank Inv. Ltd & ABLJ.J & & & \\ 
Capital Shop Cnt. Grp Plc & CSOJ.J & & & \\ 
Reinet Investments Sca & REIJ.J & & & \\
Redefine Properties Ltd & RDFJ.J & & & \\
Discovery Holdings Ltd & DSYJ.J & & & \checkmark \\
Liberty Holdings Ltd & LBHJ.J & & & \\
\hline
\end{tabular} \label{tab:fini}

%

\subsubsection{Resources / JSE-RESI (J201)}

\begin{tabular}{ | p{4cm} |c | c | c | c |}
\hline Name & RIC & \multicolumn{3}{c|}{Stock grouping}\\ \hline
Resources (JSE-RESI) & J201 & 10  & 20  & 30  \\
\hline 
BHP Billiton PLC & BILJ.J & \checkmark & \checkmark & \checkmark \\
Anglo American PLC & AGLJ.J & \checkmark & \checkmark & \checkmark \\
Sasol LTD  & SOLJ.J  & \checkmark & \checkmark & \checkmark \\
Anglo Platinum Ltd  & AMSJ.J & & \checkmark & \checkmark \\ 
Impala Platinum Hld. Ltd & IMPJ.J & \checkmark  & \checkmark & \checkmark \\ 
Anglogold Ashanti Ltd  & ANGJ.J & \checkmark & \checkmark & \checkmark \\ 
Gold Fileds Ltd  & GFIJ.J & & \checkmark & \checkmark \\ 
Exxaro Resources Ltd & EXXJ.J & & & \\
African Rainbow Minerals & ARIJ.J & & & \checkmark \\
Lonmin PLC & LONJ.J & & & \\ 
Harmony G M Co Ltd & HARJ.J & & & \\ 
Assore Ltd & ASRJ.J & & & \\ 
Northam Platinum Ltd &NHMJ.J& & & \\ 
Optimum Coal Hldgs Ltd &OPTJ.J& & & \\ 
Merafe Resources Ltd &MRFJ.J& & & \\ 
Petmin Ltd & PETJ. J & & & \\ 
Wesizwe Platinum Ltd &WEZJ.J& & & \\  
Sentula Mining Ltd &SNUJ.J& & & \\  
DRDGold Ltd &DRDJ.J& & & \\ 
Simmer and Jack Mines &SIMJ.J& & & \\ 
\hline
\end{tabular} \label{tab:resi}

\subsubsection{Industrials/JSE-INDI (J211) }

\begin{tabular}{ | p{4cm} |c | c | c | c |}
\hline Name & RIC & \multicolumn{3}{c|}{Stock grouping}\\ \hline
Industrials (JSE-INDI) & J211 & 10  & 20  & 30  \\
\hline 
SABMiller Plc & SABJ.J& \checkmark & \checkmark & \checkmark \\ 
MTN Group Ltd & MTNJ.J& \checkmark & \checkmark & \checkmark \\ 
Comp. Fin Richemont & CFRJ.J& \checkmark & \checkmark & \checkmark \\ 
Naspers Ltd & NPNnJ.J & \checkmark & \checkmark & \checkmark \\ 
Kumba Iron Ore Ltd & KIOJ.J&  &  & \\
Vodacom Group Ltd & VODJ.J&  &  &  \\
Bidvest Ltd  &BVTJ.J&  & \checkmark & \checkmark \\ 
Shoprite Hldgs Ltd & SHPJ.J&  & $\checkmark^+$ & \checkmark \\
Remgro Ltd  &REMJ.J&  & \checkmark & \checkmark \\
Aspen Pharmacare Hldgs &APNJ.J&  &  & \checkmark \\
Tiger Brands Ltd &TBSJ.J&  &  & \checkmark \\ 
ArcelorMittal SA Ltd &ACLJ.J&  &  &  \\
Steinhoff Inter. Hldgs & SHFJ.J& & \checkmark & \checkmark \\
Truworths International & TRUJ.J & & & \checkmark \\
Mediclinic International Ltd & MDCJ.J & & & \checkmark \\
Massmart Hldgs Ltd & MSMJ.J & & & $\checkmark^+$ \\
Mondi Plc & MNPJ.J & & & \\
Mondi Ltd  & MNDJ.J & & & \\ 
Imperial Hldgs Ltd  & IPLJ.J & & & \checkmark \\ 
Pik n Pay Stores Ltd  & PIKJ.J & & & \\ 
Woolworths Hldgs & WHLJ.J & & & \checkmark \\ 
The Foschini Grp Ltd & TFGJ.J & & & \\ 
Netcare Ltd & NTCJ.J & & & \\
Pretoria Port Cement & PPCJ.J & & & \\ 
Sappi Ltd  & SAPJ.J & & & \\
Telkom SA Ltd  & TKGJ.J & & & \\ 
Aveng Ltd  & AEGJ.J & & & \\
\hline
\end{tabular}\label{tab:indi}

\end{document}